\documentclass[11pt,a4paper]{article}
\pdfoutput=1

\usepackage{jheppub}
\usepackage{graphicx}
\usepackage{epstopdf}
\usepackage[T1]{fontenc}
\usepackage{extarrows}

\usepackage{dcolumn}
\usepackage{bm}
\usepackage{hyperref}
\usepackage{color}
\usepackage[utf8]{inputenc}
\usepackage{morefloats}
\usepackage{subcaption}

\title{Chiral transport in strong fields from holography}
\author[a]{Yanyan Bu,}
\author[b]{Tuna Demircik,}
\author[b]{and Michael Lublinsky}
\affiliation[a]{Department of Physics, Harbin Institute of Technology, Harbin 150001, China}
\affiliation[b]{Department of Physics, Ben-Gurion University of the Negev,
Beer-Sheva 84105, Israel}

\emailAdd{yybu@hit.edu.cn}
\emailAdd{demircik@post.bgu.ac.il}
\emailAdd{lublinm@bgu.ac.il}

\abstract{Anomaly-induced transport phenomena in presence of strong external electromagnetic  fields are  explored within a 4D field theory defined holographically as $U(1)_V\times U(1)_A$ Maxwell-Chern-Simons theory in Schwarzschild-$AdS_5$.
Two complementary studies are reported.
In the first one, we  present results on the Ohmic conductivity, diffusion constant, chiral magnetic conductivity, and additional anomaly-induced transport coefficients
as functions of external e/m fields.
Next, gradient resummation  in a constant background magnetic field is performed.
All-order resummed constitutive relations are parameterised by  four momenta-dependent transport coefficient functions (TCFs).
 A highlight of this part is a thorough study of {\it non-dissipative} chiral magnetic waves (CMW)  in strong magnetic fields.}


\arxivnumber{1903.00896}
\begin{document}
\maketitle

\flushbottom

\allowdisplaybreaks

\section{Introduction and Summary}\label{intro}

\subsection{Motivation}

We continue  exploring   transport phenomena  induced by chiral anomaly in a chiral plasma with both left- and right-handed
$U(1)$ (electrically) charged fermions.
The microscopic theory is defined holographically:
$U(1)_V \times U(1)_A$ Maxwell-Chern-Simons theory in Schwarzschild-$AdS_5$ \cite{Yee:2009vw,Gynther:2010ed}
to be introduced in Section \ref{model}. Transport phenomena for this theory have already been extensively studied by us
\cite{1608.08595,1609.09054,Bu:2018psl,Bu:2018drd},  and by other authors \cite{Yee:2009vw,Gynther:2010ed,Lin:2013sga}. Here we primarily focus on some new results
related to {\it strong} external  e/m fields ($\vec E$ and $ \vec B$).

Realistic plasmas such as  quark-gluon plasma produced in heavy ion collisions or primordial plasma in the early universe are exposed
to strong external e/m fields. Furthermore, the dynamics of these plasmas is governed by magneto-hydrodynamics (MHD) which
generates the  fields dynamically,  frequently resulting in even stronger  fields \cite{Boyarsky:2011uy,Boyarsky:2015faa,Manuel:2015zpa,Huang:2015oca,Kharzeev:2015znc,Skokov:2016yrj}. 
Chiral anomaly is known to modify the
MHD equations, turning them into {\it chiral} MHD. An essential ingredient of any hydrodynamics is
constitutive relations describing plasma medium effects. The constitutive relations for the vector current $J^{\mu}$ and axial current $J_5^{\mu}$ are of the form
\begin{equation} \label{const}
\vec J\ =\ \vec J\;[\rho, \rho_5, T, \vec E, \vec B]; \hspace{3cm}  \vec J_5=\vec J_5\;[\rho, \rho_5, T, \vec E, \vec B],
\end{equation}
where $\rho,\rho_5$ are vector and axial charge densities,  and $T$ is temperature.
The dynamics  of the plasma is governed by the ``conservation laws'' (continuity equations)
\begin{equation}\label{cont eqn}
\partial_{\mu}J^{\mu}=0, \qquad \qquad \partial_{\mu}J_5^{\mu}=12 \kappa \vec{E}\cdot \vec{B}.
\end{equation}
Note that as a result of the chiral anomaly, the global $U(1)_A$ current  is no longer conserved.
$\kappa$ is the chiral anomaly coefficient ($\kappa=eN_c/(24\pi^2)$ for $SU(N_c)$ gauge theory with a massless Dirac fermion in the fundamental representation and $e$ is the electric charge).

The constitutive relations (\ref{const}) should be derived from
the underlying microscopic theory. Yet, it is almost never feasible, even approximately. A great deal of modelling is inevitably  employed in practice, frequently  based  on (truncated) gradient expansion and/or  weak field approximations.  Both approximations, and especially the latter one,  can be inadequate. This can happen in an experimental setup, say,
in chiral materials such as Weyl semimetals, in which e/m fields $\vec E$ and $ \vec B$ can be controlled externally. Alternatively, plasma instabilities  could
generate strong fields dynamically  and thus drive the system  outside the applicability range of the constitutive relations.
In all such cases  the constitutive relations must be revised.
The necessity to properly  define chiral MHD in presence of strong  external e/m fields motivates our study.

In the hydrodynamic limit,
the gradient expansion at each order is fixed by thermodynamics and symmetries, up to a finite number of transport coefficients (TCs).
Diffusion constant, DC conductivity and shear viscosity are the most familiar examples of the lowest order TCs.
However, ``naive'' truncation of the gradient expansion explicitly breaks relativistic invariance and thus leads to serious conceptual problems such as causality violation.
Beyond conceptual issues, truncation of the gradient expansion  results in numerical instabilities rendering the entire framework unreliable.
Causality is restored  when all order gradient terms are included, in a way providing a UV completion to the ``old'' hydrodynamic effective theory. The resummation generalises the concept of TC to transport coefficient functions (TCFs), which are functionals of $\partial_t$ and  $\vec\nabla^2$ (or equivalently functions of frequency $\omega$ and three-momentum squared $\vec q^{\;2}$ in Fourier space). Therefore, TCFs contain information about infinitely many gradients and they extend the applicability of the effective theory
 beyond hydrodynamic limit of small frequency and momenta. When inverse Fourier transformed, the TCFs correspond to memory functions \cite{1502.08044,1511.08789}.  Below,
 an effective theory based on TCFs will be referred to as the \textit{all order resummed} hydrodynamics \cite{Lublinsky:2007mm,0905.4069,1406.7222,1409.3095,1502.08044,1504.01370}.

The goal of the present work is to explore the generic structure of the currents (\ref{const}), beyond the weak field limit
explored by us in \cite{1608.08595,1609.09054,Bu:2018psl,Bu:2018drd}.
Our study will be split into two complementary directions.
In the first one, we compute  various TCs as functions of constant  e/m fields, denoted as $\vec {\bf E},\vec{\bf B}$, up to first order in the gradient expansion.
 In the second part,  we consider  gradient resummation in the presence of a constant external magnetic field $\vec{\bf B}$ only.

\subsection{Summary of the  results: Part I} \label{summary_I}

First we consider the  constitutive relations (\ref{const}) at fixed order in the gradient expansion
\begin{align} \label{JJ5 gradient}
\vec J= \sum_{n=0}^\infty \lambda^n \vec J^{\;[n]},\qquad \qquad \vec J_5= \sum_{n=0}^\infty \lambda^n \vec J_5^{\;[n]},
\end{align}
where $\lambda$ is introduced via the replacement $\partial_\mu \to \lambda \partial_\mu$, and it counts the order in the gradient expansion.

\noindent{\bf Zeroth order ($n=0$)}.   The most general constitutive relations are
\begin{align}
\vec{J}^{\;[0]}&=\sigma_e^0 \vec{\bf E}+ \sigma_\chi^0 \kappa \rho_5 \vec{\bf B} + \delta \sigma_\chi^0 \kappa^2 (\vec{\bf E}\cdot \vec{\bf B})\vec{\bf B}+ \sigma_{\chi H}^0 \kappa^2 \rho \vec{\bf B}\times \vec{\bf E}+ \sigma_{\chi e}^0 \kappa^3 \rho_5 (\vec{\bf B}\cdot \vec{\bf E})\vec{\bf E}, \label{jmu 0th}\\
\vec{J}^{\;[0]}_5&=\sigma_{\chi}^0 \kappa\rho \vec{\bf B}+ \sigma_{\chi H}^0 \kappa^2 \rho_5 \vec{\bf B} \times \vec{\bf E} + \sigma_{\chi e}^0 \kappa^3 \rho(\vec{\bf B}\cdot \vec{\bf E})\vec{\bf E} +\sigma_{s}^0 \kappa^3 (\vec{\bf E}\cdot \vec{\bf B})\vec{\bf B} \times \vec{\bf E}. \label{jmu5 0th}
\end{align}
Here one recognises some familiar terms. In the vector current: the Ohmic conductivity ($\vec J\sim \vec  {\bf E}$),
the \textit{chiral magnetic effect} (CME)
($\vec J\sim \rho_5 \vec {\bf  B}$) \cite{PhysRevD.22.3080,0808.3382,0906.5044}, the chiral Hall effect ($\vec J\sim \vec {\bf  B}\times \vec {\bf E}$) \cite{Pu:2014fva}.
In the axial current one notices  the \textit{chiral separation effect} (CSE) ($\vec J_5\sim \rho \vec {\bf B}$) \cite{Son:2004tq,Metlitski:2005pr}
and the chiral electric separation effect (CESE) ($\vec J_5\sim \rho \vec {\bf E}$) \cite{Huang:2013iia}.

Apart of the last term in $\vec J^{\;[0]}_5$, which predicts separation of chiral charge along the Poynting vector $\vec {\bf S}= \vec {\bf E} \times \vec {\bf B}$,
all the terms in (\ref{jmu 0th}, \ref{jmu5 0th}) have already appeared in the literature, particularly in our previous publications \cite{1608.08595,Bu:2018psl}.
The main novelty of the present work is that we  consider all the TCs in (\ref{jmu 0th}, \ref{jmu5 0th}) as scalar functions of  external e/m fields
\begin{equation}
\sigma_e^0=\sigma_e^0[ {\bf E}^2,\, {\bf B}^2,\,(  {\vec {\bf B}\cdot \vec {\bf E}})^2]; \qquad \qquad
\sigma_\chi^0=\sigma_\chi^0[ {\bf E}^2,\, {\bf B}^2,  \,({\vec {\bf B}\cdot \vec {\bf E}})^2]; \quad  {\rm etc},
\end{equation}
without assuming any weak field approximation, in contrast to what has  been done in the past.

One might prefer an alternative representation of
 (\ref{jmu 0th}, \ref{jmu5 0th}) reflecting apparent anisotropy induced by the external fields.
\begin{align}
J^{\;[0]}_i&=\sigma_e^0 \left(\delta_{ij}- \frac{{\bf B}_i {\bf B}_j}{{\bf B}^2}\right) {\bf E}_j +\sigma_e^{0{\rm L}} \frac{{\bf B}_i {\bf B}_j} {{\bf B}^2} {\bf E}_j+\sigma_\chi^0 \kappa \rho_5 \left(\delta_{ij}- \frac{{\bf E}_i {\bf E}_j}{{\bf E}^2}\right) {\bf B}_j \nonumber\\
&+ \sigma_\chi^{0{\rm L}}\kappa \rho_5 \frac{{\bf E}_i {\bf E}_j}{{\bf E}^2} {\bf B}_j
+ \sigma_{\chi H}^0 \kappa^2 \rho (\vec{\bf B}\times \vec{\bf E})_i, \label{jmu 0th re}\\
J^{\;[0]}_{5i}&=\sigma_\chi^0 \kappa \rho \left(\delta_{ij}- \frac{{\bf E}_i {\bf E}_j}{{\bf E}^2}\right) {\bf B}_j+ \sigma_\chi^{0{\rm L}}\kappa \rho \frac{{\bf E}_i {\bf E}_j}{{\bf E}^2} {\bf B}_j+ \sigma_{\chi H}^0 \kappa^2 \rho_5 (\vec{\bf B}\times \vec{\bf E})_i \nonumber\\
&+\sigma_{s}^0 \kappa^3 (\vec{\bf E}\cdot \vec{\bf B})(\vec{\bf B} \times \vec{\bf E})_i, \label{jmu5 0th re}
\end{align}
with the longitudinal Ohmic and CME conductivities
\begin{align} \label{longitudinal_0th}
\sigma_e^{0{\rm L}}= \sigma_e^0+ \kappa^2 {\bf B}^2 \delta \sigma_\chi^0,\qquad \qquad
\sigma_\chi^{0{\rm L}}= \sigma_\chi^0 + \kappa^2 {\bf E}^2 \sigma_{\chi e}^0.
\end{align}

The constitutive relations (\ref{jmu 0th}, \ref{jmu5 0th}) are  in a sense ``off-shell'' since the charge densities $\rho$ and $\rho_5$
are treated as independent of the three-currents $\vec J$  and $\vec J_5$.
Imposing the continuity equations (\ref{cont eqn}), the constitutive relations (\ref{const}) are put ``on-shell''.
To the leading order in spatial momentum, (\ref{jmu 0th}, \ref{jmu5 0th}) result in a dispersion relation for the chiral plasma,
\begin{equation}\label{dis1}
\omega= \pm \sigma_\chi^0 \kappa \vec{q}\cdot \vec{\bf B}\pm \sigma_{\chi e}^0 \kappa^3 (\vec{\bf B}\cdot \vec{\bf E})\vec{q}\cdot \vec{\bf E} +\sigma_{\chi H}^0 \kappa^2 \vec{q} \cdot(\vec{\bf B} \times \vec{\bf E})+\mathcal{O}(q^2).
\end{equation}
There are three types of gapless modes propagating in the chiral plasma: the chiral magnetic wave (CMW)\footnote{Recently, it was claimed that \cite{Rybalka:2018uzh,Shovkovy:2018tks} if the external e/m fields are promoted into dynamical, the CMW turns into a damped diffusive mode.} \cite{Kharzeev:2010gd},
the chiral electric wave (CEW) \cite{Pu:2014fva} and the chiral Hall density wave (CHDW) \cite{Bu:2018psl,Bu:2018drd}.
The external field dependent TCs $\sigma_\chi^0$,  $\sigma_{\chi e}^0$, $\sigma_{\chi H}^0$ reflect speeds of these modes.
Note that, in contrast to CMW and CEW both propagating in two opposite directions, CHDW is directed along the Poynting vector only.

Our goal is to compute the dependence of the TCs above on the external fields.  The only analytic result that has already been known in the literature is
for  $\sigma_\chi^0[\vec{\bf E}=0]$   \cite{1609.09054,Landsteiner:2014vua,Ammon:2016fru}
\begin{align} \label{sigma_chi^0}
v_\chi\equiv \kappa {\bf B}\sigma_\chi^0= \frac{\Gamma\left[\left(3-\sqrt{1-144\kappa^2 {\bf B}^2 } \right)/4 \right] \Gamma \left[\left(3+\sqrt{1-144\kappa^2 {\bf B}^2 } \right)/4 \right]} {3\kappa{\bf B} \Gamma\left[\left(1-\sqrt{1-144\kappa^2 {\bf B}^2 } \right)/4 \right] \Gamma \left[\left(1+\sqrt{1-144\kappa^2 {\bf B}^2 } \right)/4 \right]},
\end{align}
where $\Gamma[x]$ is the Gamma function and all the units are rescaled by the temperature, $\pi T=1$. Small and large field limits are
\begin{align}
&\sigma_\chi^0\rightarrow6+216(1-2\log2)\kappa^2\mathbf{B}^2+\mathcal{O}(\mathbf{B}^4), \qquad  \text{as} \qquad\kappa\mathbf{B}\rightarrow0, \nonumber\\
&\sigma_\chi^0\rightarrow\frac{1}{\kappa\mathbf{B}}\qquad  \text{as} \qquad\kappa\mathbf{B}\rightarrow\infty. \label{schilimit}
\end{align}
Thus, in the strong magnetic field limit, the speed of CMW $v_{\chi}$ goes to that of light \cite{Kharzeev:2010gd}.

While we were not able to obtain any new analytical insights, all the TCs in (\ref{jmu 0th}, \ref{jmu5 0th}) were computed  numerically and the results are presented in section \ref{sec_a}. Here we quote some asymptotic  behaviours for large-$\kappa \mathbf B$ and large-$\kappa \mathbf E$
\begin{align}
&
\delta\sigma_\chi^0[\kappa \mathbf{B}\rightarrow\infty,\kappa \mathbf{E}=0] \to \frac{3.349}{(\kappa \mathbf{B})^{3/2}}, \qquad \sigma_{\chi H}^0[\kappa \mathbf{B} \rightarrow\infty,\kappa \mathbf{E}=0] \to -\frac{ 1}{(\kappa \mathbf{B})^2},\nonumber \\
& \sigma_{\chi e}^0[\kappa \mathbf{B}\rightarrow\infty,\kappa \mathbf{E}=0] \to \frac{0.977}{(\kappa \mathbf{B})^{3}}, \qquad  \sigma_{s}^0[\kappa \mathbf{B}\rightarrow\infty,\kappa \mathbf{E}=0]\to -\frac{6.751}{(\kappa \mathbf{B})^2}, \label{scaling_0th_E=0}
\end{align}
\begin{align}
& \delta\sigma_\chi^0[\kappa \mathbf{B}=0,\kappa \mathbf{E}\rightarrow\infty]\to -\frac{2.243}{(\kappa \mathbf{E})^{3/2}},\qquad
\sigma_{\chi H}^0[\kappa \mathbf{B}=0,\kappa \mathbf{E}\rightarrow\infty]\to -\frac{1}{(\kappa \mathbf{E})^2}, \nonumber\\
& \sigma_{\chi e}^0[\kappa \mathbf{B}=0,\kappa \mathbf{E}\rightarrow\infty] \to \frac{6.04}{(\kappa \mathbf{E})^{2}},\qquad\quad \sigma_{s}^0[\kappa \mathbf{B}=0,\kappa \mathbf{E}\rightarrow\infty] \to -\frac{3.069}{(\kappa \mathbf{E})^{5/2}}. \label{scaling_0th_B=0}
\end{align}
At large $\kappa \bf{E}$, $\sigma_\chi^0[\kappa {\bf B}=0]$ decays much faster than any other TCs, and asymptotically does not scale as a power function of $\kappa \bf{E}$.
$\sigma_e^0$ has certain asymptotic behaviour at large values of e/m fields only when $\kappa \mathbf{E}=\kappa \mathbf{B}$
\begin{equation}\label{scalese}
\sigma_e^0[\kappa \mathbf{B}=\kappa \mathbf{E}\rightarrow \infty]\simeq1+ 3.09 (\kappa \mathbf{E}\cdot\kappa\mathbf{B})^{0.27}.
\end{equation}

\noindent{\bf First order ($n=1$)}. We consider the cases of either ${\bf E}=0$ or ${\bf B}=0$ separately:
\begin{align}
\hspace*{-3cm}{\bf E}=0: \qquad
\vec J^{\;[1]}&=-\mathcal{D}_0\vec\nabla\rho+\tau_{\bar\chi} \kappa \partial_t \rho_5 \vec{\bf B} + \mathcal{D}_B^0\kappa^2 (\vec{\bf B}\cdot \vec{\nabla}\rho) \vec{\bf B}, \label{1st J_E=0} \\
\vec J_5^{\;[1]}&=-\mathcal{D}_0\vec\nabla\rho_5+\tau_{\bar\chi} \kappa \partial_t \rho \vec{\bf B} +\mathcal{D}_B^0 \kappa^2 (\vec{\bf B}\cdot \vec{\nabla}\rho_5) \vec{\bf B}. \label{1st J5_E=0}\\
\hspace*{-3cm} {\bf B}=0:  \qquad
\vec{J}^{\;[1]}&=-\mathcal{D}_0 \vec{\nabla}\rho + \sigma_{a\chi H}^0 \kappa \vec{\bf E} \times \vec{\nabla} \rho_5 +\mathcal{D}_E^0\kappa^2(\vec{\bf E}\cdot \vec\nabla\rho) \vec{\bf E}, \label{1st J_B=0}\\
\vec{J}^{\;[1]}_5&=-\mathcal{D}_0 \vec\nabla\rho_5 +\sigma_{a\chi H}^0 \kappa \vec{\bf E} \times \vec{\nabla} \rho +\mathcal{D}_E^0\kappa^2(\vec{\bf E}\cdot \vec\nabla \rho_5) \vec{\bf E}. \label{1st J5_B=0}
\end{align}

In \cite{1609.09054,Bu:2018psl}, the diffusion constant $\mathcal{D}_0$ was shown to receive {\it negative} (perturbatively small)  ${\bf E}^2$- and ${\bf B}^2$-corrections induced by the chiral anomaly. Now, we are able to study this effect for arbitrary ${\bf E}$ and ${\bf B}$ fields.
Particularly, we find that $\mathcal{D}_0$ vanishes at asymptotically large e/m fields,
see Figures \ref{grad_fig4} and \ref{grad_fig5}.

The TC $\tau_{\bar\chi}$ is the relaxation time for  CME/CSE conductivity $\sigma_\chi^0$, see (\ref{jmu 0th}, \ref{jmu5 0th}).
The anomalous chiral Hall conductivity $\sigma_{a\chi H}$ \cite{1603.03442,Bu:2018psl}  depends on the external electric field ${\bf E}$.
$\mathcal{D}_B^0$ and $\mathcal{D}_E$ are  new  TCs.  Both contribute  to the longitudinal parts of the diffusion tensor,
see (\ref{longDB}, \ref{longDE}).

The first order TCs give rise to the decay rates of the chiral plasma modes  \eqref{dis1}:
\begin{equation}\label{dis2}
\omega= \pm \sigma_\chi^0 \kappa \vec{\bf B} \cdot \vec q- i\mathcal{D}_0 q^2- i (\tau_{\bar{\chi}}\sigma_\chi^0-\mathcal{D}_B^0) \kappa^2 (\vec{\bf B}\cdot \vec q)^2+ \mathcal{O}(q^3),\qquad {\rm as} \quad {\bf E}=0.
\end{equation}
\begin{equation}\label{dis3}
\omega= - i\mathcal{D}_0 q^2+ i \mathcal{D}_E^0 \kappa^2 (\vec{\bf E}\cdot \vec q)^2+ \mathcal{O}(q^3),\qquad {\rm as} \quad {\bf B}=0.
\end{equation}
When ${\bf B}=0$ there is no propagating mode. At  ${\bf E }=0$, CMW is a propagating dissipative density wave.  One may ask the question whether
at some external magnetic field this mode becomes fully non-dissipative (real $\omega$). In our previous publication \cite{Bu:2018drd}, we were able to find such a mode in a weak external
magnetic field limit and for finite momenta (beyond $n=1$ approximation).  It is obviously interesting to explore the effect more rigorously, and
confirm the finding of \cite{Bu:2018drd}  beyond the weak field approximation. This is one of the subjects of our second study.

All TCs in (\ref{1st J_E=0}, \ref{1st J5_E=0}, \ref{1st J_B=0}, \ref{1st J5_B=0}) are computed numerically, see section \ref{sec_b} and \ref{sec_c}.
Here, we present their large-$\kappa \mathbf B$ and large-$\kappa \mathbf E$ behavior:
\begin{align}
& \mathcal D_0[\kappa \mathbf{B}\rightarrow\infty,\kappa \mathbf{E}=0]\to \frac{0.083}{(\kappa \mathbf{B})},\qquad \tau_{\bar \chi}[\kappa \mathbf{B}\rightarrow\infty,\kappa \mathbf{E}=0]\to -\frac{0.36}{(\kappa \mathbf{B})^{3/2}},\nonumber\\
&\mathcal D_B^0[\kappa \mathbf{B}\rightarrow\infty,\kappa \mathbf{E}=0]\to -\frac{0.269}{(\kappa \mathbf{B})^{5/2}}. \label{scaling_1st_E=0}
\end{align}
\begin{align}
\sigma_{a \chi H}^0[\kappa \mathbf{B}=0,\kappa \mathbf{E}\rightarrow\infty] \to -\frac{0.141}{(\kappa \mathbf{E})^2}, \quad
\mathcal D_E^0[\kappa \mathbf{B}=0,\kappa \mathbf{E}\rightarrow\infty] \to -\frac{0.298}{(\kappa \mathbf{E})^{2}}.  \label{scaling_1st_B=0}
\end{align}
Just like $\sigma_\chi^0[\kappa {\bf B}=0]$, at large-$\kappa \bf{E}$, $\mathcal D_0[\kappa {\bf B}=0]$ decays much faster than other TCs and asymptotically does not scale as a power function of $\kappa \bf{E}$.

%

\subsection{Summary of the  results: Part II} \label{summary_II}

In this part,  we perform the gradient resummation taking into account inhomogeneity and time-dependence of the charge densities $\rho,\rho_5$.
That is, the constitutive relations will be constructed to all orders in $\lambda$.
The resummation technique was developed in \cite{Lublinsky:2007mm,0905.4069,1406.7222,1409.3095,1502.08044,1504.01370}. To this goal we
neglect the terms {\it nonlinear} in $\rho,\rho_5$.
Consequently, the currents are parameterised by four TCFs (only the case $\vec{\bf E}=0$ will be considered here),
\begin{align}
&\vec J= -\mathcal{D}\vec\nabla\rho+ \mathcal{D}_B\kappa^2\vec{\bf B}(\vec{\bf B}\cdot \vec\nabla\rho)+ \bar{\sigma}_{\bar\chi} \kappa \vec{\bf B}\rho_5 + \mathcal{D}_\chi \kappa (\vec{\bf B}\cdot \vec\nabla)\vec\nabla \rho_5, \label{resumjj5}\\
&\vec J_5= -\mathcal{D}\vec\nabla\rho_5+ \mathcal{D}_B\kappa^2\vec{\bf B}(\vec{\bf B} \cdot \vec\nabla\rho_5)+ \bar{\sigma}_{\bar\chi} \kappa \vec{\bf B}\rho + \mathcal{D}_\chi \kappa (\vec{\bf B}\cdot \vec\nabla)\vec\nabla \rho. \label{resumjj5b}
\end{align}
The TCFs $\mathcal{D}[\omega,q^2]$, $\bar{\sigma}_{\bar{\chi}}[\omega,q^2]$ and $\mathcal{D}_\chi[\omega,q^2]$ were introduced in
\cite{1511.08789,1608.08595,Bu:2018drd}. Here they are promoted into
$\vec{\bf B}$-dependent,  $\mathcal{D}[\omega,q^2; {\bf B}^2,\, (\vec q \cdot\vec{\bf B})^2]$, etc. $\mathcal{D}_B$ is a new TCF,  and the relevant TC appeared in \cite{Bu:2018psl} at  finite (third) order only.
In section \ref{resummation}, these TCFs are calculated numerically for generic values of frequency/momenta and magnetic field.

Beyond some critical value of $\kappa{\bf B}\gtrsim0.5$, these TCFs display singular behavior at certain values of \textit{real} $\omega$,  identified with
quasi-normal modes (QNMs). These QNMs become real at large  $\bf B$  \cite{Ammon:2016fru,Ammon:2017ded}, and their   $\bf B$ dependence is that of a Landau level.
These  real modes lead  to a phenomenon of \textit{anomalous resonance}  \cite{Haack:2018ztx}.

The resummed constitutive relations (\ref{resumjj5}, \ref{resumjj5b}) give rise to exact  dispersion relations  for the CMW, beyond small frequency/momenta approximations of
(\ref{dis1}, \ref{dis2}, \ref{dis3}),
\begin{align}\label{disprel}
\omega= \pm \left(\bar{\sigma}_{\bar{\chi}}- q^2 \mathcal{D}_\chi \right)\kappa \vec{\bf B} \cdot \vec q-i\left(q^2\mathcal{D}- \mathcal{D}_B (\kappa\vec{\bf B} \cdot \vec q)^2 \right).
\end{align}
Since the TCFs are complex functions of the frequency/momenta,  (\ref{disprel}) is expected to  have infinitely many  solutions, including many gapped modes.
In section \ref{resummation}, we will demonstrate that there are  purely real, and thus non-dissipative  solutions to (\ref{disprel}),
similar to the ones discovered by us in \cite{Bu:2018drd} based on  weak magnetic field analysis.

The rest of this paper is structured as follows. Section \ref{model} presents the holographic setup.
Section \ref{expansion} introduces  calculational details for Part I  and displays numerical results for the TCs mentioned above.
Section \ref{resummation} presents the results for Part II. Section \ref{conclusion} contains  concluding remarks.

\section{Holographic setup: $U(1)_V\times U(1)_A$}\label{model}

The holographic model is a Maxwell-Chern-Simons theory with two $U(1)$ fields in the Schwarzschild-$AdS_5$.
A more detailed presentation of the model could be found
in \cite{1608.08595,1609.09054,Bu:2018psl,Bu:2018drd,Yee:2009vw,Gynther:2010ed}.
The bulk action is
\begin{equation}
S=\int d^5x \sqrt{-g}\mathcal{L}+S_{\textrm{c.t.}},
\end{equation}
where
\begin{equation}\label{LPVA}
\begin{split}
\mathcal{L}=&-\frac{1}{4} (F^V)_{MN} (F^V)^{MN}-\frac{1}{4} (F^a)_{MN} (F^a)^{MN} +\frac{\kappa\,\epsilon^{MNPQR}}{2\sqrt{-g}}\\
&\times\left[3 A_M (F^V)_{NP} (F^V)_{QR} + A_M (F^a)_{NP}(F^a)_{QR}\right],
\end{split}
\end{equation}
and the counter-term action $S_{\textrm{c.t.}}$ is
\begin{equation}\label{ct VA}
S_{\textrm{c.t.}}=\frac{1}{4}\log r \int d^4x \sqrt{-\gamma}\left[(F^V)_{\mu\nu} (F^V)^{\mu\nu} +(F^a)_{\mu\nu}(F^a)^{\mu\nu}\right].
\end{equation}
Above, $(F^V)_{MN}$ and $(F^a)_{MN}$ denote the field strengths for the vector $V$ and axial $A$ gauge fields in the bulk, respectively. $\kappa$ is
Chern-Simons coupling.  The terms proportional to $\kappa$ mimic chiral anomaly of the boundary field theory.

In the ingoing Eddington-Finkelstein coordinate, the line element of the metric of Schwarzschild-$AdS_5$ is
\begin{equation}
ds^2=g_{_{MN}}dx^Mdx^N=2dtdr-r^2f(r)dt^2+r^2\delta_{ij}dx^idx^j,
\end{equation}
where the blackening factor is $f(r)=1-1/r^4$. The Hawking temperature, identified as temperature of the boundary theory, is normalised to $\pi T=1$.

The bulk equations of motion are
\begin{equation}\label{EV}
\textrm{EV}^M\equiv \nabla_N(F^V)^{NM}+\frac{3\kappa  \epsilon^{MNPQR}} {\sqrt{-g}} (F^a)_{NP} (F^V)_{QR}\,=\,0,
\end{equation}
\begin{equation}\label{EA}
\textrm{EA}^M\equiv \nabla_N(F^a)^{NM} +\frac{3\kappa \epsilon^{MNPQR}} {2\sqrt{-g}} \left[(F^V)_{NP} (F^V)_{QR}+  (F^a)_{NP} (F^a)_{QR}\right]\,=\,0,
\end{equation}
where the radial and boundary components of (\ref{EV}, \ref{EA}) correspond to dynamical and constraint equations, respectively. The boundary currents read
\begin{equation} \label{current definition}
J^\mu\equiv \lim_{r\to\infty}\frac{\delta S}{\delta V_\mu},~~~~~~~~~~~~~
J^\mu_5\equiv \lim_{r\to\infty}\frac{\delta S}{\delta A_\mu}.
\end{equation}
The radial gauge $V_r=A_r=0$ is imposed.
The  boundary currents (\ref{current definition}) can be determined by solving the dynamical equations only,  leaving the constraints aside.
The dynamical equations are sufficient to  fix all the TCs/TCFs while
the constraints translate into the continuity equations for the boundary theory currents (\ref{cont eqn}).
We use the following ansatz for the bulk gauge fields \cite{1511.08789,1608.08595},
%
\begin{equation} \label{corrections}
\begin{split}
V_\mu(r,x_\alpha)=\mathcal{V}_\mu(x_\alpha)-\frac{\rho(x_\alpha)}{2r^2}\delta_{\mu t}+ \mathbb{V}_\mu(r,x_\alpha),~~
A_\mu(r,x_\alpha)=
-\frac{\rho_{_5}(x_\alpha)}{2r^2}\delta_{\mu t} + \mathbb{A}_\mu(r,x_\alpha).
\end{split}
\end{equation}
Here $\mathcal{V}_\mu$ is a gauge potential of the external e/m fields $\vec {E}$ and $\vec{B}$,
\begin{equation}
E_i=\mathcal{F}_{it}^V=\partial_i\mathcal{V}_t-\partial_t \mathcal{V}_i,~~~~~~~~~~~~ B_i=\frac{1}{2}\epsilon_{ijk}\mathcal{F}_{jk}^V=\epsilon_{ijk}\partial_{j}\mathcal{V}_k.
\end{equation}
Both $\mathcal{V}_\mu$  and $\rho,\,\rho_5$ are assumed to be known functions of the boundary coordinates, while the dynamical equations of motion are solved for
$\mathbb{V}_\mu,\mathbb{A}_\mu$ as functionals of $\mathcal{V}_\mu$  and $\rho,\,\rho_5$.
Boundary conditions are classified into three types. First, $\mathbb{V}_\mu$ and $\mathbb{A}_\mu$ are regular over the domain $r\in [1,\infty)$. Second, at the  boundary $r=\infty$, we require
\begin{equation}\label{AdS constraint}
\mathbb{V}_\mu\to 0,~~~~~~\mathbb{A}_\mu \to 0~~~~~~~\textrm{as}~~~~~~r\to \infty.
\end{equation}
Additional integration constants will be fixed by employing the Landau frame convention for the boundary currents
\begin{equation}\label{Landau frame}
J^t=\rho(x_\alpha),~~~~~~~~~~~J^t_5=\rho_{_5}(x_\alpha).
\end{equation}

According to the holographic dictionary,  the boundary currents are determined in terms  of near boundary ($r=\infty$) pre-asymptotic expansion of the bulk gauge fields:
\begin{equation}\label{bdry currents}
\begin{split}
J^{\mu}	=\eta^{\mu\nu}(2V_{\nu}^{(2)}+2V^{\textrm{L}}_{\nu}+\eta^{\sigma t} \partial_{\sigma} \mathcal{F}_{t\nu}^V),\qquad \qquad
J_{5}^{\mu}= \eta^{\mu\nu}2A_{\nu}^{(2)},
\end{split}
\end{equation}
where $4V_\mu^{\textrm{L}}=\partial^\nu \mathcal{F}_{\mu\nu}^V$. $V_\mu^{(2)}$ and $A_\mu^{(2)}$ are the coefficients of $r^{-2}$ terms  in the near boundary expansion of  the bulk vector $V_\mu$ and axial $A_\mu$ gauge fields, respectively.

Thus, our program boils down to integrating the dynamical bulk equations, from the horizon to the boundary,  for the bulk gauge fields and determining their
near boundary asymptotic behaviour encoded in the coefficients $V_\mu^{(2)}$ and $A_\mu^{(2)}$. This has to be done for  our specific setup of
constant e/m fields corresponding to linear potential  $\mathcal{V}$ of arbitrary strength.

\section{Part I: Gradient expansion in external e/m fields} \label{expansion}

In this section we study the non-perturbative $\vec{\bf E}$-, $\vec{\bf B}$-dependencies of the TCs introduced in Section \ref{summary_I}, (up to first order in the gradient expansion).
The background fields $\vec{\bf E}$ and $\vec{\bf B}$ are treated as {\it zeroth} order in the gradient expansion, as opposed to our previous studies
\cite{1608.08595,Bu:2018drd}.
 Introducing $\lambda$ as a gradient expansion parameter  (by $\partial_\mu\to \lambda \partial_\mu$),
the bulk fields  $\mathbb{V}_\mu$ and $\mathbb{A}_\mu$  are expandable in powers of $\lambda$,
\begin{equation}
\mathbb{V}_\mu =\sum_{n=0}^\infty \lambda^n \mathbb{V}_\mu^{[n]},\qquad \qquad \qquad \mathbb{A}_\mu = \sum_{n=0}^\infty \lambda^n \mathbb{A}_\mu^{[n]}.
\end{equation}
Hence the currents $\vec J$ and $\vec J_5$ are expanded in $\lambda$ too, see (\ref{JJ5 gradient}).
In what follows, we  compute $\vec J^{\;[0]}$, $\vec J_5^{\;[0]}$, $\vec J^{\;[1]}$ and $\vec J^{\;[1]}_5$.

\subsection{ Constitutive relations at zeroth order}\label{sec_a}

The dynamical equations for $\mathbb{V}_t^{[0]}$ and $\mathbb{A}_t^{[0]}$ are
\begin{equation}\label{Vt0_1}
0=\partial_r\left(r^3\partial_r \mathbb{V}_t^{[0]}\right)+12\kappa \partial_r \mathbb{A}_k^{[0]} {\bf B}_k \Longrightarrow r^3 \partial_r\mathbb{V}_t^{[0]}+12\kappa \mathbb{A}_k^{[0]} {\bf B}_k =0,
\end{equation}
\begin{equation}\label{At0_1}
0=\partial_r\left(r^3\partial_r \mathbb{A}_t^{[0]}\right)+12\kappa \partial_r \mathbb{V}_k^{[0]} {\bf B}_k\Longrightarrow r^3 \partial_r\mathbb{A}_t^{[0]} +12\kappa \mathbb{V}_k^{[0]} {\bf B}_k =0,
\end{equation}
where the Landau frame convention (\ref{Landau frame}) has been used to fix one integration constant in both $\mathbb{V}_t^{[0]}$ and $\mathbb{A}_t^{[0]}$. The dynamical equations for the remaining components $\mathbb{V}_i^{[0]}$ and $\mathbb{A}_i^{[0]}$ read
\begin{equation}
\begin{split}
0=&(r^5-r)\partial_r^2\mathbb{V}_i^{[0]}+(3r^4+1) \partial_r \mathbb{V}_i^{[0]} -r^2
{\bf E}_i +\frac{12\kappa}{r}{\bf B}_i \left(\rho_5- 12\kappa \mathbb{V}_k^{[0]} {\bf B}_k \right) \\
&+12\kappa r^2\epsilon^{ijk} \partial_r \mathbb{A}_j^{[0]} {\bf E}_k,
\end{split}
\end{equation}
\begin{equation}
\begin{split}
0=(r^5-r)\partial_r^2\mathbb{A}_i^{[0]}+(3r^4+1) \partial_r \mathbb{A}_i^{[0]} + \frac{12\kappa}{r}{\bf B}_i \left(\rho- 12\kappa \mathbb{A}_k^{[0]} {\bf B}_k \right) +12\kappa r^2\epsilon^{ijk} \partial_r \mathbb{V}_j^{[0]} {\bf E}_k.
\end{split}
\end{equation}
Solutions for $\mathbb{V}_i^{[0]}$ and $\mathbb{A}_i^{[0]}$ take the form:
\begin{equation} \label{sol_Vi}
\mathbb{V}_i^{[0]}= C_1^{(0)} {\bf E}_i + C_2^{(0)} \kappa \rho_5 {\bf B}_i+ C_3^{(0)} \kappa^2 (\vec{\bf E} \cdot \vec{\bf B}) {\bf B}_i +C_4^{(0)} \kappa^2 \rho (\vec{\bf B} \times \vec{\bf E})_i + C_5^{(0)} \kappa^3\rho_5 (\vec{\bf B}\cdot \vec{\bf E}){\bf E}_i,
\end{equation}
\begin{equation} \label{sol_Ai}
\mathbb{A}_i^{[0]}= C_2^{(0)} \kappa \rho {\bf B}_i +C_4^{(0)} \kappa^2 \rho_5 (\vec{\bf B} \times \vec{\bf E})_i + C_5^{(0)} \kappa^3\rho (\vec{\bf B}\cdot \vec{\bf E}){\bf E}_i+ C_6^{(0)} \kappa^3 (\vec{\bf B}\cdot \vec{\bf E})(\vec{\bf B}\times \vec{\bf E})_i,
\end{equation}
where the decomposition coefficients $C_i^{(0)}(i=1-6)$ are functions of $r$ only, satisfying a system of coupled ordinary differential equations (ODEs). These ODEs could be grouped into two partially decoupled sub-sectors:\\
$\{C_1^{(0)},C_3^{(0)},C_6^{(0)}\}$:
\begin{equation}\label{c10}
0= (r^5-r)\partial_r^2C_1^{(0)} +(3r^4+1)\partial_r C_1^{(0)}-r^2 +12\kappa^4 r^2 (\vec{\bf E} \cdot \vec{\bf B})^2 \partial_r { C}_6^{(0)},
\end{equation}
\begin{equation}
0=(r^5-r)\partial_r^2C_3^{(0)} +(3r^4+1)\partial_r C_3^{(0)}-\frac{144}{r} (C_1^{(0)}+\kappa^2 {\bf B}^2 C_3^{(0)})-12r^2\kappa^2 {\bf E}^2 \partial_r {C}_6^{(0)},
\end{equation}
\begin{equation}
0=(r^5-r)\partial_r^2 C_6^{(0)} +(3r^4+1) \partial_r C_6^{(0)} + 12r^2 \partial_r C_3^{(0)}.
\end{equation}
$\{C_2^{(0)},C_4^{(0)},C_5^{(0)}\}$:
\begin{equation}
0= (r^5-r)\partial_r^2 C_2^{(0)} +(3r^4+1) \partial_r C_2^{(0)} +\frac{12}{r}-\frac{144} {r} \left(\kappa^2{\bf B}^2 C_2^{(0)}+\kappa^4(\vec{\bf E}\cdot \vec{\bf B})^2 C_5^{(0)} \right) - 12 r^2\kappa^2 {\bf E}^2 \partial_r C_4^{(0)},
\end{equation}
\begin{equation}
0= (r^5-r)\partial_r^2 C_4^{(0)} +(3r^4+1) \partial_r C_4^{(0)} + 12r^2 \partial_r C_2^{(0)},
\end{equation}
\begin{equation}\label{c50}
0=(r^5-r)\partial_r C_5^{(0)} +(3r^4+1)\partial_rC_5^{(0)} +12r^2 \partial_r C_4^{(0)}.
\end{equation}
%
%
When the solutions (\ref{sol_Vi}, \ref{sol_Ai}) are substituted into holographic expression (\ref{bdry currents}) for the currents, one obtains the
zeroth order constitutive relations (\ref{jmu 0th}, \ref{jmu5 0th}) with the TCs given by the near boundary expansion of the decomposition coefficients $C_i^{(0)}$
\begin{equation} \label{ci0 bdry}
\sigma_e^0=2c_1^{(0)},\quad \sigma_\chi^0=2c_2^{(0)},\quad \delta\sigma_\chi^0= 2c_3^{(0)}, \quad \sigma_{\chi H}^0=2c_4^{(0)}, \quad \sigma_{\chi e}^0= 2c_5^{(0)}, \quad \sigma_{s}^0= 2c_6^{(0)}.
\end{equation}
Here $c_i^{(0)}$ in \eqref{ci0 bdry} are the coefficients of $1/r^2$ behaviour of $C_i^{(0)}$ near the boundary.

When $\vec{\bf E}=0$, $C_2^{(0)}$ was found analytically in \cite{1609.09054,Landsteiner:2014vua,Ammon:2016fru}, yielding an analytic expression (\ref{sigma_chi^0}) for $\sigma_\chi^0$.
%
We are able to compute the remaining TCs in (\ref{jmu 0th},\ref{jmu5 0th}) numerically only.

Prior to demonstrating new results for the field dependent TCs, we quote  their values at vanishing e/m fields
(see \cite{Bu:2018psl} and the references therein):
\begin{align}
&\sigma_e^0({\bf E}={\bf B}=0)=1,\qquad\qquad\qquad\;\qquad \sigma_\chi^0({\bf E}={\bf B}=0)= 6, \nonumber\\
& \delta\sigma_\chi^0({\bf E}={\bf B}=0)= 18(\pi-2\log2), \qquad
\sigma_{\chi H}^0({\bf E}={\bf B}=0)=-36\log2, \nonumber\\
&\sigma_{\chi e}^0({\bf E}={\bf B} =0) = 9\pi^2,\qquad\qquad\qquad\;\;\; \sigma_s^0({\bf E}={\bf B}=0)=12\left(\frac{3}{8}\pi^2-18 \mathcal{C} \right),
\end{align}
where $\mathcal{C}\approx 0.915966$ is the Catalan's constant.

\begin{figure}
    \centering
    \begin{subfigure}[h]{0.485\textwidth}
        \includegraphics[width=\textwidth]{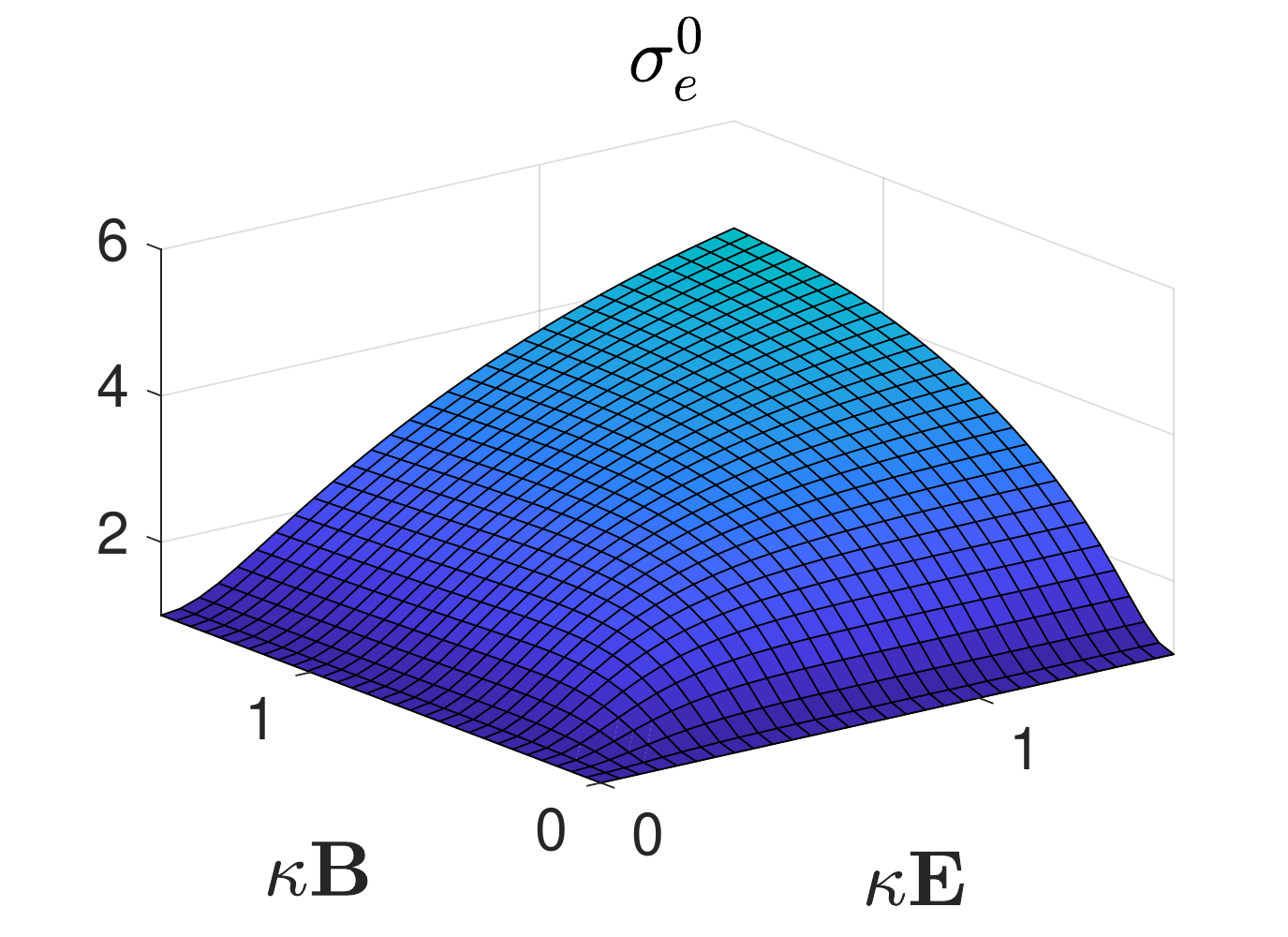}
        \caption{}
        \label{grad_fig1_a}
    \end{subfigure}
    ~ 
      \begin{subfigure}[h]{0.485\textwidth}
        \includegraphics[width=\textwidth]{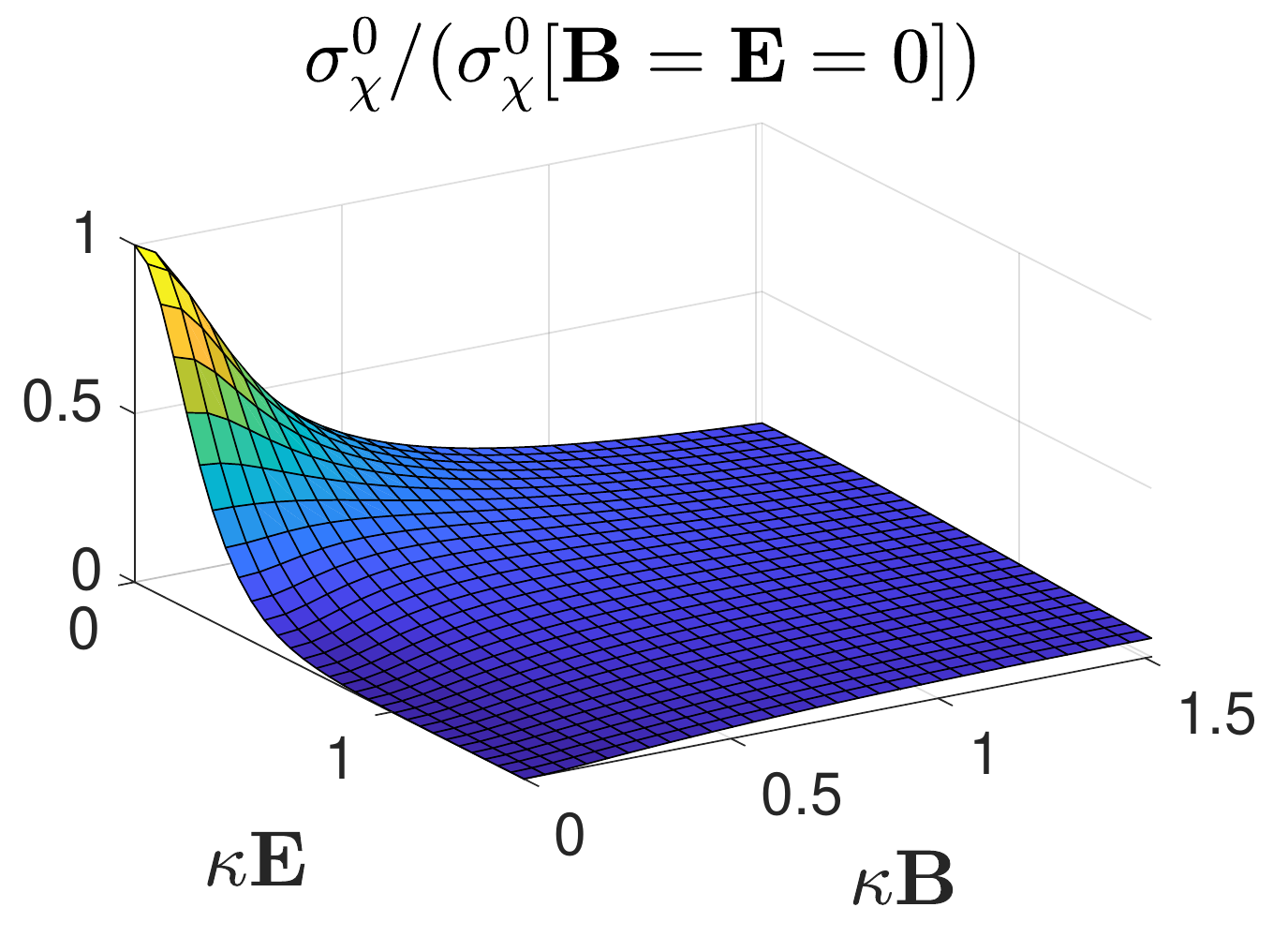}
        \caption{}
        \label{grad_fig1_b}
    \end{subfigure}
    ~ 
      \begin{subfigure}[h]{0.485\textwidth}
        \includegraphics[width=\textwidth]{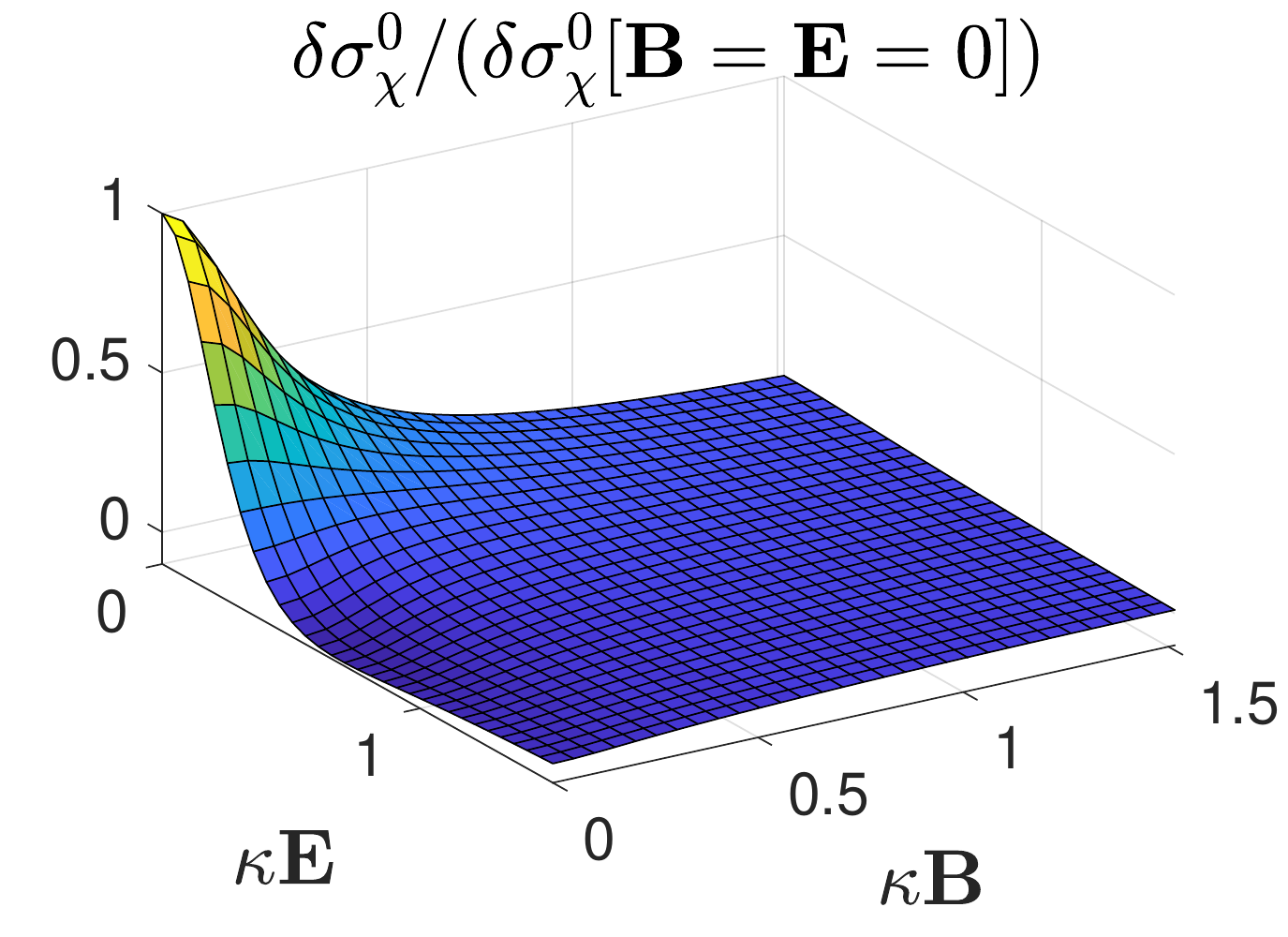}
        \caption{}
        \label{grad_fig1_c}
    \end{subfigure}
    ~ 
    \begin{subfigure}[h]{0.485\textwidth}
        \includegraphics[width=\textwidth]{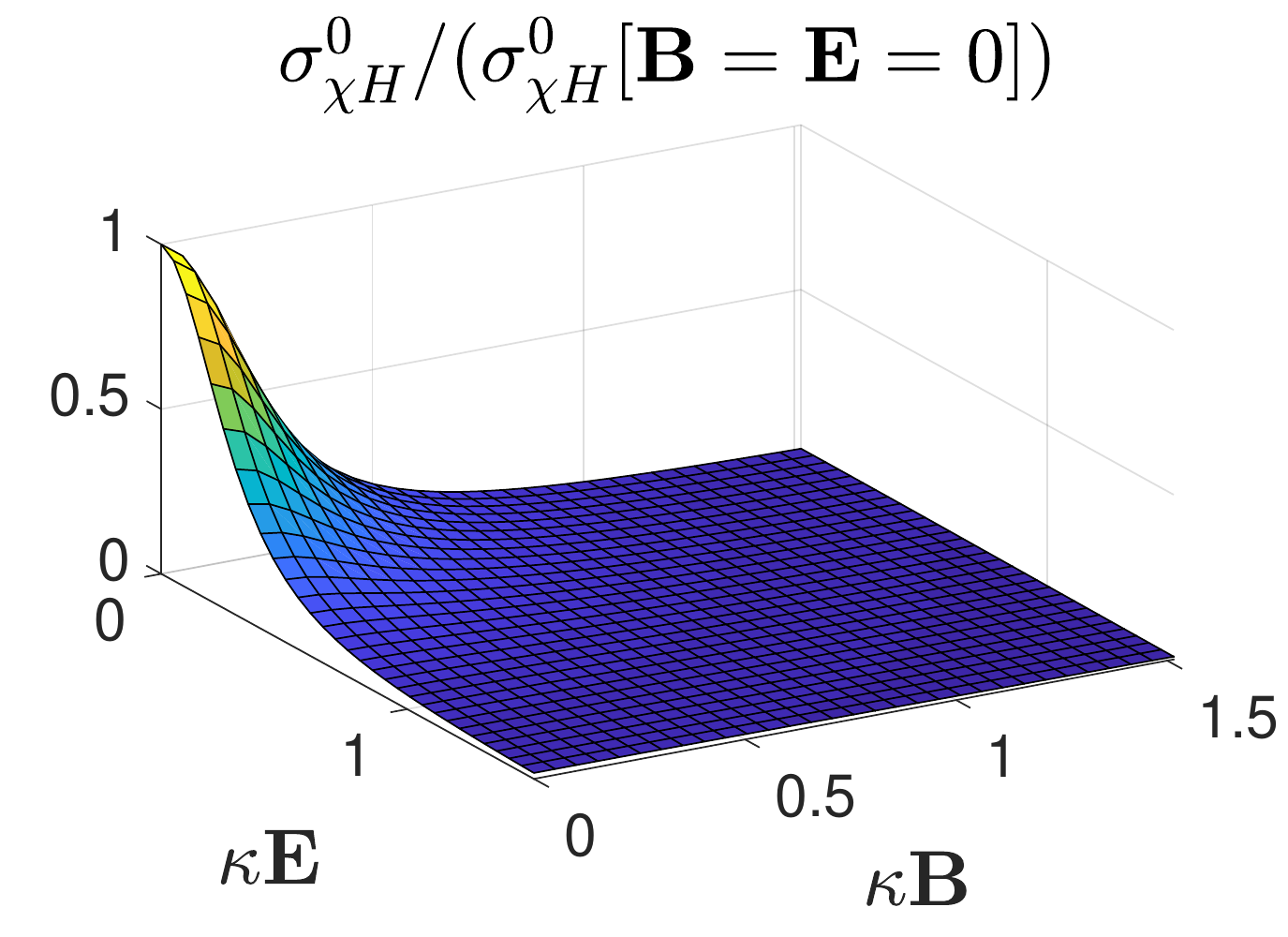}
        \caption{}
        \label{grad_fig1_d}
    \end{subfigure}
    ~ 
        \begin{subfigure}[h]{0.485\textwidth}
        \includegraphics[width=\textwidth]{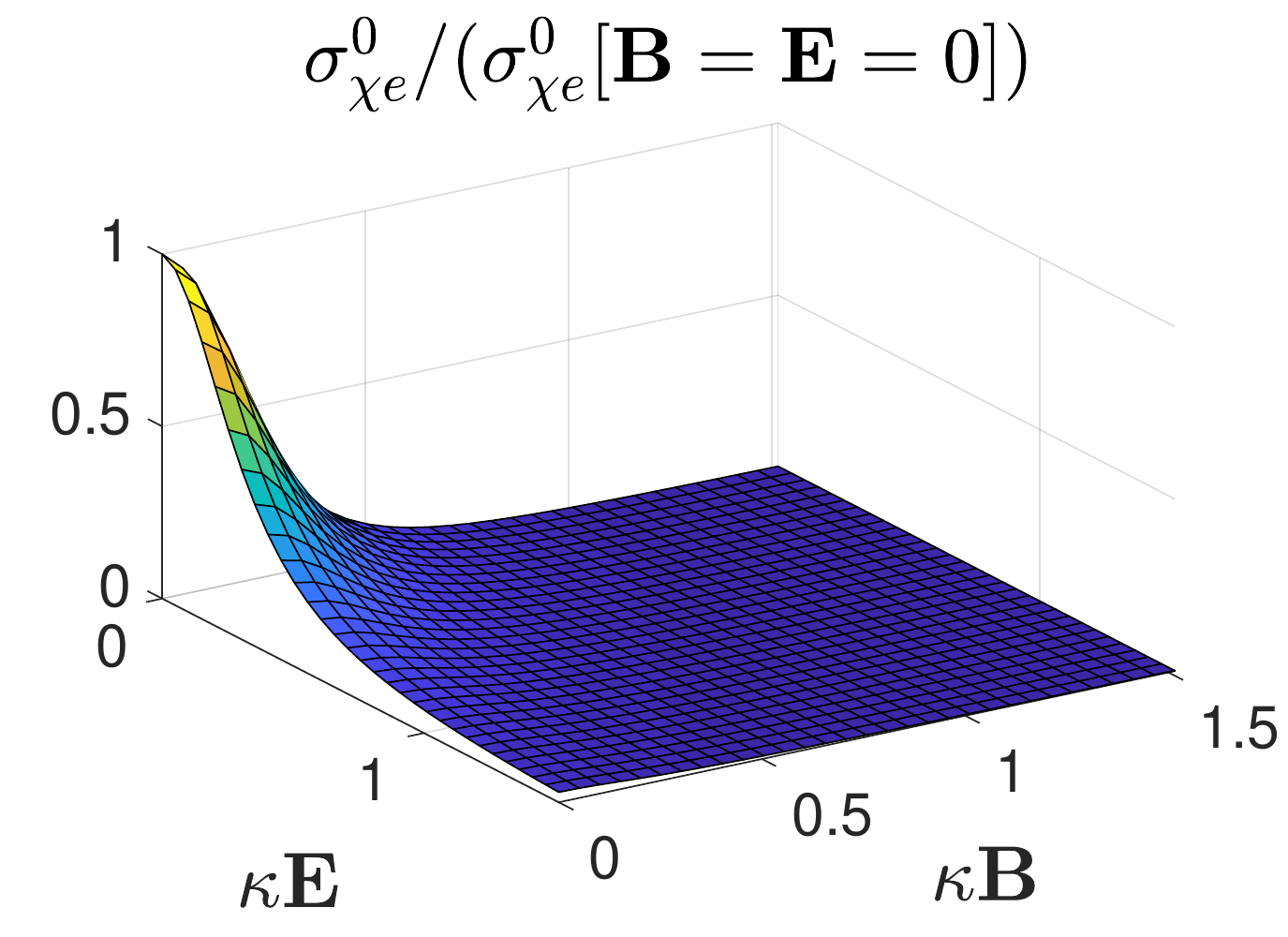}
        \caption{}
        \label{grad_fig1_e}
    \end{subfigure}
        \begin{subfigure}[h]{0.485\textwidth}
        \includegraphics[width=\textwidth]{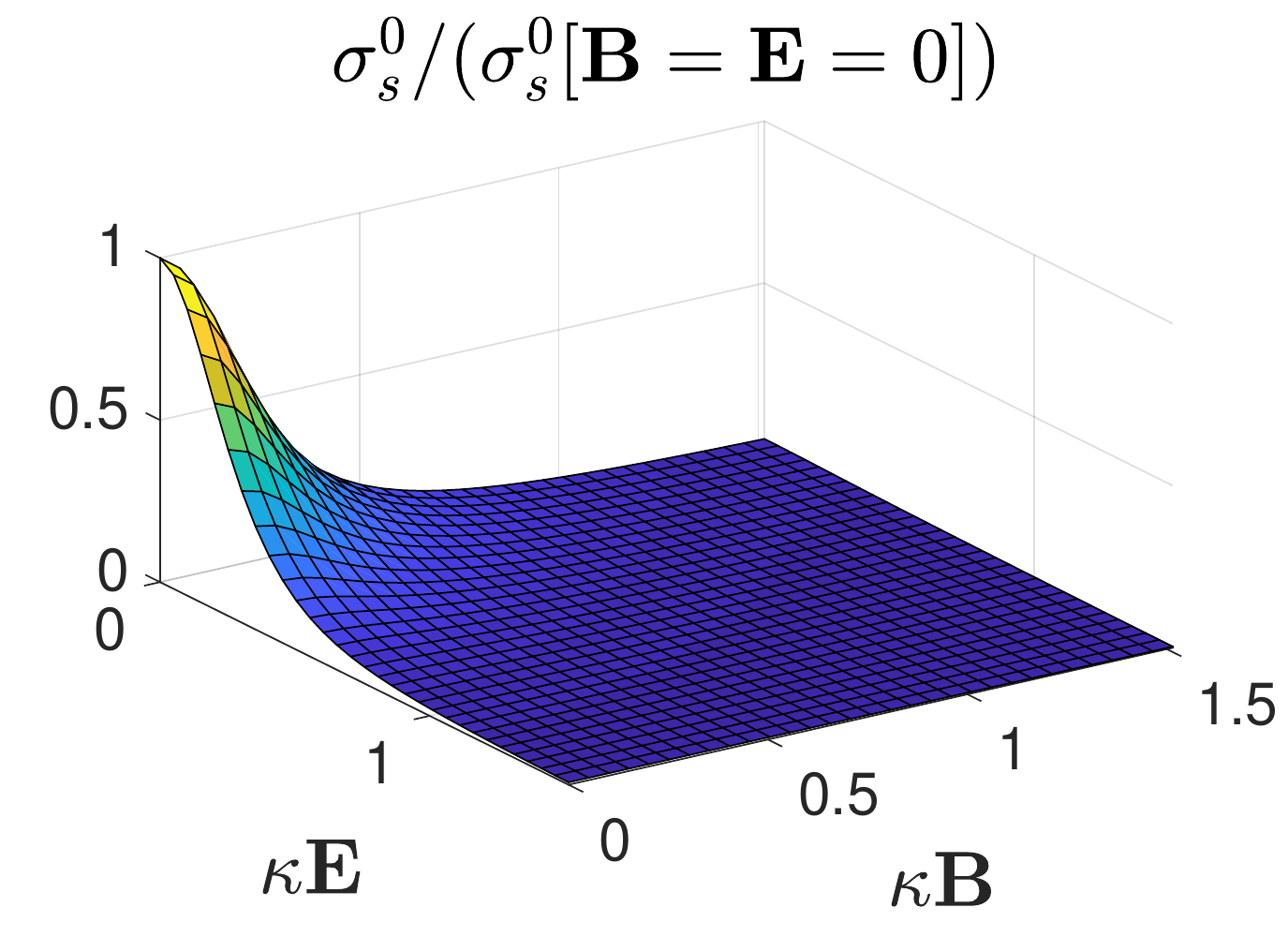}
        \caption{}
        \label{grad_fig1_f}
    \end{subfigure}
    \caption{Zeroth order TCs as functions of ${\bf E}$ and ${\bf B}$ when $\vec{\bf E}\parallel \vec{\bf B}$.
    }\label{grad_fig1}
\end{figure}

In Figure \ref{grad_fig1},  3D plots for all the TCs in (\ref{jmu 0th}, \ref{jmu5 0th}) are shown as functions of $\kappa {\bf E}$ and $\kappa {\bf B}$,  first focusing on the case of $\vec{\bf E}\parallel \vec{\bf B}$. Influence of the relative angle $\theta$ between $\vec{\bf E}, \vec{\bf B}$ will be discussed later (see Figure \ref{grad_fig2}).
Figure \ref{grad_fig1_a} displays $\kappa {\bf E}$- and $\kappa {\bf B}$-dependence of the Ohmic conductivity $\sigma_e^0$.   In our previous publications \cite{1511.08789,1608.08595,1609.09054,Bu:2018drd,Bu:2018psl}, due to weak field assumption, the Ohmic conductivity $\sigma_e^0$ did not depend on  external e/m fields at all. Here, we observe that $\sigma_e^0$ gets enhancement when $\vec {\bf E} \cdot \vec {\bf B}\ne 0$.
While the dependence of Ohmic conductivity on external electric field was already considered in holography \cite{Horowitz:2013mia,Zeng:2016api,Zeng:2016gqj},
to the best of our knowledge, anomaly-induced corrections to $\sigma_e^0$ found here have not been reported before.
In \cite{Horowitz:2013mia}, the nonlinear conductivity emerged from
a gravitational back-reaction effect, while in \cite{Zeng:2016api,Zeng:2016gqj} it emerged from a coupling between the bulk gauge fields and an additional charged scalar. Overall, the nonlinear conductivities of \cite{Horowitz:2013mia,Zeng:2016api,Zeng:2016gqj} decrease with strengthening of the electric field,  while, as demonstrated  in the present work,
the anomaly-induced effect has totally opposite signature.

%
%

All the other TCs decrease dramatically with increasing $\kappa {\bf E}, \kappa {\bf B}$, and vanish asymptotically.
The results are presented normalised  with respect to their values at vanishing $\kappa{\bf E}$ and $\kappa{\bf B}$.
The TC $\delta\sigma_\chi^0$ turns negative in a certain region of $\kappa {\bf E}$ and $\kappa\bf{B}$ (Figure \ref{grad_fig1_c}).



\begin{figure}
    \centering
    \begin{subfigure}[h]{0.485\textwidth}
        \includegraphics[width=\textwidth]{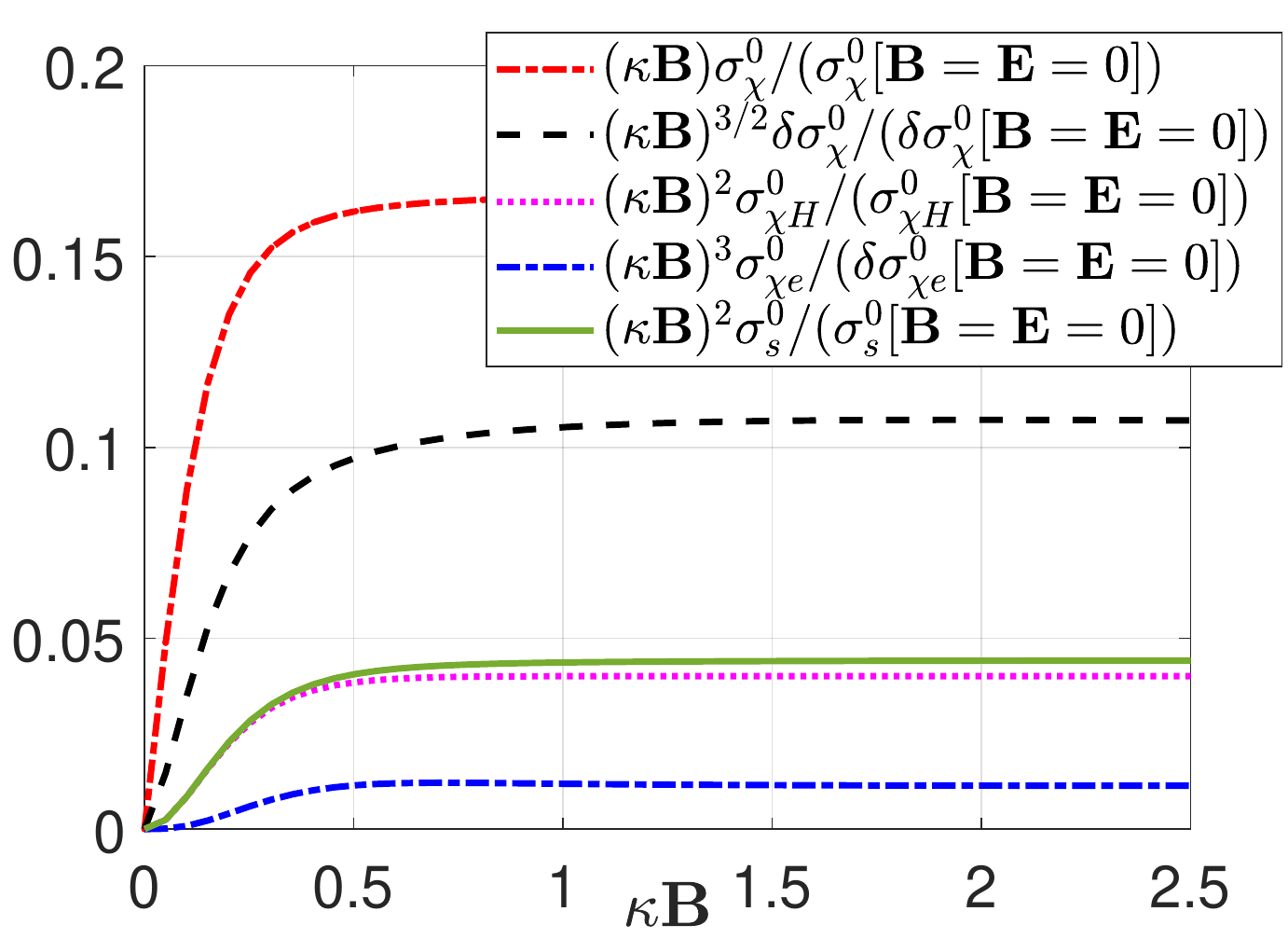}
        \caption{}
        \label{grad_fig3_a}
    \end{subfigure}
    ~ 
      \begin{subfigure}[h]{0.485\textwidth}
        \includegraphics[width=\textwidth]{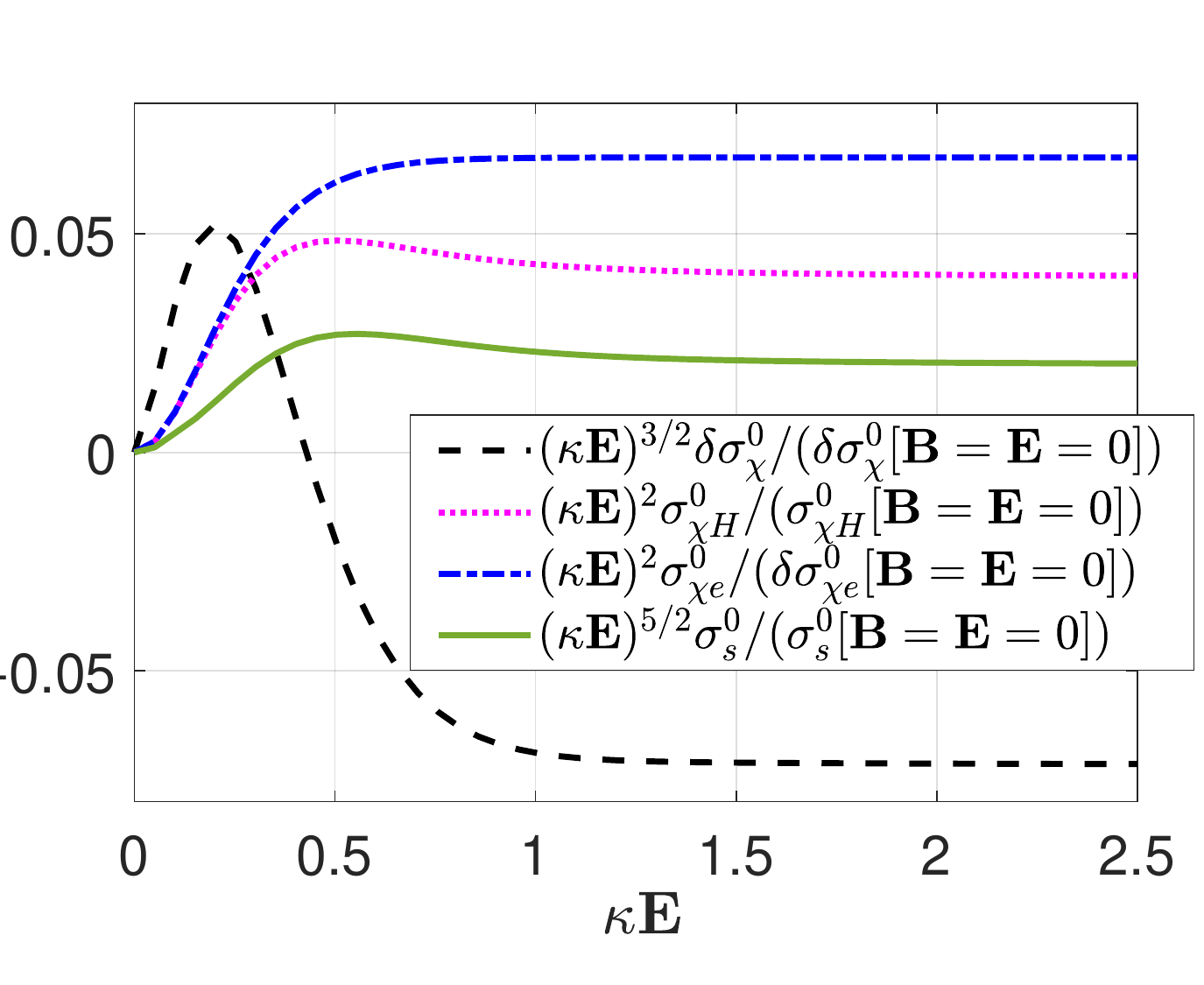}
        \caption{}
        \label{grad_fig3_b}
    \end{subfigure}
    \caption{Normalised zeroth order TCs: (a) $\kappa \mathbf E=0$  and (b) $\kappa \mathbf B=0$.}\label{grad_fig3}
\end{figure}

\begin{figure}
\centering
\includegraphics[width=0.48\textwidth]{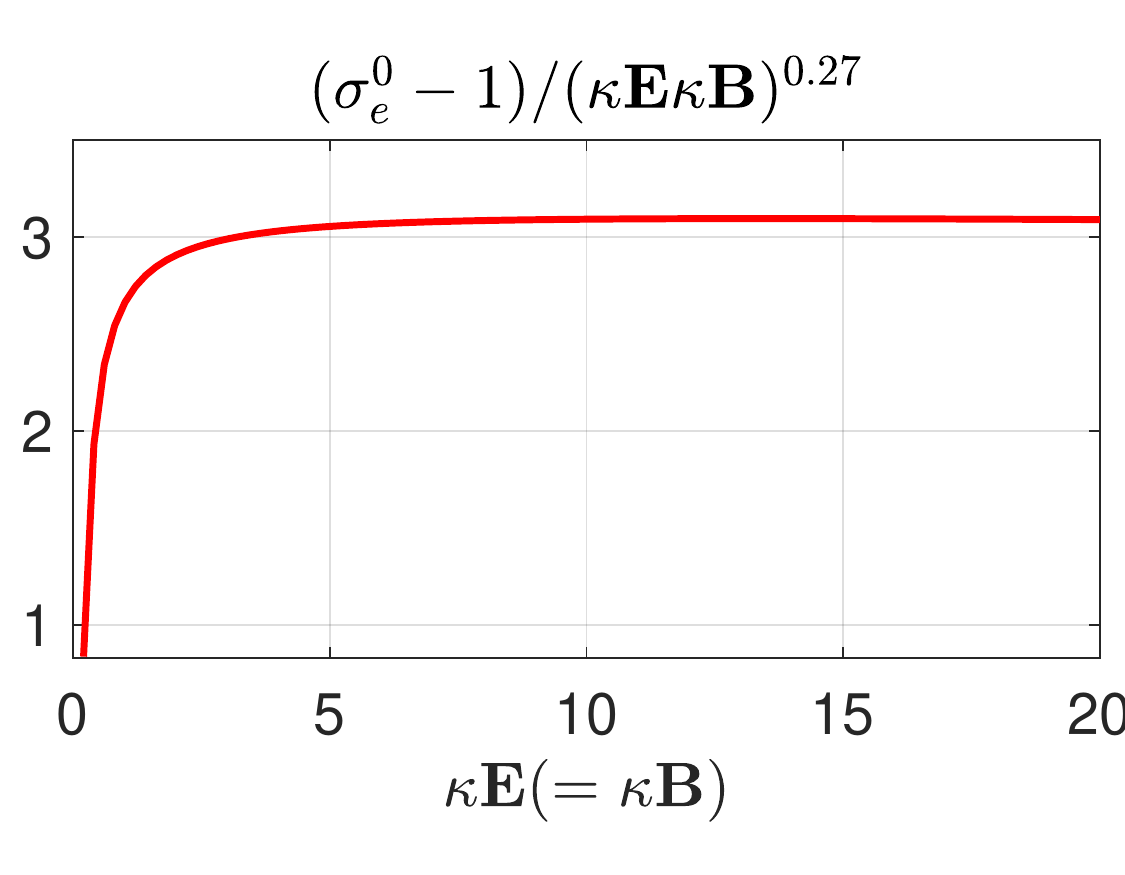}
\caption{$\sigma_e^0$ functions of $\kappa {\bf B}$ and $\kappa {\bf E}$ (on the slice of $\kappa {\bf B}=\kappa {\bf E}$) .}\label{grad_fig_se}
\end{figure}

\begin{figure}
    \centering
    \begin{subfigure}[h]{0.485\textwidth}
        \includegraphics[width=\textwidth]{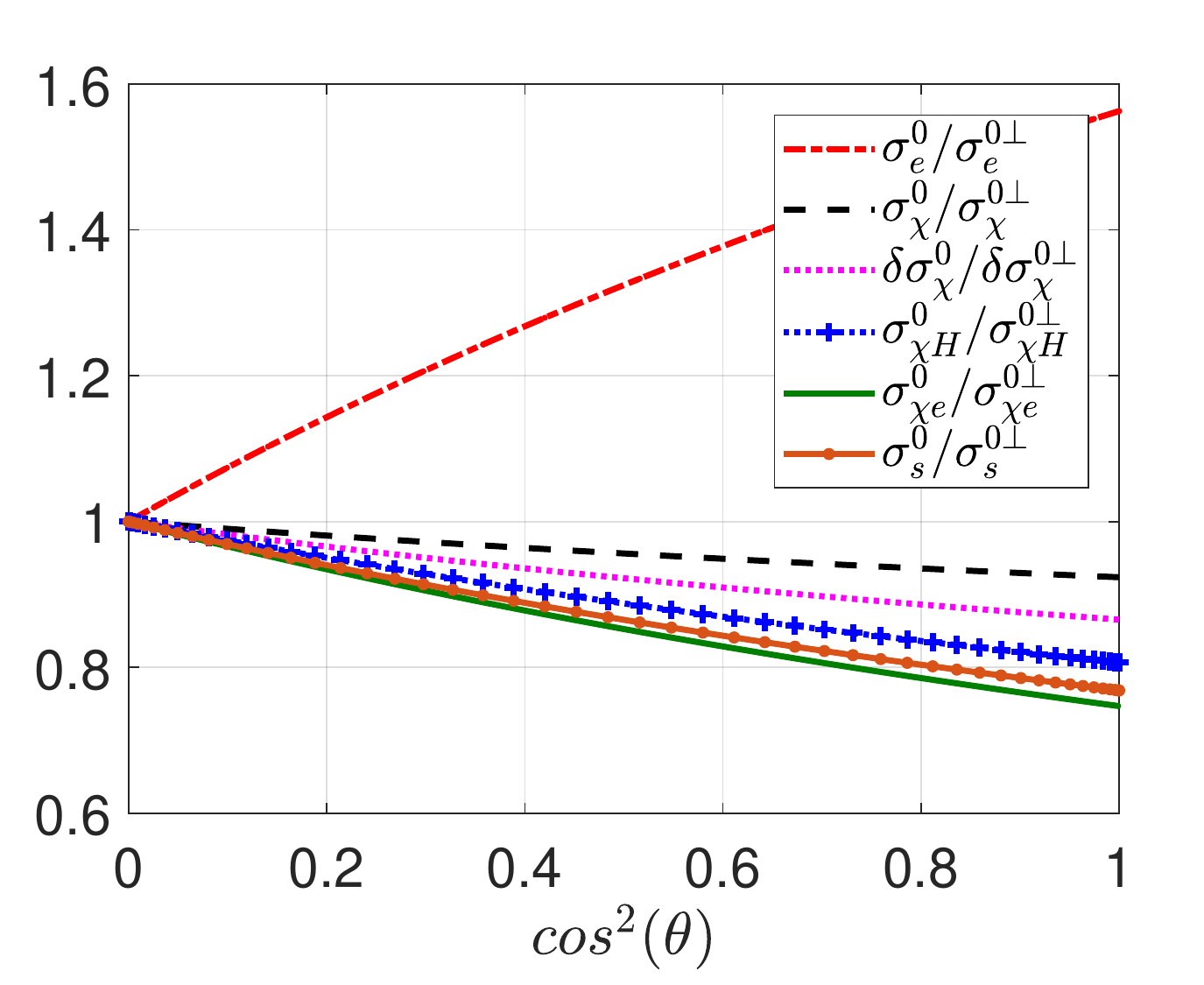}
        \caption{}
        \label{grad_fig2_a}
    \end{subfigure}
    ~ 
      \begin{subfigure}[h]{0.485\textwidth}
        \includegraphics[width=\textwidth]{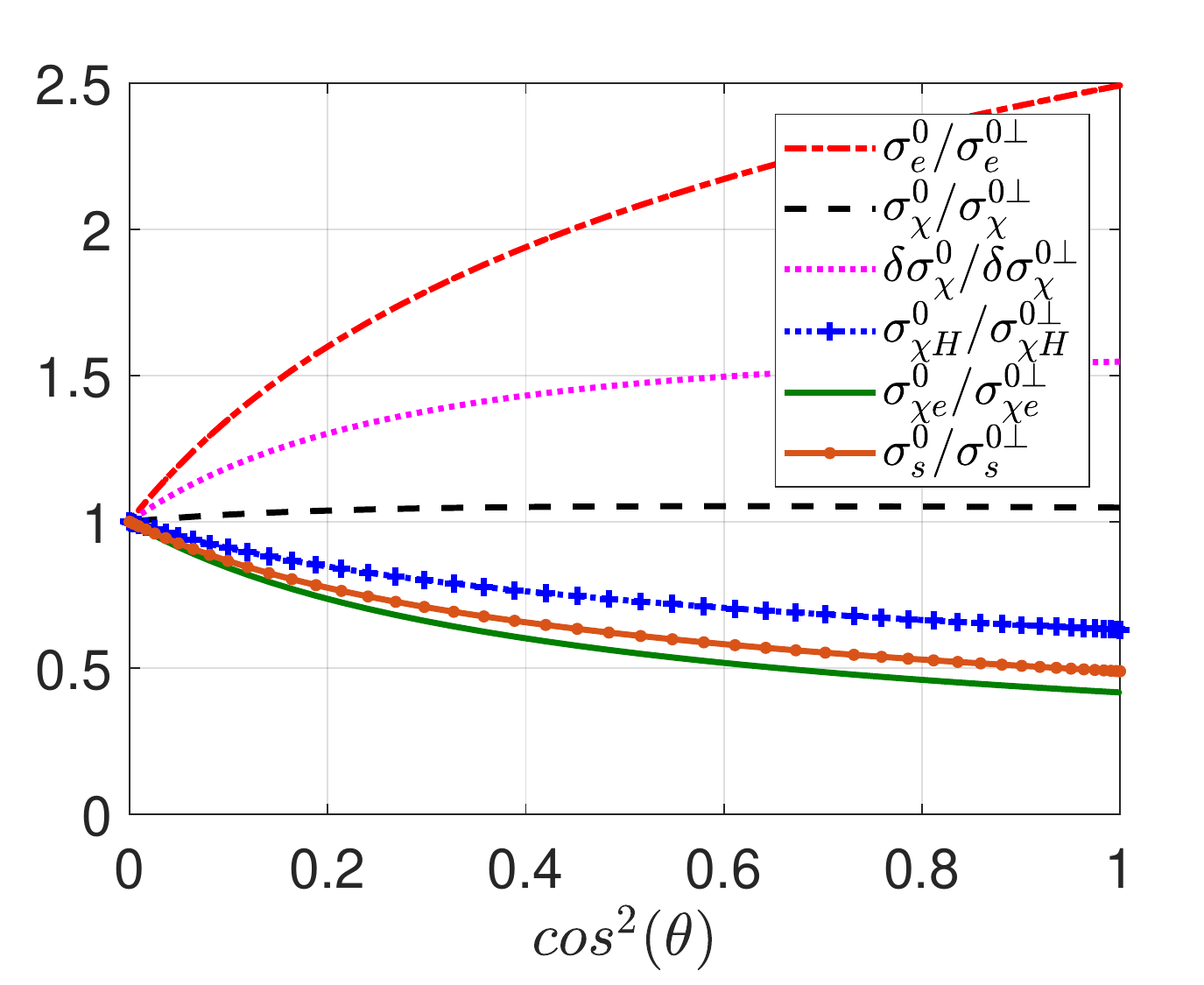}
        \caption{}
        \label{grad_fig2_b}
    \end{subfigure}
    \caption{$\theta$-dependence of the TCs: (a) $\kappa \mathbf B=\kappa \mathbf E=0.25$ and (b) $\kappa \mathbf B=\kappa \mathbf E=0.5$.}\label{grad_fig2}
\end{figure}

Figure \ref{grad_fig3} helps to extract the asymptotic behaviour of the TCs   when either $\kappa{\bf E}$ or $\kappa{\bf B}$ is very large.
%
The TCs are rescaled so that the asymptotic  scaling summarised in (\ref{scaling_0th_E=0}, \ref{scaling_0th_B=0}) becomes transparent.
Figure \ref{grad_fig_se} displays rescaled $\sigma_e^0$ on ($\kappa \mathbf{E}=\kappa \mathbf{B}$)-slice of Figure \ref{grad_fig1_a}.

Finally, in Figure \ref{grad_fig2}, we examine the effect of the relative angle $\theta$ between $\vec{\bf E}$ and $\vec{\bf B}$ on the TCs in (\ref{jmu 0th},\ref{jmu5 0th}).
In  Figure \ref{grad_fig2}, the TCs are normalised with respect to their values when  $\vec{\mathbf E} \perp \vec{\mathbf B}$, denoted as $\sigma_e^{0\perp}$, $\sigma_\chi^{0\perp}$, $\delta\sigma_{\chi}^{0 \perp}$, $\sigma_{\chi H}^{0 \perp}$, $\sigma_{\chi e}^{0 \perp}$ and $\sigma_{s}^{0 \perp}$. Two  values of the e/m fields are considered: $\kappa {\bf B}=\kappa {\bf E}=0.25$ (Figure \ref{grad_fig2_a}) and  $\kappa {\bf B}=\kappa {\bf E}=0.5$ (Figure \ref{grad_fig2_b}). We observe that the stronger the fields
the more pronounced the dependence of the TCs on $cos^2\theta$.  It is also very clear that while some TCs become weaker others become stronger.

Constitutive relations for $\vec J$ and $\vec J_5$ at  first order are derived in next two Subsections \ref{sec_b} and \ref{sec_c}. To simplify the algebra, two cases are explored separately,
either $\vec{\bf E} =0$ or $\vec{\bf B}=0$.  Furthermore, only the terms linear in both $\rho$ and $\rho_5$ are considered.  It is important to stress that
 smallness of the charge densities  is a self-consistent approximation: when $\vec{\bf E} \cdot\vec{\bf B}=0$,
 there is no pumping of the axial charge into the system through the continuity equation (\ref{cont eqn}).
  %

\subsection{Constitutive relations at first order --- $\vec{\bf B}\neq 0, \, \vec{\bf E}=0$ }\label{sec_b}

When $\vec{\bf E}=0$, the dynamical equations for $\mathbb{V}_{\mu}^{[1]}$ and $\mathbb{A}_{\mu}^{[1]}$ are
\begin{equation}
0=r^3\partial_r^2 \mathbb{V}_t^{[1]}+3r^2 \partial_r \mathbb{V}_t^{[1]} +r \partial_r C_2^{(0)} \kappa \vec{\bf B}\cdot \vec\nabla\rho_5 +12 \kappa {\bf B}_k\partial_r \mathbb{A}_k^{[1]},
\end{equation}
\begin{equation}
0=r^3\partial_r^2 \mathbb{A}_t^{[1]}+3r^2 \partial_r \mathbb{A}_t^{[1]} +r \partial_r C_2^{(0)} \kappa \vec{\bf B}\cdot \vec\nabla\rho +12 \kappa {\bf B}_k\partial_r \mathbb{V}_k^{[1]},
\end{equation}
\begin{equation}
\begin{split}
0=&(r^5-r)\partial_r^2\mathbb{V}_i^{[1]}+(3r^4+1)\partial_r\mathbb{V}_i^{[1]}+ 2r^3 \partial_r \partial_t \mathbb{V}_i^{[0]}-r^3\partial_r\partial_i\mathbb{V}_t^{[0]} +r^2(\partial_t\mathbb{V}_i^{[0]}-\partial_i\mathbb{V}_t^{[0]}) \\
&-\frac{1}{2}\partial_i\rho+12r^2\kappa {\bf B}_i \partial_r\mathbb{A}_t^{[1]},
\end{split}
\end{equation}
\begin{equation}
\begin{split}
0=&(r^5-r)\partial_r^2\mathbb{A}_i^{[1]}+(3r^4+1)\partial_r\mathbb{A}_i^{[1]}+ 2r^3 \partial_r \partial_t \mathbb{A}_i^{[0]}-r^3\partial_r\partial_i\mathbb{A}_t^{[0]} +r^2(\partial_t\mathbb{A}_i^{[0]}-\partial_i\mathbb{A}_t^{[0]}) \\
&-\frac{1}{2}\partial_i\rho_5+12r^2\kappa {\bf B}_i \partial_r\mathbb{V}_t^{[1]}.
\end{split}
\end{equation}
The solutions are
\begin{equation}
\mathbb{V}_t^{[1]}= C_1^{(1)}\kappa \vec{\bf B}\cdot \vec\nabla\rho_5 + C_2^{(1)} \partial_t \rho, \qquad \mathbb{V}_i^{[1]}=C_3^{(1)}\partial_i\rho+C_4^{(1)} \kappa \partial_t \rho_5 {\bf B}_i + C_5^{(1)} \kappa^2 (\vec{\bf B}\cdot \vec\nabla)\rho {\bf B}_i,
\end{equation}
\begin{equation}
\mathbb{A}_t^{[1]}=C_1^{(1)}\kappa \vec{\bf B}\cdot \vec\nabla\rho + C_2^{(1)} \partial_t \rho_5,\qquad \mathbb{A}_i^{[1]}=C_3^{(1)}\partial_i\rho_5+C_4^{(1)} \kappa \partial_t \rho {\bf B}_i + C_5^{(1)} \kappa^2 (\vec{\bf B}\cdot \vec\nabla)\rho_5 {\bf B}_i.
\end{equation}
The decomposition coefficients  $C_i^{(1)}$ obey partially decoupled ODEs, which we group into two decoupled sub-sectors:\\
$\{C_1^{(1)},C_3^{(1)},C_5^{(1)}\}$
\begin{equation}\label{dec1astart}
0=r^3\partial_r^2C_1^{(1)}+3r^2 \partial_rC_1^{(1)} +r \partial_rC_2^{(0)} +12(\partial_r C_3^{(1)}+\kappa^2{\bf B}^2 \partial_rC_5^{(1)}),
\end{equation}
\begin{equation}
0=(r^5-r)\partial_r^2 C_3^{(1)} +(3r^4+1)\partial_rC_3^{(1)} -r^3\partial_r F_1-r^2F_1-\frac{1}{2},
\end{equation}
\begin{equation}
0=(r^5-r)\partial_r^2C_5^{(1)} +(3r^4+1) \partial_rC_5^{(1)} +12r^2 \partial_rC_1^{(1)},
\end{equation}
where $F_1$ parameterises solutions to $V_t^{[0]}, A_t^{[0]}$ (see (\ref{Vt0_1}, \ref{At0_1})):
%
%
\begin{equation}
r^3\partial_r F_1+12\kappa^2{\bf B}^2 C_2^{(0)}=0,\qquad F_1(r\to \infty)=0.
\end{equation}
$\{C_2^{(1)},C_4^{(1)}\}$
\begin{equation}
0=r^3\partial_r^2C_2^{(1)}+3r^2 \partial_rC_2^{(1)}+12\kappa^2 {\bf B}^2 \partial_r C_4^{(1)},
\end{equation}
\begin{equation}\label{dec1afin}
0=(r^5-r)\partial_r^2C_4^{(1)}+(3r^4+1)\partial_rC_4^{(1)} +2r^3\partial_rC_2^{(0)} +r^2 C_2^{(0)} +12 r^2\partial_r C_2^{(1)}.
\end{equation}

Near the boundary ($r=\infty$), the asymptotic expansion of the decomposition coefficients is
\begin{equation}\label{nb1a}
C_i^{(1)}=\frac{c_i^{(1)}}{r^2}+\mathcal{O}\left(\frac{1}{r^3}\right),\qquad i=1,2,\cdots, 5.
\end{equation}
Plugging (\ref{nb1a}) into (\ref{bdry currents}),  the first order constitutive relations (\ref{1st J_E=0}, \ref{1st J5_E=0}) are obtained with
the TCs related to the boundary data as
\begin{equation}
\mathcal{D}_0=-2c_3^{(1)},\qquad \tau_{\bar\chi}=2c_4^{(1)},\qquad \mathcal{D}_B^0= 2c_5^{(1)}.
\end{equation}
Alternatively, (\ref{1st J_E=0}, \ref{1st J5_E=0}) can be re-expressed by collecting  terms involving $\vec \nabla \rho$, $\vec \nabla \rho_5$ into transverse and longitudinal parts (with respect to $\vec{\bf B}$):
\begin{align} \label{1st J_E=0 re}
J_i^{[1]}= -\mathcal{D}_0^{\rm T} \left(\delta_{ij}- \frac{{\bf B}_i {\bf B}_j}{{\bf B}^2 } \right)\partial_j \rho- \mathcal{D}_0^{\rm L} \frac{{\bf B}_i {\bf B}_j}{{\bf B}^2} \partial_j \rho+ \tau_{\bar{\chi}}\kappa {\bf B}_i \partial_t \rho_5,\\
J_{5i}^{[1]}= -\mathcal{D}_0^{\rm T} \left(\delta_{ij}- \frac{{\bf B}_i {\bf B}_j}{{\bf B}^2 } \right)\partial_j \rho_5- \mathcal{D}_0^{\rm L} \frac{{\bf B}_i {\bf B}_j}{{\bf B}^2} \partial_j \rho_5+ \tau_{\bar{\chi}}\kappa {\bf B}_i \partial_t \rho, \label{1st J5_E=0 re}
\end{align}
where
\begin{align}\label{longDB}
\mathcal{D}_0^{\rm T} =\mathcal{D}_0, \qquad \qquad \mathcal{D}_0^{\rm L}= \mathcal{D}_0 - \kappa^2 {\bf B}^2 \mathcal{D}_B^0.
\end{align}



\begin{figure}
\centering
\includegraphics[width=0.485\textwidth]{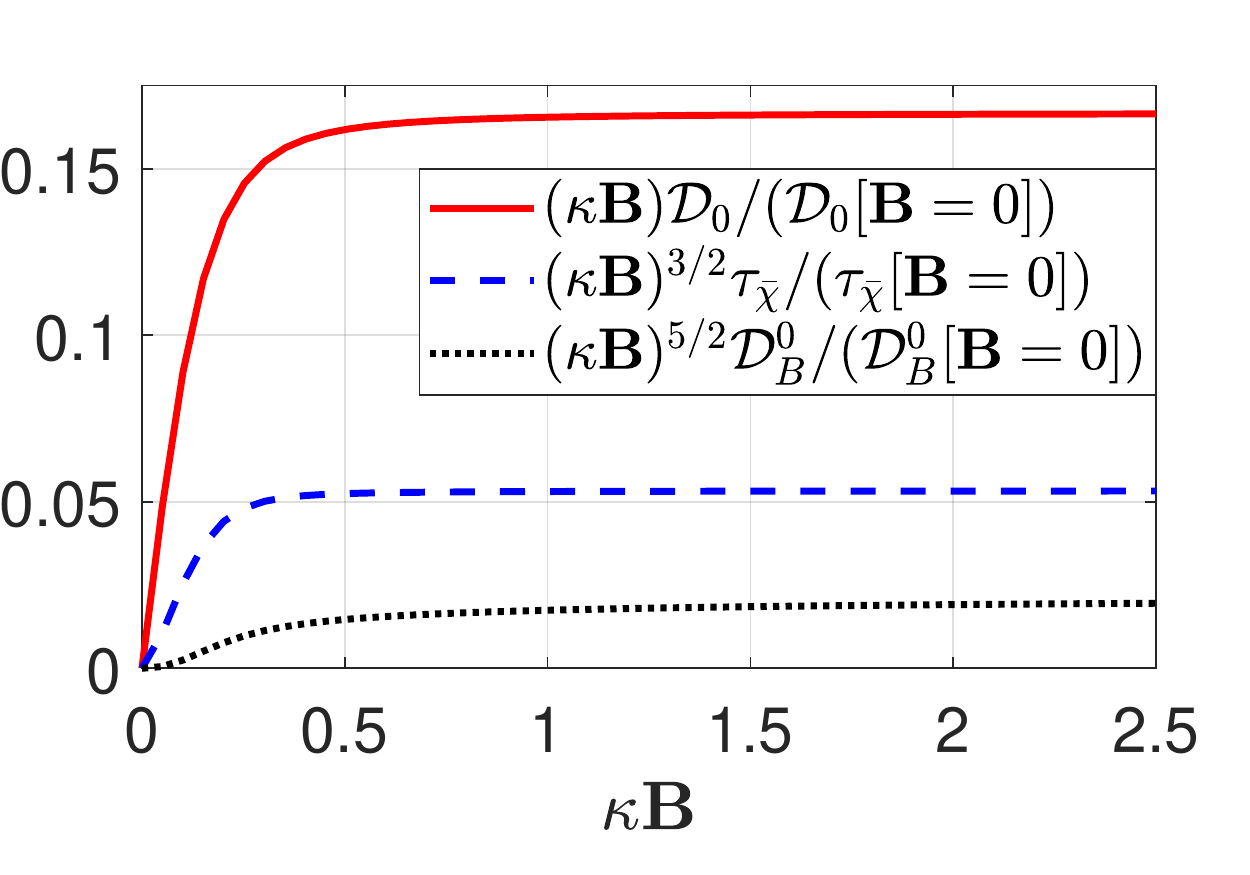}
\caption{Normalised first order TCs as functions of $\kappa {\bf B}$  when $\kappa {\bf E} =0$.}\label{grad_fig4}
\end{figure}

For generic values of $\kappa {\bf B}$, we are able to solve the ODEs (\ref{dec1astart}-\ref{dec1afin}) numerically only. The results are presented in Figure \ref{grad_fig4}.
The TCs are normalised to their values at vanishing e/m fields  (see \cite{Bu:2018psl}) quoted below and also rescaled by their asymptotic behaviour:
\begin{align}
\mathcal D_0[{\bf B}=0]= \frac{1}{2}, \qquad \tau_{\bar \chi}[{\bf B}=0]=-\frac{3}{2}\left(\pi+2\log2\right), \qquad \mathcal D_B^0[{\bf B}=0]= -9\left(\pi-2\log2\right).
\end{align}
Obviously, all the TCs in (\ref{1st J_E=0}, \ref{1st J5_E=0}) decrease when $\kappa {\bf B}$ becomes stronger, and vanish asymptotically.
Large-$\kappa \mathbf B$ asymptotic behavior for all the TCs in (\ref{1st J_E=0}, \ref{1st J5_E=0}) is summarised in (\ref{scaling_1st_E=0}).


\subsection{Constitutive relations at first order --- $\vec{\bf E}\neq 0, \, \vec{\bf B}=0$}\label{sec_c}

The equations for $\mathbb{V}_t^{[1]}$ and $\mathbb{A}_t^{[1]}$ are {\it homogeneous}, so both  $\mathbb{V}_t^{[1]}$ and $\mathbb{A}_t^{[1]}$ vanish.
Consequently, the equations for $\mathbb{V}_i^{[1]}$ and $\mathbb{A}_i^{[1]}$ are
\begin{equation}
0=(r^5-r)\partial_r^2\mathbb{V}_i^{[1]}+ (3r^4+1) \partial_r \mathbb{V}_i^{[1]} -\frac{1}{2}\partial_i \rho+ 12\kappa r^2\epsilon^{ijk}\partial_r\mathbb{A}_j^{[1]} {\bf E}_k -6\kappa \epsilon^{ijk}\partial_r\mathbb{V}_j^{[0]}\partial_k\rho_5,
\end{equation}
\begin{equation}
0=(r^5-r)\partial_r^2\mathbb{A}_i^{[1]}+ (3r^4+1) \partial_r \mathbb{A}_i^{[1]} -\frac{1}{2}\partial_i \rho_5+ 12\kappa r^2\epsilon^{ijk}\partial_r\mathbb{V}_j^{[1]} {\bf E}_k -6\kappa \epsilon^{ijk}\partial_r\mathbb{V}_j^{[0]}\partial_k\rho.
\end{equation}
The solutions are
\begin{equation}
\begin{split}
&\mathbb{V}_i^{[1]}=C_6^{(1)}\partial_i\rho+C_7^{(1)} \kappa (\vec{\bf E}\times \vec\nabla \rho_5)_i+C_8^{(1)} \kappa^2(\vec{\bf E}\cdot \vec\nabla\rho){\bf E}_i,\\
&\mathbb{A}_i^{[1]}=C_6^{(1)}\partial_i\rho_5+C_7^{(1)} \kappa (\vec{\bf E}\times \vec\nabla \rho)_i+C_8^{(1)} \kappa^2(\vec{\bf E}\cdot \vec\nabla\rho_5){\bf E}_i,
\end{split}
\end{equation}
where the decomposition coefficients satisfy coupled ODEs:
\begin{equation}\label{dec1bstart}
0=(r^5-r)\partial_r^2C_6^{(1)}+(3r^4+1)\partial_rC_6^{(1)} -\frac{1}{2} +12\kappa^2{\bf E}^2 r^2 \partial_rC_7^{(1)},
\end{equation}
\begin{equation}
0=(r^5-r)\partial_r^2C_7^{(1)}+(3r^4+1)\partial_rC_7^{(1)}-12 r^2\partial_rC_6^{(1)} -6\partial_r C_1^{(0)},
\end{equation}
\begin{equation}\label{dec1bfin}
0=(r^5-r)\partial_r^2C_8^{(1)}+(3r^4+1)\partial_rC_8^{(1)} -12r^2\partial_rC_7^{(1)}.
\end{equation}
Near $r=\infty$, the asymptotic expansion for $C_i^{(1)}$ is,
\begin{equation}\label{nb1b}
C_i^{(1)}=\frac{c_i^{(1)}}{r^2}+\mathcal{O}\left(\frac{1}{r^3}\right),\qquad i=6,7,8,
\end{equation}
which via (\ref{bdry currents}) helps to derive the first order constitutive relations (\ref{1st J_B=0}, \ref{1st J5_B=0}) with the TCs given by
\begin{equation}
\mathcal{D}_0=-2c_6^{(1)},\qquad \sigma_{a\chi H}^0=2c_7^{(1)},\qquad \mathcal{D}_E^0=2c_8^{(1)}.
\end{equation}

As done for (\ref{1st J_E=0}, \ref{1st J5_E=0}), we combine the $\vec\nabla\rho, \vec\nabla\rho_5$-terms in (\ref{1st J_B=0}, \ref{1st J5_B=0}) into longitudinal and transverse parts (with respect to $\vec{\bf E}$),
\begin{align}
J_i^{[1]}= - \mathcal{D}_0^{\rm T} \left(\delta_{ij}- \frac{{\bf E}_i {\bf E}_j} {{\bf E}^2 }\right)\partial_j\rho- \mathcal{D}_0^{\rm L}\frac{{\bf E}_i {\bf E}_j}{{\bf E}^2} \partial_j \rho+\sigma_{a\chi H}^0\epsilon^{ijk} \kappa E_j \partial_k \rho_5, \label{1st J_B=0 re}\\
J_{5i}^{[1]}= - \mathcal{D}_0^{\rm T} \left(\delta_{ij}- \frac{{\bf E}_i {\bf E}_j} {{\bf E}^2 }\right)\partial_j\rho_5- \mathcal{D}_0^{\rm L}\frac{{\bf E}_i {\bf E}_j} {{\bf E}^2} \partial_j \rho_5+\sigma_{a\chi H}^0\epsilon^{ijk} \kappa E_j \partial_k \rho,  \label{1st J5_B=0 re}
\end{align}
where
\begin{align}\label{longDE}
\mathcal{D}_0^{\rm T}=\mathcal{D}_0,\qquad \qquad \mathcal{D}_0^{\rm L}= \mathcal{D}_0- \kappa^2{\bf E}^2 \mathcal{D}_E^0,
\end{align}

\begin{figure}
\centering
\includegraphics[width=0.485\textwidth]{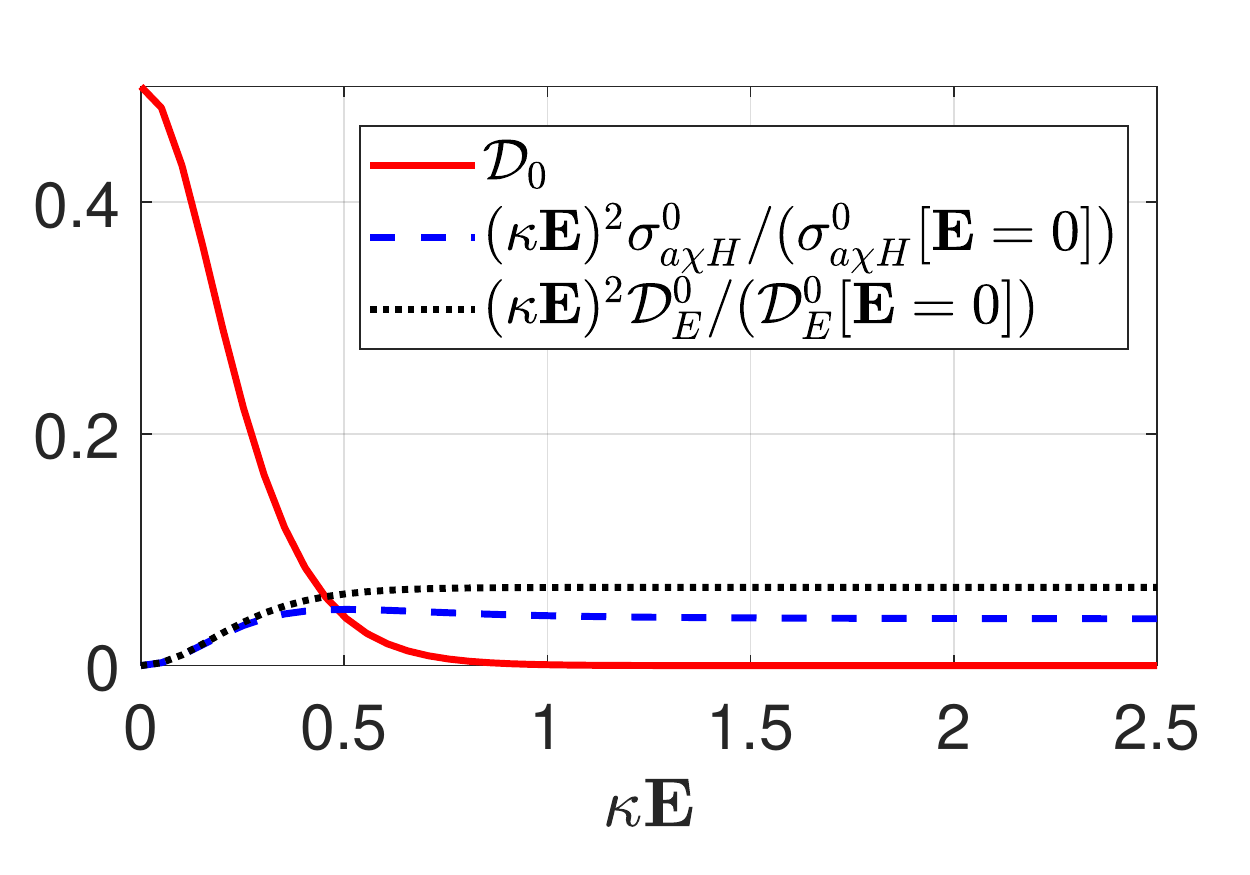}
\caption{ The first order TCs as function of $\kappa {\bf E}$ when ${\bf B}=0$.} \label{grad_fig5}
\end{figure}

For generic value of $\kappa {\bf E}$,  the ODEs (\ref{dec1bstart}-\ref{dec1bfin}) are solved numerically and the results are displayed in Figure \ref{grad_fig5}.
Analytical results  for each TCs when ${\bf E}={\bf B}=0$ are quoted for completeness (see \cite{Bu:2018psl}):
\begin{align}
\sigma_{a\chi H}^0[{\bf E}=0]=-3\log2 , \qquad \mathcal D_E^0[\mathbf{E}=0]= -\frac{3}{4}\pi^2.
\end{align}
From Figure \ref{grad_fig5}, one can read off the large-$\kappa \mathbf E$ behavior for the TCs in (\ref{1st J_B=0}, \ref{1st J5_B=0}), as summarised in (\ref{scaling_1st_B=0}).
We note that large-$\kappa \bf{E}$ behavior of $\mathcal D_0$ does not scale as a power function and it decays faster.

\section{Part II: gradient resummation in external magnetic field} \label{resummation}

In this section, we focus on all-order resummation of  gradient terms that are {\it linear} in both $\rho$ and $\rho_5$ when $\vec{\bf B}$ is taken as arbitrary in amplitude.
External electric field $\vec{\bf E}$ will be turned off throughout this section.
In order to keep the terms linear in the charge densities only,  we introduce a parameter $\epsilon$,
\begin{equation}\label{sch1}
\rho\to \epsilon \rho,\qquad \rho_5\to \epsilon \rho_5,\qquad \vec{\bf B}\sim \mathcal{O}(\epsilon^0).
\end{equation}
We solve the dynamical equations up to $\mathcal{O}(\epsilon^1)$:
\begin{equation}\label{sch2}
\mathbb{V}_\mu=\mathbb{V}_\mu^{(0)}+ \epsilon \mathbb{V}_\mu^{(1)}+\cdots,\qquad \qquad \mathbb{A}_\mu=\mathbb{A}_\mu^{(0)}+ \epsilon \mathbb{A}_\mu^{(1)}+\cdots.
\end{equation}

\subsection{Derivation of the all-order resummed constitutive relations}

Since the hydrodynamic variables $\rho,\rho_5$
 are  $\mathcal{O}(\epsilon^1)$, the zeroth order corrections $\mathbb{V}_\mu^{(0)}$ and $\mathbb{A}_\mu^{(0)}$ can depend on
  $\vec{\bf B}$ only and are independent of the boundary coordinates $x^\alpha$.
  Therefore, at $\mathcal{O}(\epsilon^0)$ the dynamical equations  are
\begin{align}\label{eom Vt0-alt}
0=r^3\partial_r^2 \mathbb{V}_t^{(0)}+3r^2 \partial_r \mathbb{V}_t^{(0)}+ 12\kappa \mathbf{B}_k \partial_r \mathbb{A}_k^{(0)},
\end{align}
\begin{align}\label{eom Vi0-alt}
0=(r^5-r)\partial_r^2 \mathbb{V}_i^{(0)}+(3r^4+1)\partial_r \mathbb{V}_i^{(0)}+12\kappa r^2 \mathbf{B}_i \partial_r \mathbb{A}_t^{(0)},
\end{align}
\begin{align}\label{eom At0-alt}
0=r^3\partial_r^2 \mathbb{A}_t^{(0)}+3r^2 \partial_r \mathbb{A}_t^{(0)}+ 12\kappa \mathbf{B}_k \partial_r \mathbb{V}_k^{(0)},
\end{align}
\begin{align}\label{eom Ai0-alt}
0=(r^5-r)\partial_r^2 \mathbb{A}_i^{(0)}+(3r^4+1)\partial_r \mathbb{A}_i^{(0)}
+12\kappa r^2\mathbf{B}_i \partial_r \mathbb{V}_t^{(0)}.
\end{align}
The equations (\ref{eom Vt0-alt}, \ref{eom At0-alt}) could be  integrated over $r$:
\begin{align}
r^3\partial_r \mathbb{V}_t^{(0)}+12\kappa {\bf B}_k \mathbb{A}_k^{(0)}=0,\qquad \qquad
r^3\partial_r \mathbb{A}_t^{(0)}+12\kappa {\bf B}_k \mathbb{V}_k^{(0)}=0,
\end{align}
where the Landau frame convention (\ref{Landau frame}) has been used to fix the integration constants.
Hence, the equations (\ref{eom Vi0-alt}, \ref{eom Ai0-alt}) become {\it homogeneous} ODEs.
Combined with the boundary conditions specified in section \ref{model}, the equations (\ref{eom Vi0-alt}, \ref{eom Ai0-alt}) do not have any nontrivial solution.
Thus, we conclude that
\begin{equation}
\mathbb{V}_\mu^{(0)}= \mathbb{A}_\mu^{(0)}=0.
\end{equation}

At $\mathcal{O}(\epsilon^1)$, the dynamical equations  become
\begin{equation}\label{eom Vt1}
0=r^3\partial_r^2 \mathbb{V}_t^{(1)}+3r^2 \partial_r \mathbb{V}_t^{(1)}+r\partial_r \partial_k \mathbb{V}_k^{(1)}+ 12\kappa \mathbf{B}_k\partial_r \mathbb{A}_k^{(1)},
\end{equation}
\begin{align}\label{eom Vi1}
0=&(r^5-r)\partial_r^2 \mathbb{V}_i^{(1)}+(3r^4+1)\partial_r \mathbb{V}_i^{(1)} +2r^3 \partial_r \partial_t \mathbb{V}_i^{(1)}-r^3 \partial_r\partial_i \mathbb{V}_t^{(1)}+ r^2\left(\partial_t \mathbb{V}_i^{(1)}- \partial_i \mathbb{V}_t^{(1)}\right) \nonumber\\
&+r(\partial^2 \mathbb{V}_i^{(1)} - \partial_i \partial_k \mathbb{V}_k^{(1)}) -\frac{1}{2}\partial_i \rho +12\kappa r^2\mathbf{B}_i\left(\frac{1}{r^3}\rho_{5} +\partial_r \mathbb{A}_t^{(1)} \right),
\end{align}
\begin{equation}\label{eom At1}
0=r^3\partial_r^2 \mathbb{A}_t^{(1)}+ 3r^2 \partial_r \mathbb{A}_t^{(1)}+r\partial_r \partial_k \mathbb{A}_k^{(1)}+ 12\kappa \mathbf{B}_k\partial_r \mathbb{V}_k^{(1)},
\end{equation}
\begin{align}\label{eom Ai1}
0=&(r^5-r)\partial_r^2 \mathbb{A}_i^{(1)}+(3r^4+1)\partial_r \mathbb{A}_i^{(1)}+2r^3 \partial_r \partial_t \mathbb{A}_i^{(1)}-r^3 \partial_r\partial_i \mathbb{A}_t^{(1)}+ r^2\left(\partial_t \mathbb{A}_i^{(1)}- \partial_i \mathbb{A}_t^{(1)}\right) \nonumber\\
&+r(\partial^2 \mathbb{A}_i^{(1)} - \partial_i\partial_k\mathbb{A}_k^{(1)}) -\frac{1}{2} \partial_i \rho_5 +12\kappa r^2\mathbf{B}_i\left(\frac{1}{r^3}\rho +\partial_r \mathbb{V}_t^{(1)} \right).
\end{align}
The corrections $\mathbb{V}_\mu^{(1)}$ and $\mathbb{A}_\mu^{(1)}$ are decomposed in terms of basic structures built from the external magnetic field $\vec {\bf B}$ and the charge densities $\rho,\rho_5$,
\begin{align}\label{vt1}
\mathbb{V}_t^{(1)}=S_1 \rho +S_2 \kappa\mathbf{B}_k \partial_k \rho_5,
\end{align}
\begin{align}
\mathbb{V}_i^{(1)}&=V_1 \partial_i \rho +V_2 \kappa^2\mathbf{B}_i \mathbf{B}_k \partial_k \rho+V_3 \kappa\mathbf{B}_i \rho_5+V_4 \kappa\mathbf{B}_k \partial_i \partial_k \rho_5 \nonumber \\
&+V_5\epsilon^{ijk}\kappa\mathbf{B}_j \partial_k \rho +V_6\kappa^2 \epsilon^{ijk} \mathbf{B}_j \mathbf{B}_l\partial_l\partial_k\rho_5,
\end{align}
\begin{align}
\mathbb{A}_t^{(1)}=\bar{S}_1 \rho_5 +\bar{S}_2 \kappa\mathbf{B}_k \partial_k \rho,
\end{align}
\begin{align}\label{ai1}
\mathbb{A}_i^{(1)}=&\bar{V}_1 \partial_i \rho_5 +\bar{V}_2 \kappa^2\mathbf{B}_i \mathbf{B}_k \partial_k \rho_5+\bar{V}_3 \kappa\mathbf{B}_i \rho+\bar{V}_4 \kappa \mathbf{B}_k \partial_i \partial_k\rho \nonumber \\
&+\bar{V}_5\epsilon^{ijk}\kappa \mathbf{B}_j \partial_k \rho_5 +\bar{V}_6\epsilon^{ijk} \kappa^2 \mathbf{B}_j\mathbf{B}_l \partial_l \partial_k\rho,
\end{align}
where, in contrast to our previous publications \cite{1608.08595,Bu:2018drd}, the decomposition coefficients $S_i$, $\bar S_i$, $V_i$ and $\bar V_i$ now depend on $\vec{\bf B}$ non-linearly. Particularly, they are scalar functionals of $\partial_t$, $\vec\nabla^2$ and $(\kappa\vec{\mathbf{B}}\cdot\vec{\nabla})^2$; and  functions of $r$ and $(\kappa \mathbf{B})^2 $. In Fourier space via $(\partial_t, \vec \nabla) \to (-i\omega, i\vec q)$, they turn into scalar functions
%
%
\begin{align}
&S_i\left(\partial_t,\vec{\nabla}^2,(\kappa\vec{\mathbf{B}}\cdot\vec{\nabla})^2;r,
(\kappa\mathbf{B})^2\right)\rightarrow S_i\left(\omega,q^2,(\kappa\vec{\mathbf{B}} \cdot\vec{q}\;)^2,r,(\kappa\mathbf{B})^2\right),\label{S_iB}\\
&V_i\left(\partial_t,\vec{\nabla}^2,(\kappa\vec{\mathbf{B}}\cdot\vec{\nabla})^2;r,
(\kappa\mathbf{B})^2\right)\rightarrow V_i\left(\omega,q^2,(\kappa\vec{\mathbf{B}}
\cdot\vec{q}\;)^2,r,(\kappa\mathbf{B})^2\right),\\
&\bar{S}_i\left(\partial_t,\vec{\nabla}^2,(\kappa\vec{\mathbf{B}}\cdot\vec{\nabla})^2;r,
(\kappa\mathbf{B})^2\right)\rightarrow \bar{S}_i\left(\omega,q^2,(\kappa\vec{\mathbf{B}}
\cdot\vec{q}\;)^2,r,(\kappa\mathbf{B})^2\right),\\
&\bar{V}_i\left(\partial_t,\vec{\nabla}^2,(\kappa\vec{\mathbf{B}}\cdot\vec{\nabla})^2;r,
(\kappa\mathbf{B})^2\right)\rightarrow \bar{V}_i\left(\omega,q^2,(\kappa\vec{\mathbf{B}}
\cdot\vec{q}\;)^2,r,(\kappa\mathbf{B})^2\right).\label{V_iB}
\end{align}
Accordingly, the PDEs (\ref{eom Vt1}-\ref{eom Ai1}) give rise to a set of partially decoupled ODEs for the decomposition coefficients
 in (\ref{vt1}-\ref{ai1}), which we collect as\\
\noindent(i):~$\left\{S_1,\bar{S}_2,V_1,V_2,\bar{V}_3,\bar{V}_4\right\}$
\begin{equation} \label{eom S1}
0=r^3\partial_r^2S_1+3r^2\partial_rS_1-q^2r\partial_rV_1-(\kappa\vec{\mathbf{B}}\cdot\vec{q} \;)^2r\partial_rV_2+12 (\kappa \mathbf{B})^2 \partial_r\bar{V}_3-12(\kappa\vec{\bf B} \cdot \vec{q}\;)^2 \partial_r\bar{V}_4,
\end{equation}
\begin{equation}
0=r^3\partial_r^2\bar{S}_2+3r^2\partial_r\bar{S}_2+r\partial_r\bar{V}_3+12\partial_r V_1 +12(\kappa\mathbf{B})^2\partial_rV_2-q^2r\partial_r\bar{V}_4,
\end{equation}
\begin{equation}
\begin{split}
0=&(r^5-r)\partial_r^2V_1+(3r^4+1-2i\omega r^3)\partial_r V_1-i\omega r^2V_1-r^2(S_1+r\partial_r S_1)\\
&+(\kappa\vec{\mathbf{B}}\cdot\vec{q}\;)^2 r V_2-1/2,
\end{split}
\end{equation}
\begin{equation}
\begin{split}
0=&(r^5-r)\partial_r^2V_2+(3r^4+1-2i\omega r^3)\partial_r V_2-i\omega r^2V_2-q^2rV_2+12r^2\partial_r \bar{S}_2,
\end{split}
\end{equation}
\begin{equation}
\begin{split}
0=&(r^5-r)\partial_r^2\bar{V}_3+(3r^4+1-2i\omega r^3)\partial_r \bar{V}_3-i\omega r^2\bar{V}_3-q^2r\bar{V}_3+12r^2\partial_r S_1+12/r,
\end{split}
\end{equation}
\begin{equation} \label{eom Vb4}
0=(r^5-r)\partial_r^2\bar{V}_4+(3r^4+1-2i\omega r^3)\partial_r \bar{V}_4-i\omega r^2\bar{V}_4-r^2(\bar{S}_2+r\partial_r \bar{S}_2)-r\bar{V}_3.
\end{equation}

\noindent(ii):~$\left\{V_5\right\}$
\begin{equation} \label{V56=0}
0=(r^5-r)\partial_r^2V_5+(3r^4+1-2i\omega r^3)\partial_r V_5-i\omega r^2V_5-q^2rV_5.
\end{equation}

The sub-sector  $\left\{\bar{S}_1,S_2,\bar{V}_1,\bar{V}_2,V_3,V_4\right\}$ satisfies the same equations as the sub-sector (i):
$\left\{S_1, \bar{S}_2, V_1, V_2, \bar{V}_3, \bar{V}_4\right\}$.  Given that they obey the same boundary conditions,
we conclude that
\begin{equation}
\left\{\bar{S}_1,S_2,\bar{V}_1,\bar{V}_2,V_3,V_4\right\}= \left\{S_1,\bar{S}_2,V_1,V_2,\bar{V}_3,\bar{V}_4\right\}.
\end{equation}
The remaining functions $\bar{V}_5$, $V_6$ and $\bar{V}_6$ satisfy the same ODEs as $V_5$.
Note that the ODEs for $V_5,V_6,\bar V_5,\bar V_6$ are {\it homogeneous}, they do not have any non-trivial solutions due to the regularity
requirement (at $r=1$) and the vanishing boundary condition (at $r=\infty$):
\begin{equation}\label{V56}
V_5=V_6=\bar V_5=\bar V_6=0.
\end{equation}
Indeed, the conclusion $V_5= \bar V_5=0$ is in perfect agreement with \cite{Bu:2018psl, Bu:2018drd},
where $V_5,\bar{V}_5$ were found to depend on $\rho,\rho_5$ {\it nonlinearly}.

Solving (\ref{eom S1}-\ref{eom Vb4}) near the boundary $r=\infty$ gives rise to the currents' constitutive relations
(\ref{resumjj5}, \ref{resumjj5b}) with the TCFs related to the near boundary expansion of $V_i$'s:
%
\begin{equation}
\mathcal{D}=-2v_1,\qquad \mathcal{D}_B=2v_2,\qquad \bar{\sigma}_{\bar{\chi}}= 2v_3,\qquad \mathcal{D}_\chi= 2v_4,
\end{equation}
where $v_i$'s denote the coefficients of $1/r^2$ in the  expansion of $V_i$'s.
In contrast to our previous publications \cite{1608.08595,Bu:2018drd}, the TCFs $\mathcal{D}$, $\mathcal{D}_B$,
$\bar{\sigma}_{\bar\chi}$ and $\mathcal{D}_\chi$ also depend on the external magnetic field $\vec{\bf B}$ non-perturbatively,
\begin{align} \label{TCFs}
\mathcal{D}=\mathcal{D}(\omega,q^2,(\kappa\vec{\bf B}\cdot \vec q)^2; (\kappa{\bf B})^2), \qquad \qquad \mathcal{D}_B=\mathcal{D}_B(\omega,q^2,(\kappa\vec{\bf B}\cdot \vec q)^2; (\kappa{\bf B})^2),\\
\bar{\sigma}_{\bar{\chi}}=\bar{\sigma}_{\bar{\chi}}(\omega,q^2,(\kappa\vec{\bf B}\cdot \vec q)^2; (\kappa{\bf B})^2), \qquad \qquad \mathcal{D}_\chi=\mathcal{D}_\chi(\omega,q^2,(\kappa\vec{\bf B}\cdot \vec q)^2; (\kappa {\bf B})^2).
\end{align}
Determination of these TCFs is one of the novel results of the present work.
Since the external magnetic field is  constant, the TCFs in (\ref{resumjj5})  measure a response of the chiral medium to inhomogeneity and time-dependence of the
charge densities $\rho,\rho_5$.


\subsection{The TCFs}

%

We are not able  to solve the
ODEs (\ref{eom S1}-\ref{eom Vb4}) analytically and thus resort to numerical techniques.  The results are summarised in several plots below.
%
%
First, we fix the magnetic field $\bf B$ and its relative angle $\alpha$ with respect to $\vec q$, and study the $\omega,q^2$-dependencies of the TCFs in (\ref{resumjj5}, \ref{resumjj5b}).
Then, we examine the effects due to  the magnetic field $\bf B$  and   angle $\alpha$ variations.

\begin{figure}
    \centering
        \begin{subfigure}[h]{0.485\textwidth}
        \includegraphics[width=\textwidth]{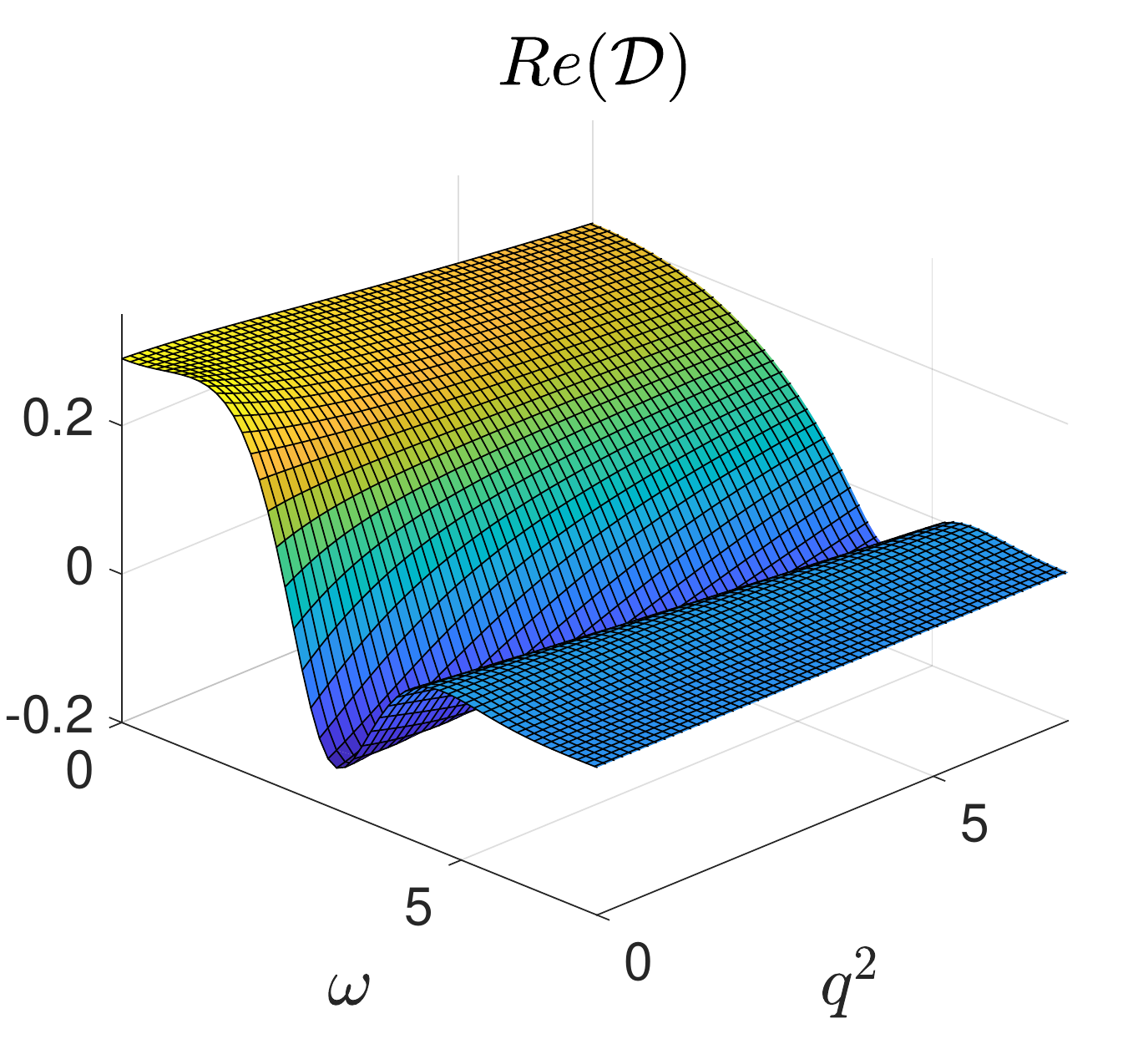}
        \caption{}
        \label{resum_Re_D_kB_025_alph_pi0}
    \end{subfigure}
    ~ 
    \begin{subfigure}[h]{0.485\textwidth}
        \includegraphics[width=\textwidth]{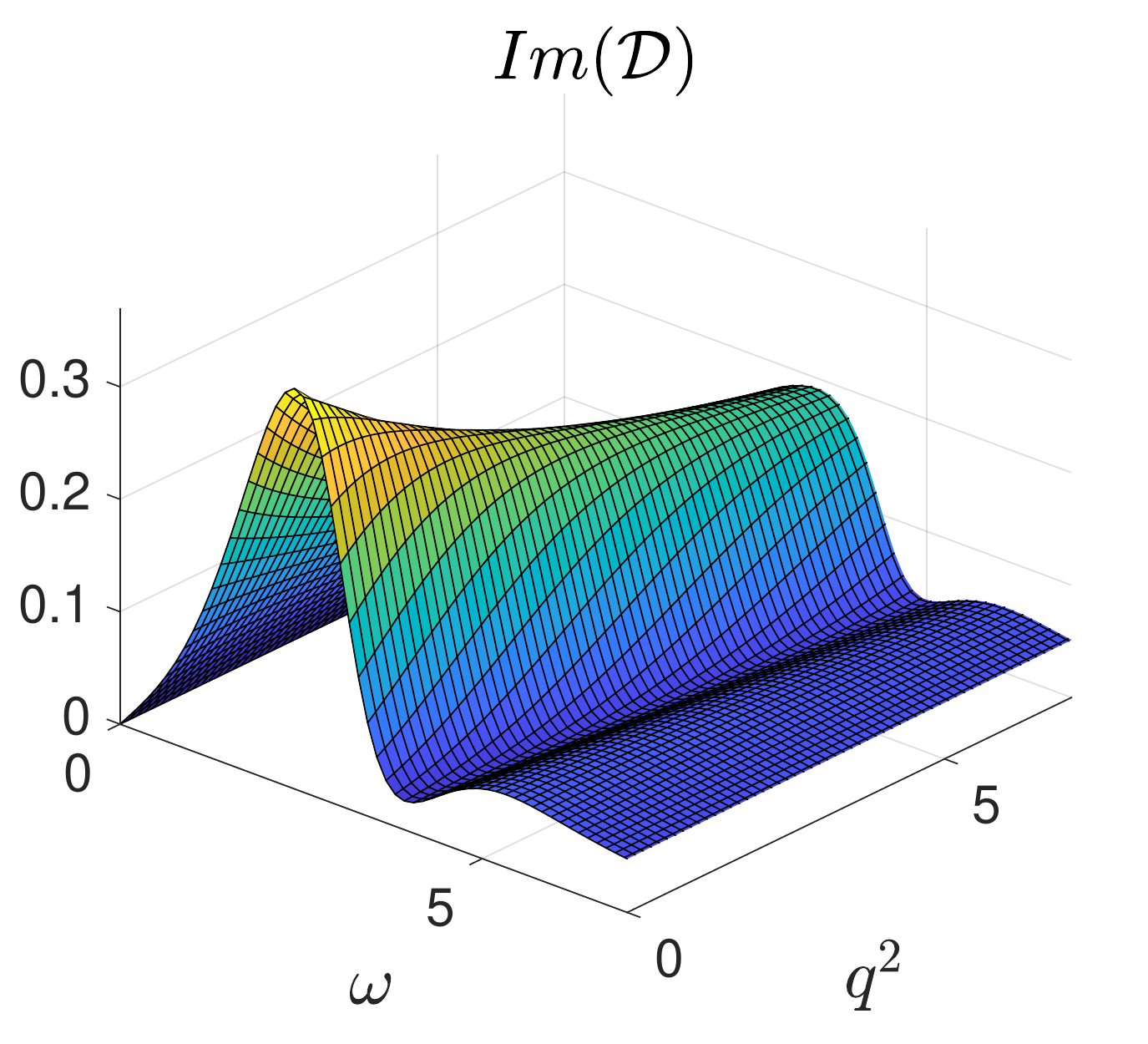}
        \caption{}
        \label{resum_Im_D_kB_025_alph_pi0}
    \end{subfigure}
    ~ 
    \begin{subfigure}[h]{0.485\textwidth}
        \includegraphics[width=\textwidth]{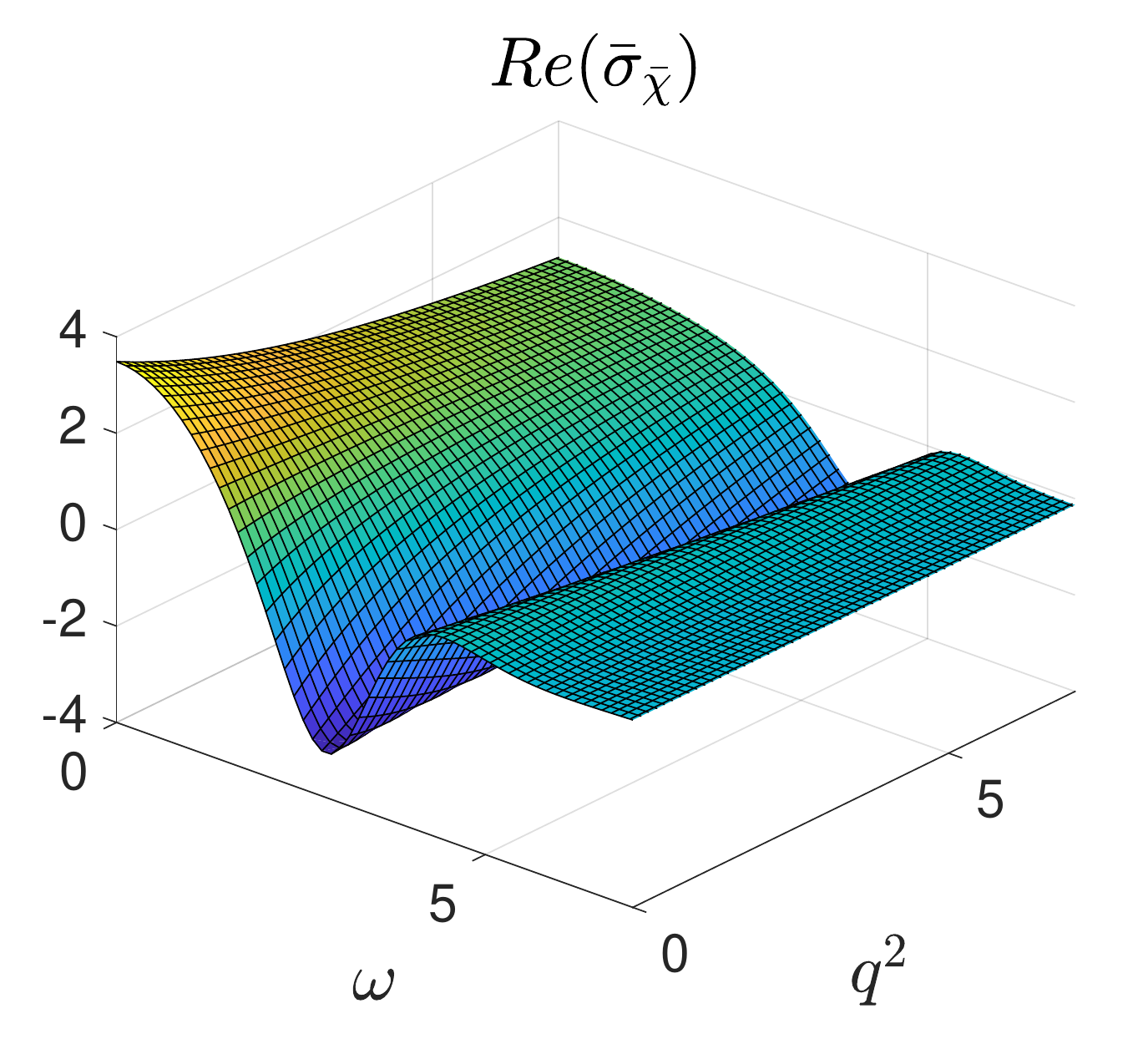}
        \caption{}
        \label{resum_Re_s_barchi_kB_025_alph_pi0}
    \end{subfigure}
 ~ 
    \begin{subfigure}[h]{0.485\textwidth}
        \includegraphics[width=\textwidth]{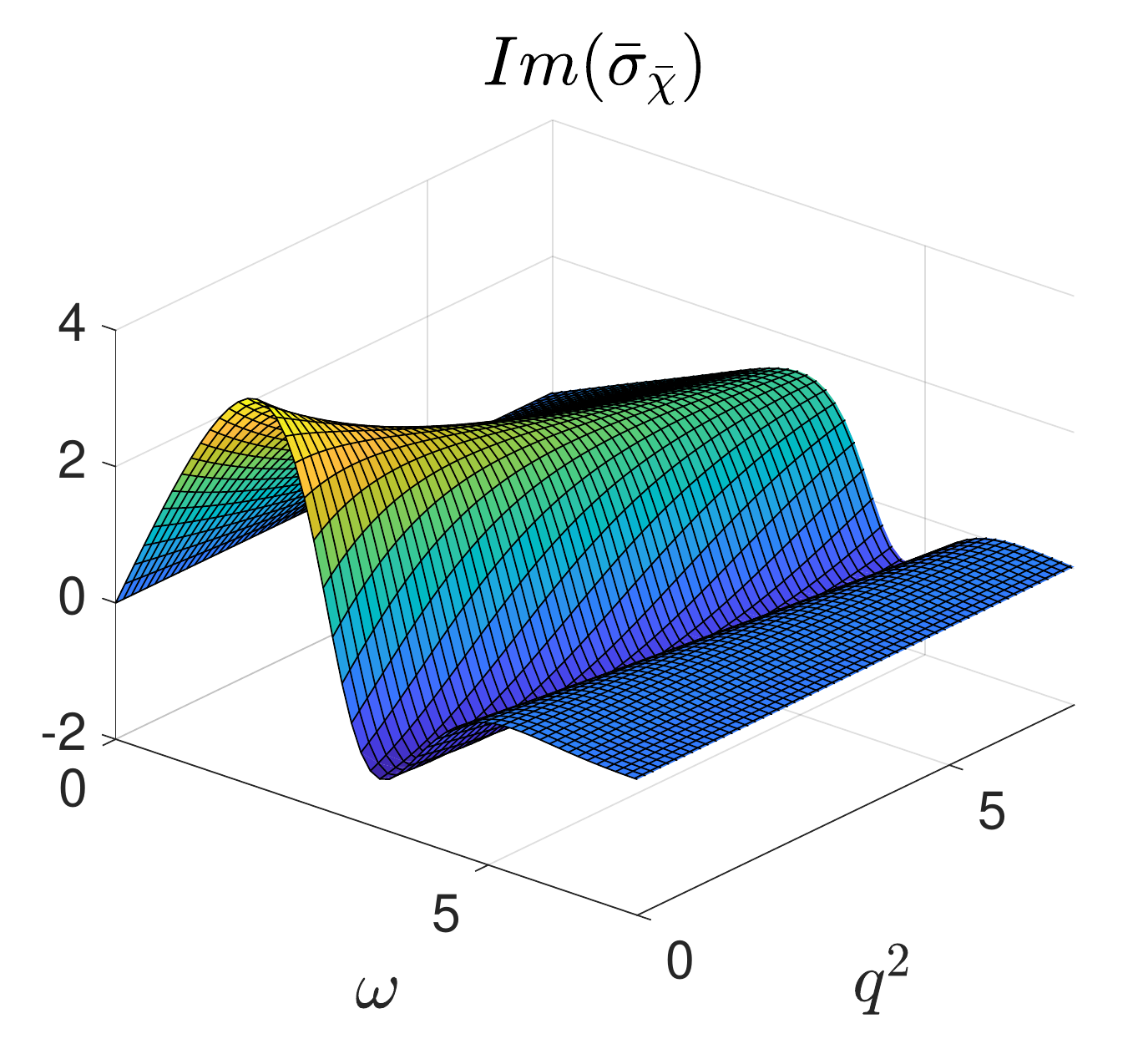}
        \caption{}
        \label{resum_Im_s_barchi_kB_025_alph_pi0}
    \end{subfigure}
    ~ 
    \caption{Diffusion function $\mathcal{D}$ and generalised CME/CSE conductivity $\bar{\sigma}_{\bar{\chi}}$ as functions of $\omega$ and $q^2$ when $\kappa {\bf B}=0.25$ and $\alpha=0$. }\label{resum_fig1}
\end{figure}

\begin{figure}
    \centering
    \begin{subfigure}[h]{0.485\textwidth}
        \includegraphics[width=\textwidth]{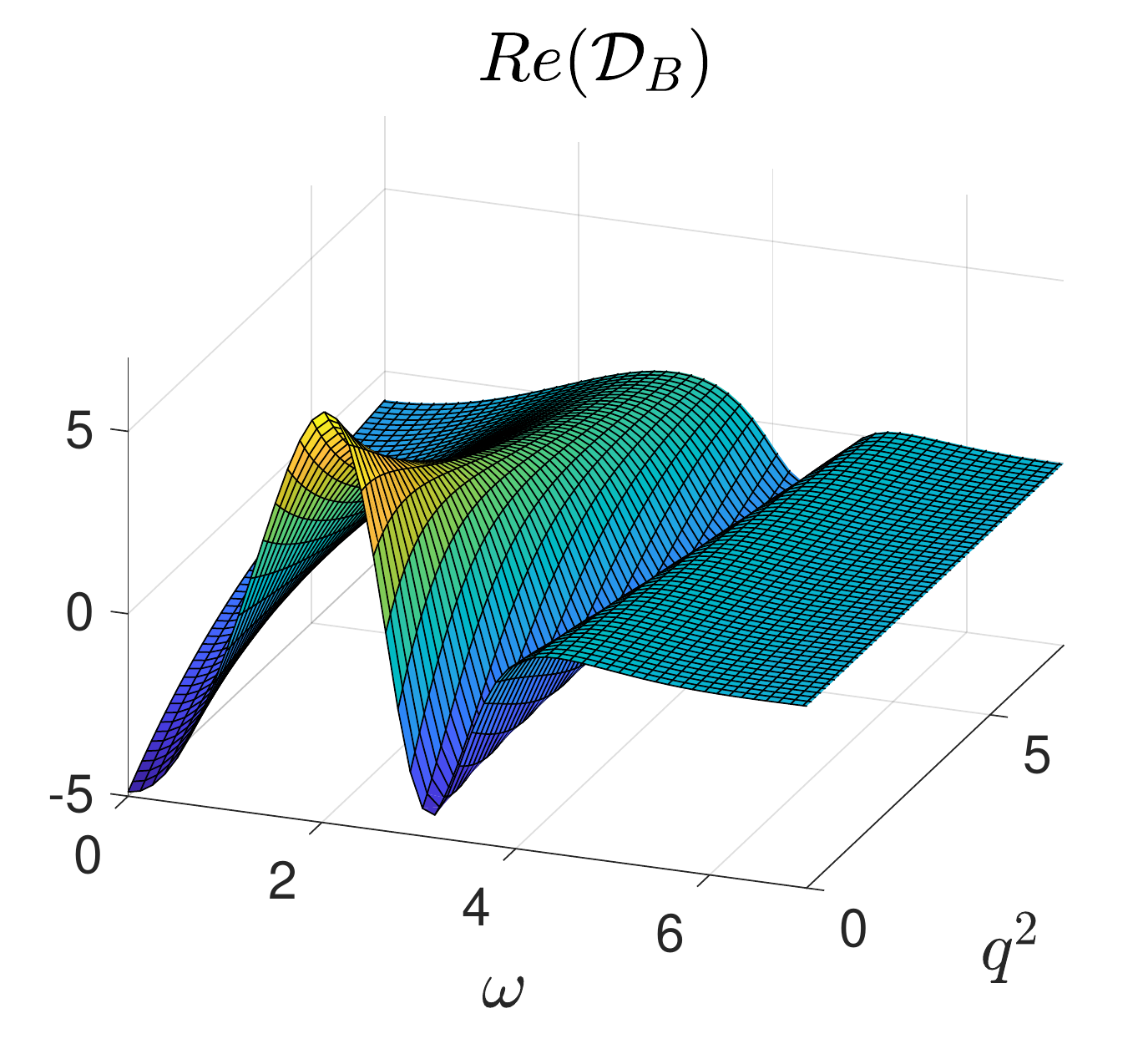}
        \caption{}
        \label{resum_Re_DB_kB_03_alph_pi2}
    \end{subfigure}
    ~ 
    \begin{subfigure}[h]{0.485\textwidth}
        \includegraphics[width=\textwidth]{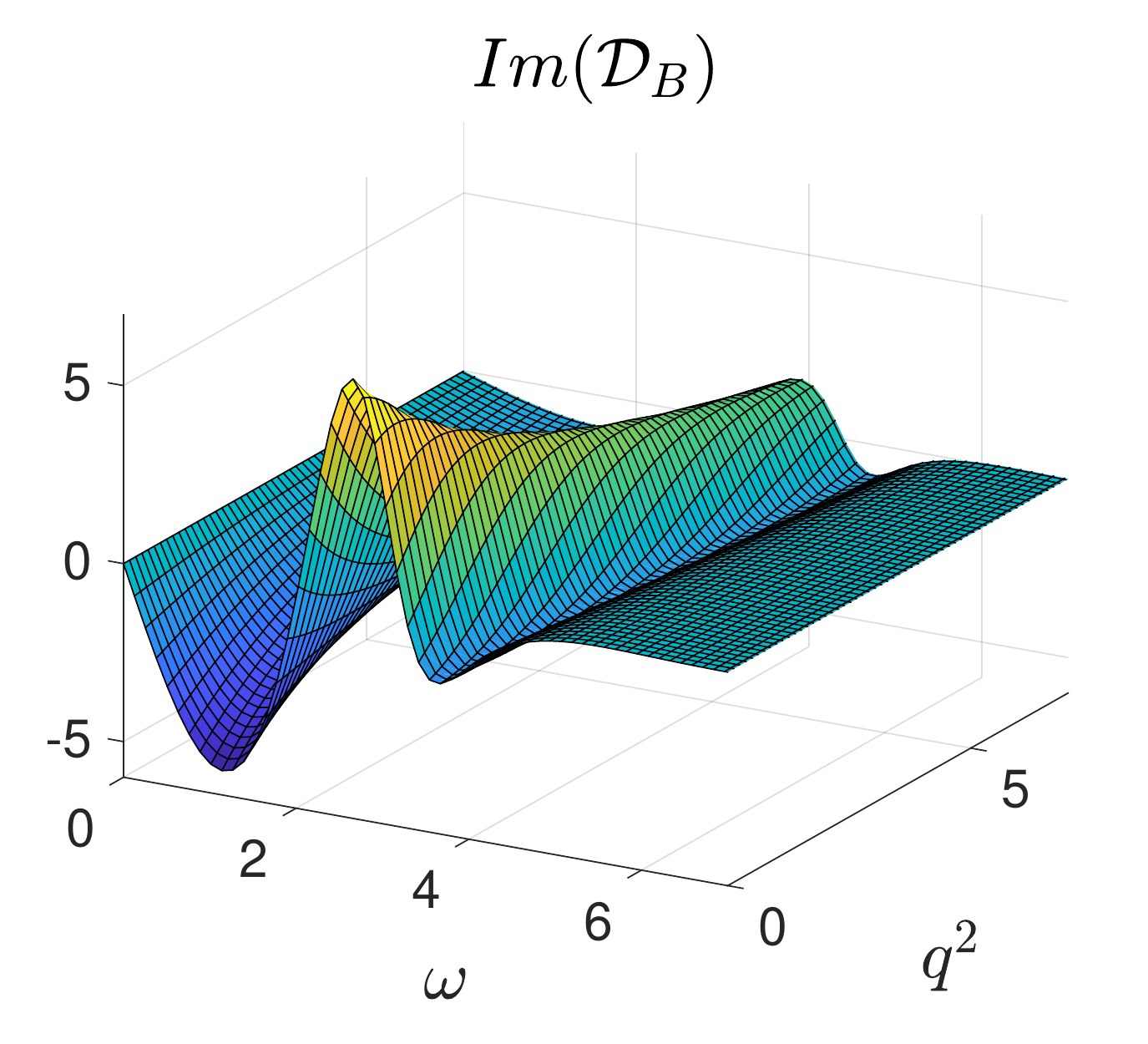}
        \caption{}
        \label{resum_Im_DB_kB_03_alph_pi2}
    \end{subfigure}
    ~ 
        \begin{subfigure}[h]{0.485\textwidth}
        \includegraphics[width=\textwidth]{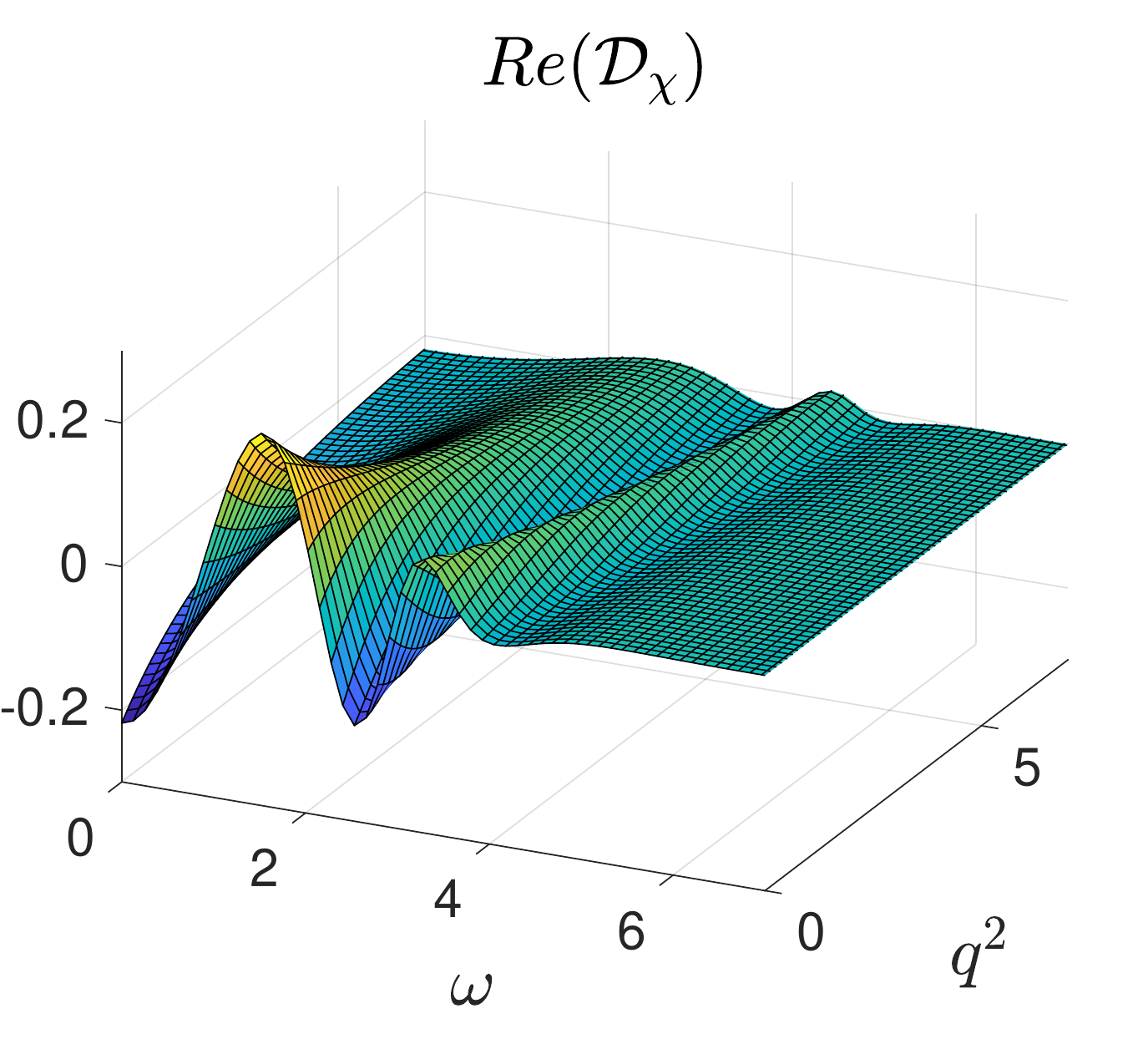}
        \caption{}
        \label{resum_Re_Dchi_kB_03_alph_pi2}
    \end{subfigure}
    ~ 
    \begin{subfigure}[h]{0.485\textwidth}
        \includegraphics[width=\textwidth]{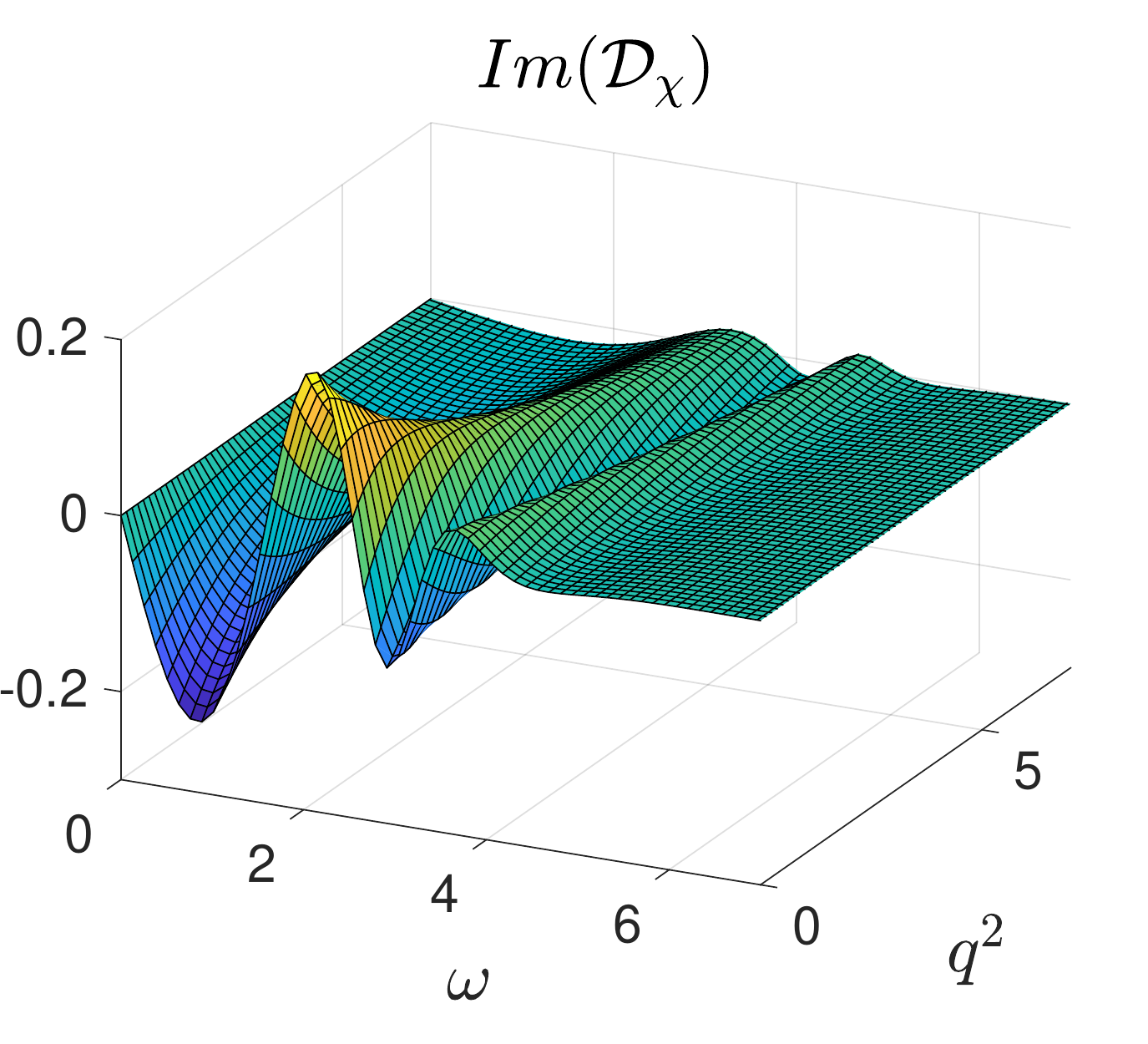}
        \caption{}
        \label{resum_Im_Dchi_kB_03_alph_pi2}
    \end{subfigure}
    ~ 
    \caption{TCFs $\mathcal D_B$ and $\mathcal D_\chi$ as functions of $\omega$ and $q^2$ when $\kappa \mathbf{B}=0.25$ and $\alpha=0$. }\label{resum_fig2}
\end{figure}

In Figures \ref{resum_fig1} and \ref{resum_fig2}, selecting representative values $\kappa{\bf B} =0.25$ and $\alpha=0$, we display 3D plots for all the TCFs in (\ref{resumjj5}, \ref{resumjj5b}) as functions of $\omega$ and $q^2$.  All the TCFs exhibit a similar behavior: relatively weak dependence on the spatial momentum squared $q^2$, reflecting spatial quasi-locality of relevant transport processes; damped oscillations in the frequency $\omega$, towards asymptotic region where all the TCFs essentially vanish.

%
%

\begin{figure}
    \centering
    \begin{subfigure}[h]{0.485\textwidth}
        \includegraphics[width=\textwidth]{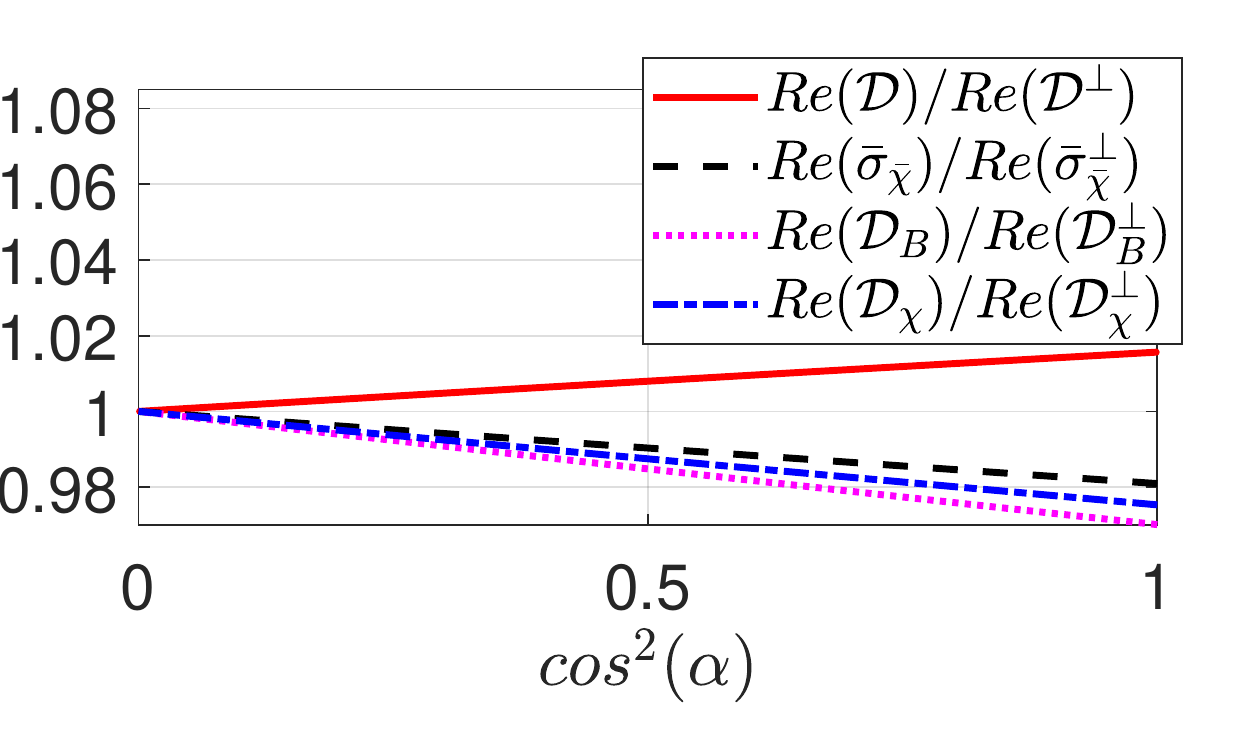}
        \caption{}
        \label{f3a}
    \end{subfigure}
    ~ 
    \begin{subfigure}[h]{0.485\textwidth}
        \includegraphics[width=\textwidth]{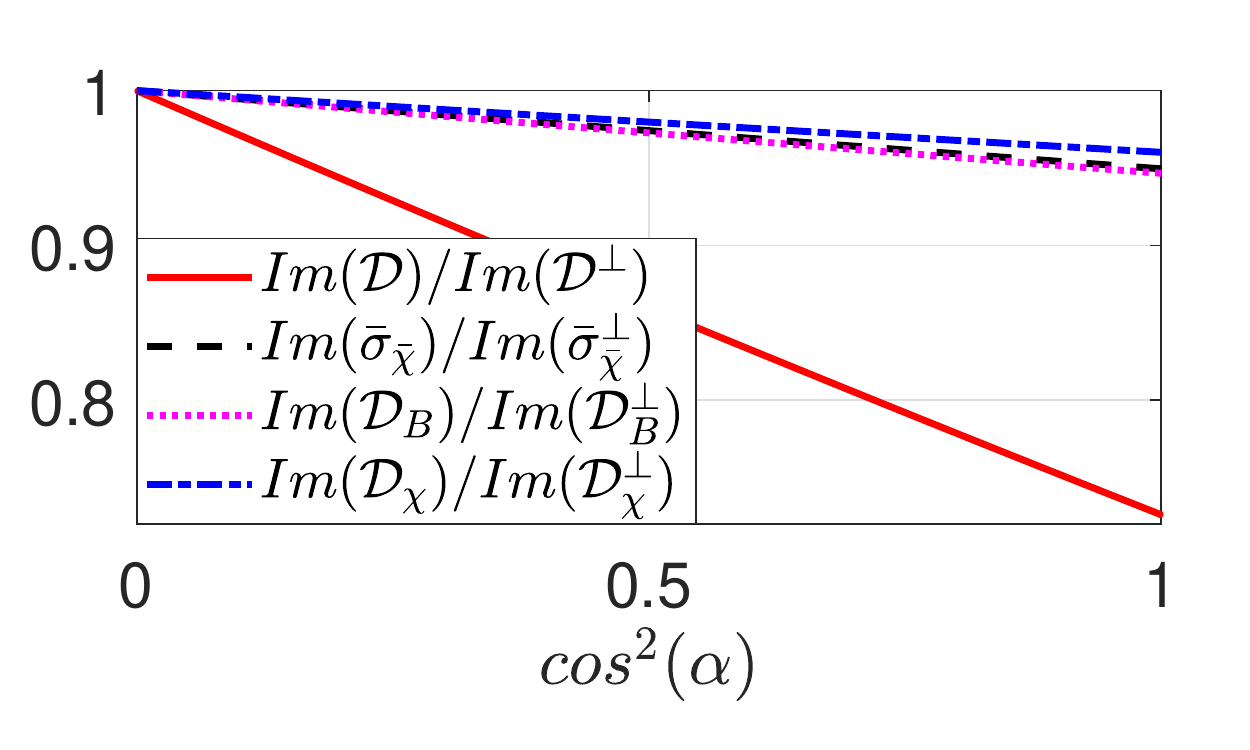}
        \caption{}
        \label{f3b}
    \end{subfigure}
    ~ 
    \caption{Normalised TCFs as functions of $cos^2(\alpha)$ when $\kappa{\bf B}=0.5$ and $\omega=q^2=0.1$. }\label{resum_fig6}
\end{figure}

Figure \ref{resum_fig6} reveals the effect of variation of the angle $\alpha$ between $\vec{\bf B}$ and $\vec q$ on the TCFs.
The results are normalised with respect to their values at $\alpha=\pi/2$ denoted as $\mathcal D^\perp$, $\bar{\sigma}_{\bar\chi}^\perp$, $\mathcal D_B^\perp$, $\mathcal D_\chi^\perp$. In Figure \ref{resum_fig6}, $\kappa{\bf B}=0.5$ and $\omega=q^2=0.1$. All the TCFs display a relatively mild dependence on $cos^2(\alpha)$.


%

\begin{figure}
    \centering
    \begin{subfigure}[h]{0.485\textwidth}
        \includegraphics[width=\textwidth]{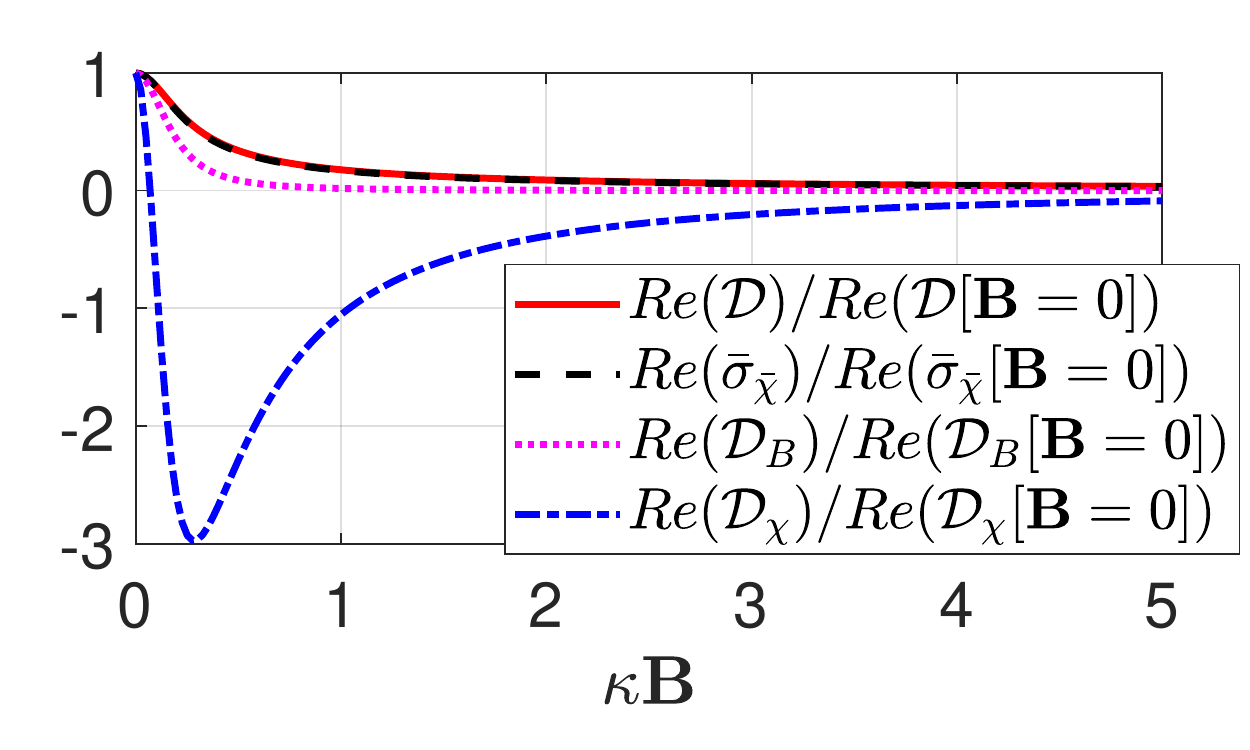}
        \caption{}
        \label{f3a}
    \end{subfigure}
    ~ 
    \begin{subfigure}[h]{0.485\textwidth}
        \includegraphics[width=\textwidth]{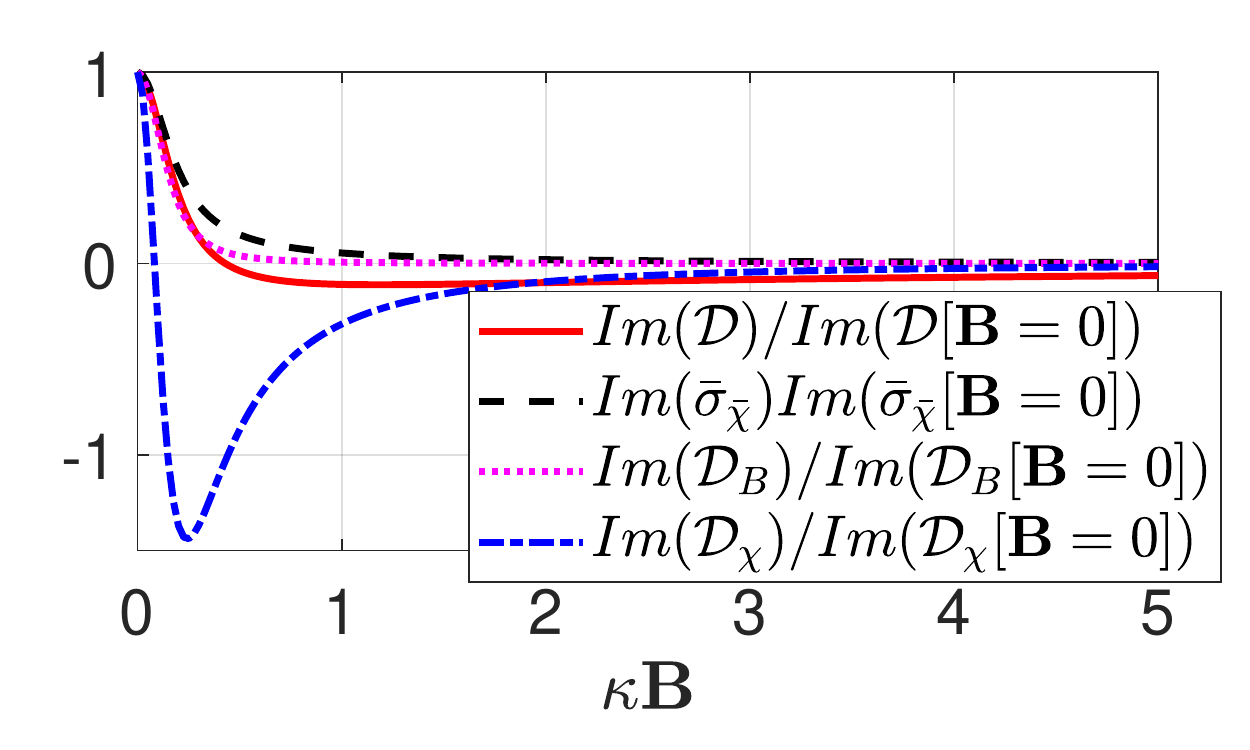}
        \caption{}
        \label{f3b}
    \end{subfigure}
    ~ 
        \begin{subfigure}[h]{0.485\textwidth}
        \includegraphics[width=\textwidth]{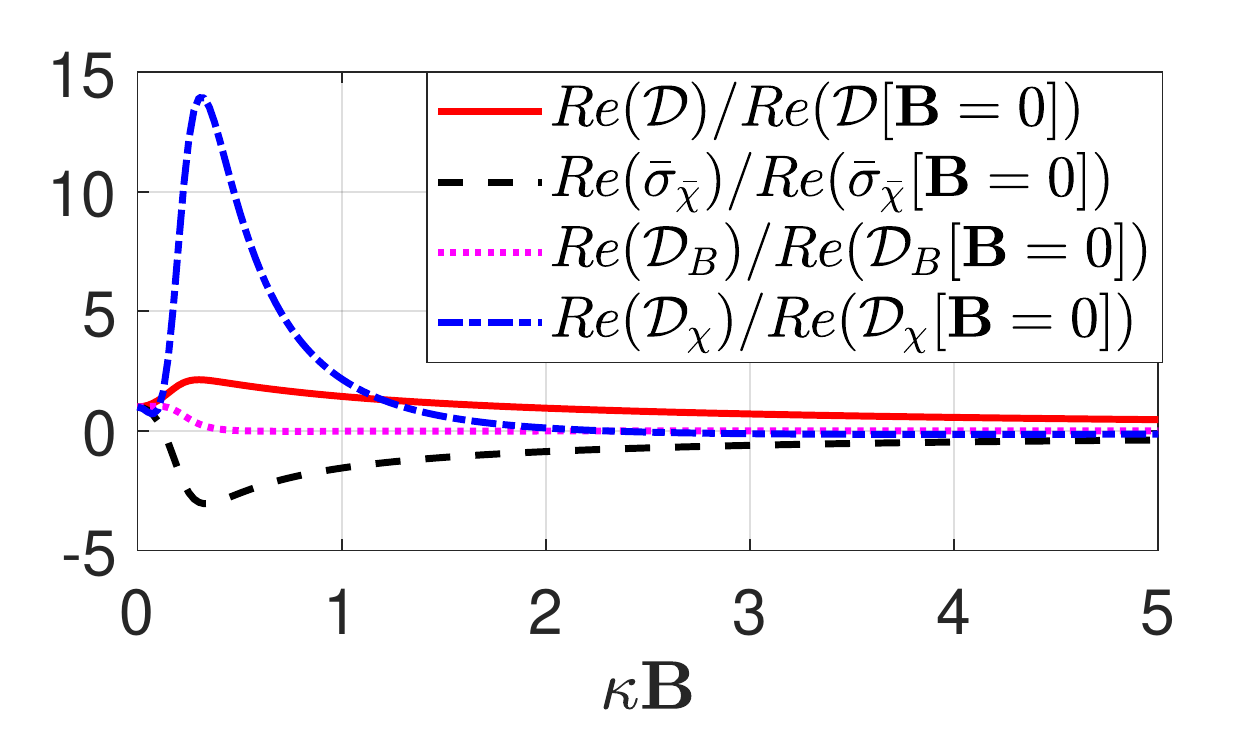}
        \caption{}
        \label{f3c}
    \end{subfigure}
    ~ 
    \begin{subfigure}[h]{0.485\textwidth}
        \includegraphics[width=\textwidth]{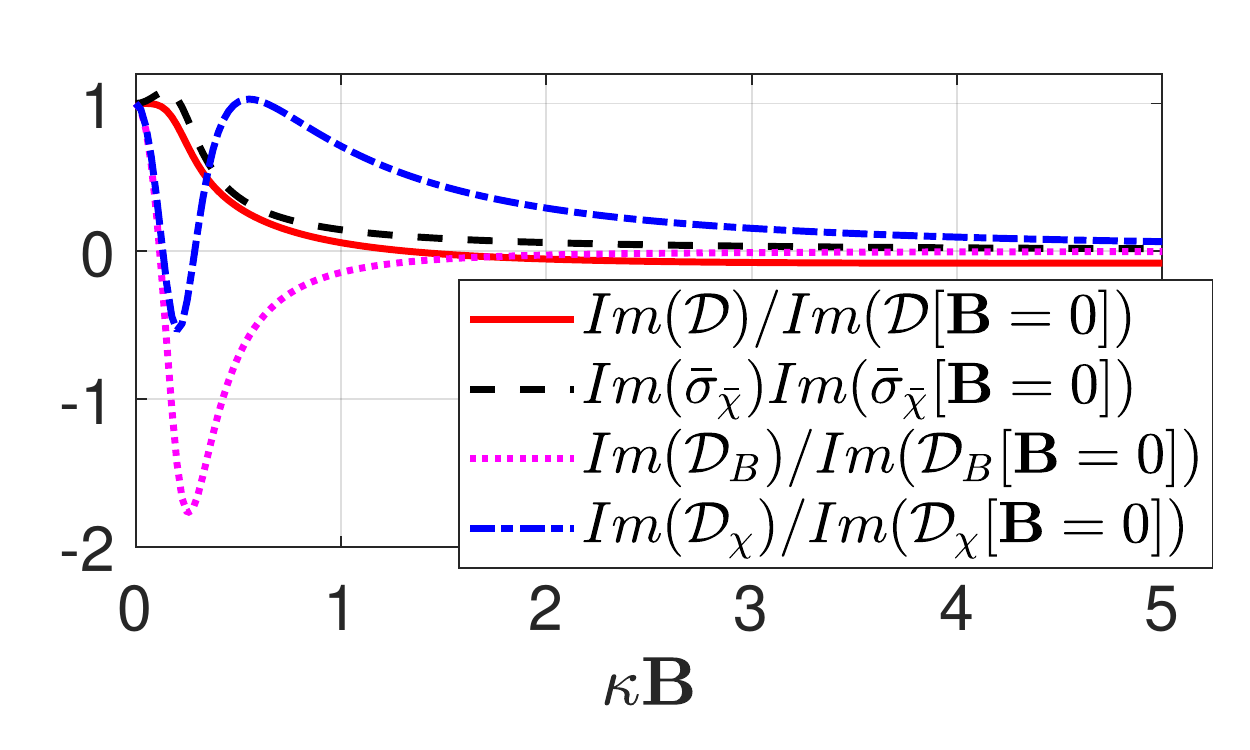}
        \caption{}
        \label{f3d}
    \end{subfigure}
    ~ 
    \caption{Real and imaginary parts of the TCFs as a function of $\kappa \mathbf B$ for (a), (b) $\omega=q^2=0.1$ and (c), (d) $\omega=q^2=2$. }\label{resum_fig3}
\end{figure}

In Figure \ref{resum_fig3} we show the TCFs as functions of $\kappa {\bf B}$ while fixing all the rest of the parameters $\omega,q^2$ and $\alpha$.
Since the latter effect is weak, we take $\alpha=0$ in the subsequent discussion. Regarding $\omega$ and $q^2$, we make two different choices to implement a comparative study: $\omega=q^2=0.1$ versus $\omega=q^2=2$. In Figure \ref{resum_fig3}, the TCFs are normalized with respect to their values when $\kappa\mathbf B=0$  with the same $\omega$ and $q^2$ values (which are obtained from our numerical results).
All the TCFs are found to approach zero  at large magnetic field. When $\omega,q^2$ are increased (i.e., with more non-hydrodynamic modes included), the asymptotic regime is  shifted towards  larger values of $\kappa {\bf B}$.

\begin{figure}
    \centering
        \begin{subfigure}[h]{0.3\textwidth}
        \includegraphics[width=\textwidth]{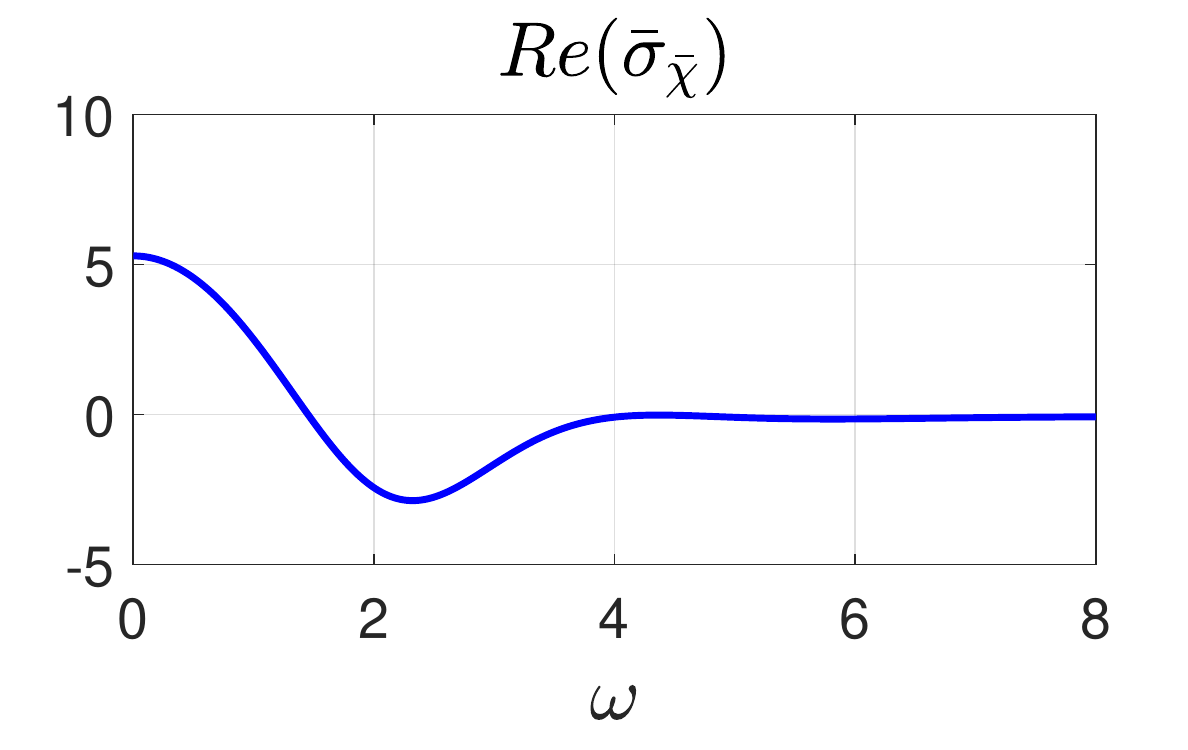}
        \caption{}
        \label{f4a}
    \end{subfigure}
    ~ 
    \begin{subfigure}[h]{0.3\textwidth}
        \includegraphics[width=\textwidth]{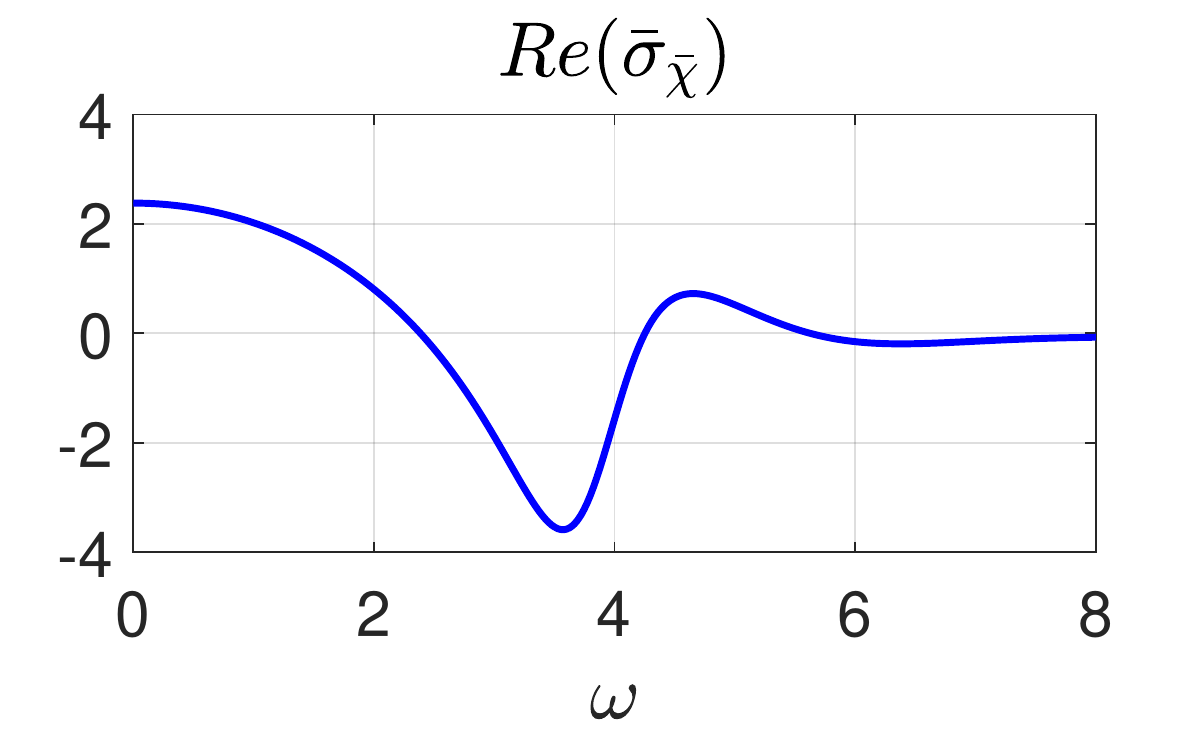}
        \caption{}
        \label{f4b}
    \end{subfigure}
    ~ 
     \begin{subfigure}[h]{0.3\textwidth}
        \includegraphics[width=\textwidth]{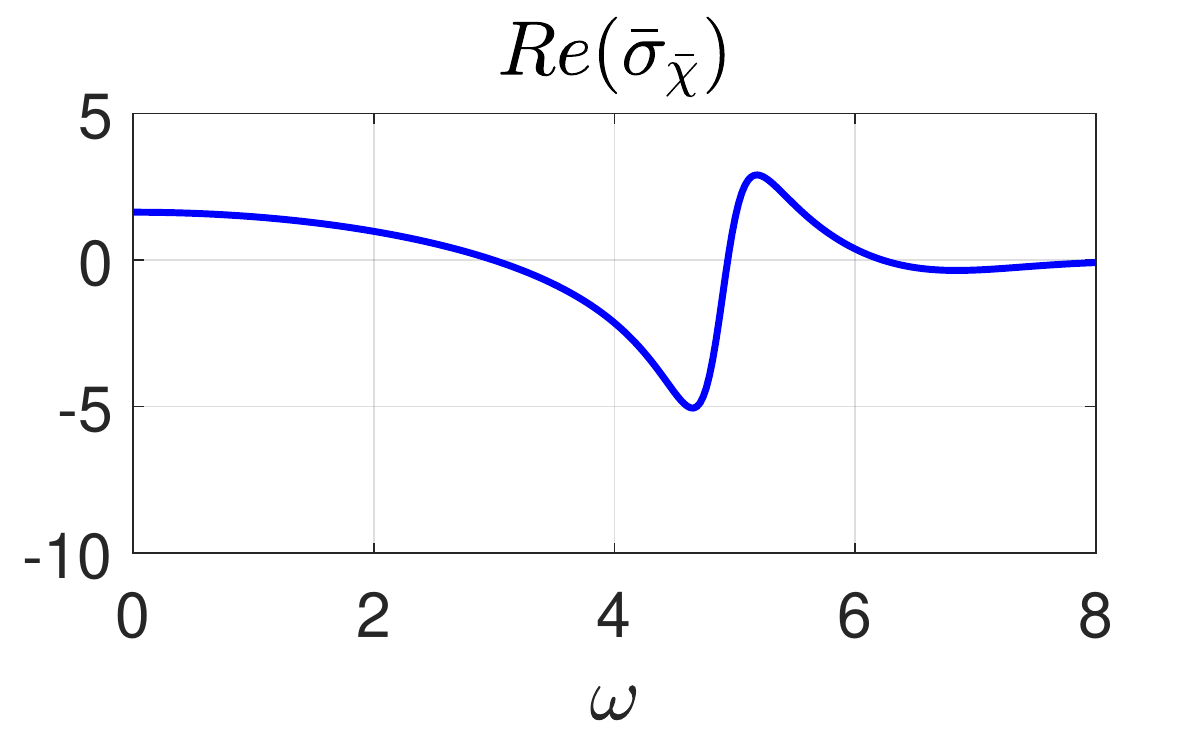}
        \caption{}
        \label{f4c}
    \end{subfigure}
    \begin{subfigure}[h]{0.3\textwidth}
        \includegraphics[width=\textwidth]{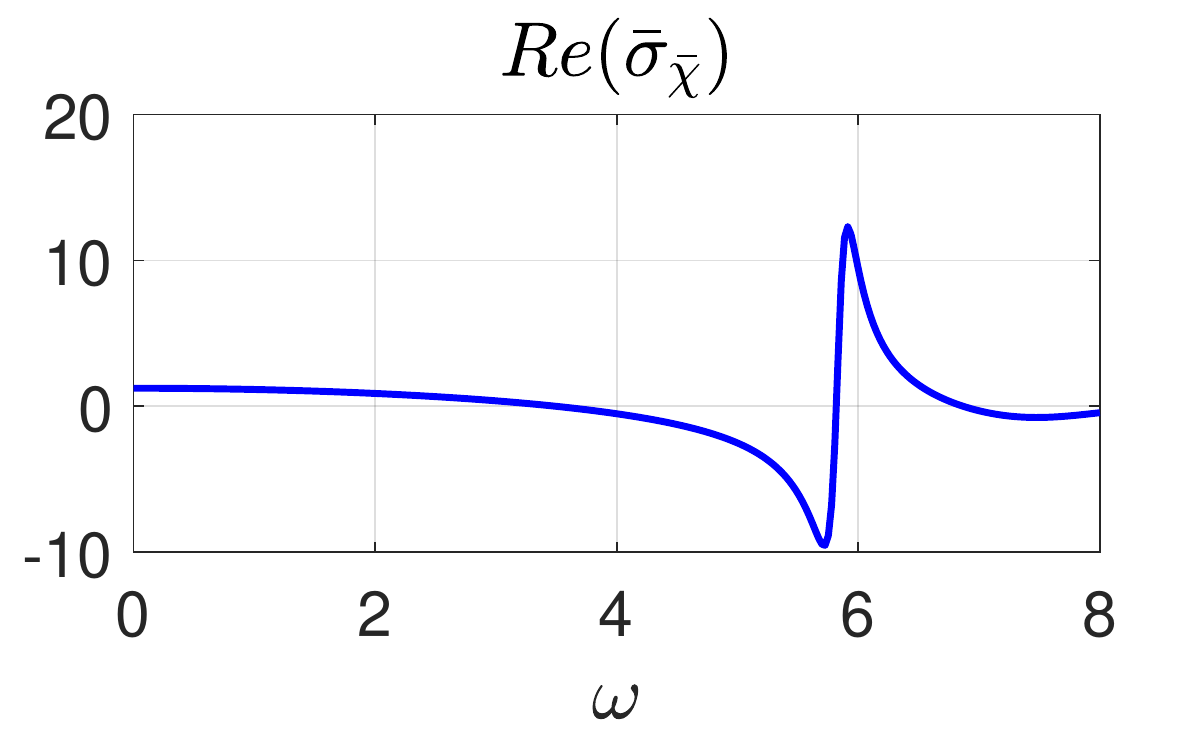}
        \caption{}
        \label{f4d}
    \end{subfigure}
    ~ 
     \begin{subfigure}[h]{0.3\textwidth}
        \includegraphics[width=\textwidth]{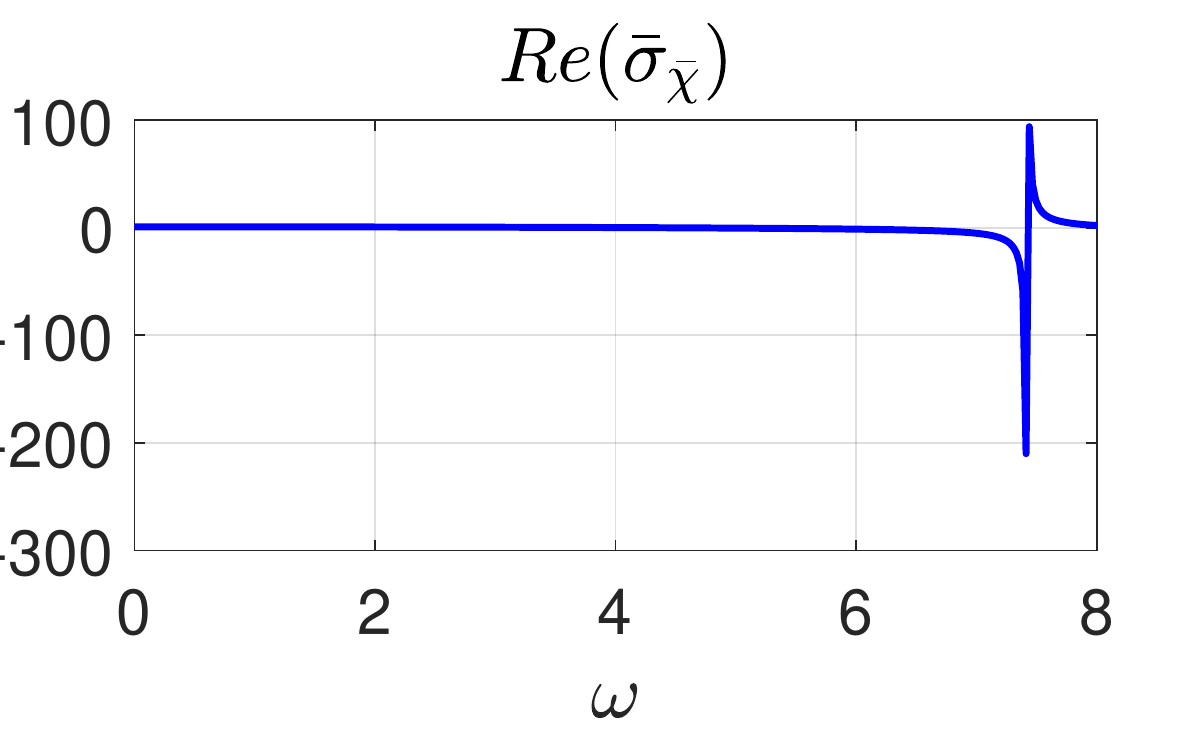}
        \caption{}
        \label{f4e}
    \end{subfigure}
     \begin{subfigure}[h]{0.3\textwidth}
        \includegraphics[width=\textwidth]{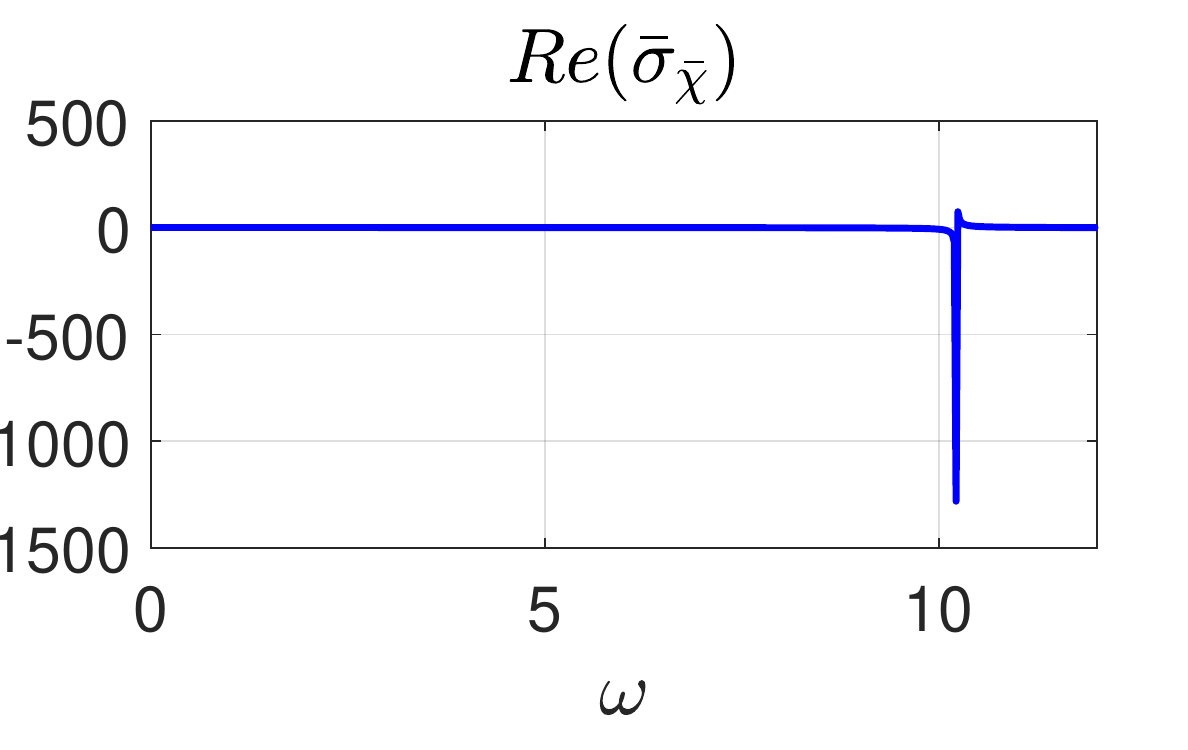}
        \caption{}
        \label{f4f}
    \end{subfigure}
    \caption{$Re(\bar{\sigma}_{\bar\chi})$ as a function of $\omega$ when $q=0$ and (a) $\kappa \mathbf{B}=0.1$, (b) $\kappa \mathbf{B}=0.4$, (c) $\kappa \mathbf{B}=0.6$, (d) $\kappa \mathbf{ B}=0.8$,(e) $\kappa \mathbf{B}=1.2$, (f) $\kappa \mathbf{B}=2.2$. }\label{res_fig4}
\end{figure}

An interesting new phenomenon emerges at finite $\kappa {\bf B}$: when the magnetic field is large enough, say $\kappa{\bf B}\gtrsim0.5$, the TCFs develop singularity as functions of $\omega$, at real values of $\omega$ depending on  $\kappa {\bf B}$ (see Figure \ref{res_fig4}  as illustration).
Moreover,  all the TCFs (both real and imaginary parts) exhibit singularity at the same value of $\omega$.
In Figure \ref{res_fig4},  $Re[\bar{\sigma}_{\bar{\chi}}]$ is taken as an example demonstrating the phenomenon at $q=0$.
 If one continues to increase $\kappa {\bf B}$ and considers larger $\omega$, additional singularities  emerge at larger values of $\omega$.
 The  locations of these singularities $\omega_n$ are symmetric around the origin.
These  $\omega_n$ are presumably the quasi-normal modes (QNM) in the bulk and here we find that they become real at some values of magnetic field.
Within essentially the same holographic model, similar observation was made recently in  \cite{Haack:2018ztx}.


\begin{figure}[htbp]
\centering
\includegraphics[width=0.4\textwidth]{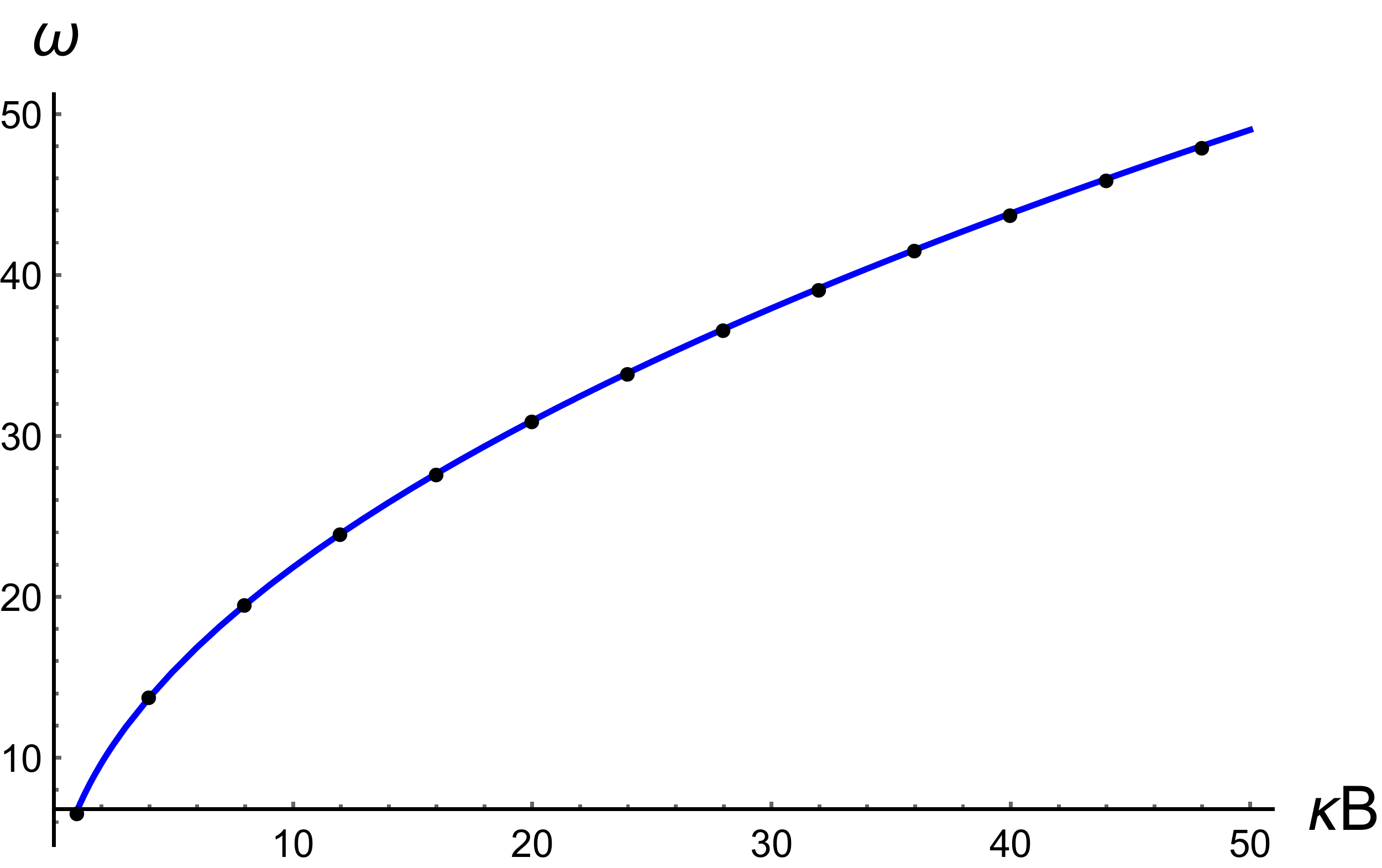}
\caption{Location of the singularity as a function of $\kappa \bf B$: numerical result (black dots)
is best fitted by $\omega=-0.155157 + 6.95306 \sqrt{\kappa \bf B}$ (blue curve).}\label{fig5}
\end{figure}

To conclude this part, we present the dependence of the lowest QNM mode on $\kappa \mathbf B$, see Figure \ref{fig5}.
The numerical results (black dots) are best fitted by:
\begin{align} \label{QNM fit}
\omega=-0.155157+6.95306 \sqrt{\kappa \mathbf B},
\end{align}
which is plotted as a continuous (blue) curve in Figure \ref{fig5}.  The mode with
$Re(\omega)\propto \sqrt{\kappa \mathbf B}$ while $Im(\omega)\rightarrow 0$ is a manifestation of the Landau level behavior \cite{Ammon:2016fru,Ammon:2017ded}.
Magnetic field dependence of the QNMs  in a holographic model was first studied in \cite{Janiszewski:2015ura, Demircik:2016nhr} but without any anomaly effects included.

\subsection{Non-dissipative CMW modes}

The original CMW is a dissipative wave at small momenta \cite{Kharzeev:2010gd}. In \cite{Bu:2018drd}, we asked the question if it can happen that, beyond the hydrodynamic limit,
the dissipative (imaginary) part of the CMW vanishes. The answer is within the all order dispersion relation.
We indeed found that the CMW possesses a discrete spectrum of non-dissipative modes when the magnetic field is larger than  a critical value.
Yet, the results of \cite{Bu:2018drd} were based on a ``weak'' magnetic field approximation. Below we  reexamine the very same question,
but now without any approximations involved.

The  constitutive relations (\ref{resumjj5}, \ref{resumjj5b}), combined with the continuity equations (\ref{cont eqn}), result in the exact CMW dispersion relation
%
\begin{align}\label{cmwexact}
\omega= \pm \left(\bar{\sigma}_{\bar{\chi}}- q^2 \mathcal{D}_\chi \right)\kappa \vec{\bf B} \cdot \vec q-i\left(q^2\mathcal{D}- \mathcal{D}_B (\kappa\vec{\bf B} \cdot \vec q)^2 \right),
\end{align}
The last term is new compared to \cite{Bu:2018drd}. Furthermore, all the TCFs in (\ref{cmwexact}) are functions of the magnetic field.

The procedure for finding a purely real solution was devised  in \cite{Bu:2018drd}. To this goal, we first split the dispersion relation (\ref{disprel}) into real and imaginary parts (assuming  $\vec{\mathbf{B}}\parallel \vec q$)
\begin{equation}\label{PRPI}
\begin{split}
&\phi_R(\omega, q^2, \kappa \mathbf B)\equiv-\omega\pm  \text{Re}[(\bar{\sigma}_{\bar\chi}-q^2 \mathcal D_\chi)\kappa q \mathbf B ]+\text{Im}[q^2 \mathcal D-\mathcal D_B (\kappa q\mathbf B )^2],\\
&\phi_I(\omega, q^2, \kappa \mathbf B)\equiv\pm  \text{Im}[(\bar{\sigma}_{\bar\chi}-q^2 \mathcal D_\chi)\kappa q \mathbf B ]-\text{Re}[q^2 \mathcal D-\mathcal D_B (\kappa q\mathbf B )^2].
\end{split}
\end{equation}
Then we search for a real $\omega$ solution of two equations $\phi_R=0$ and $\phi_I=0$.
\begin{figure}
    \centering
        \begin{subfigure}[h]{0.485\textwidth}
        \includegraphics[width=\textwidth]{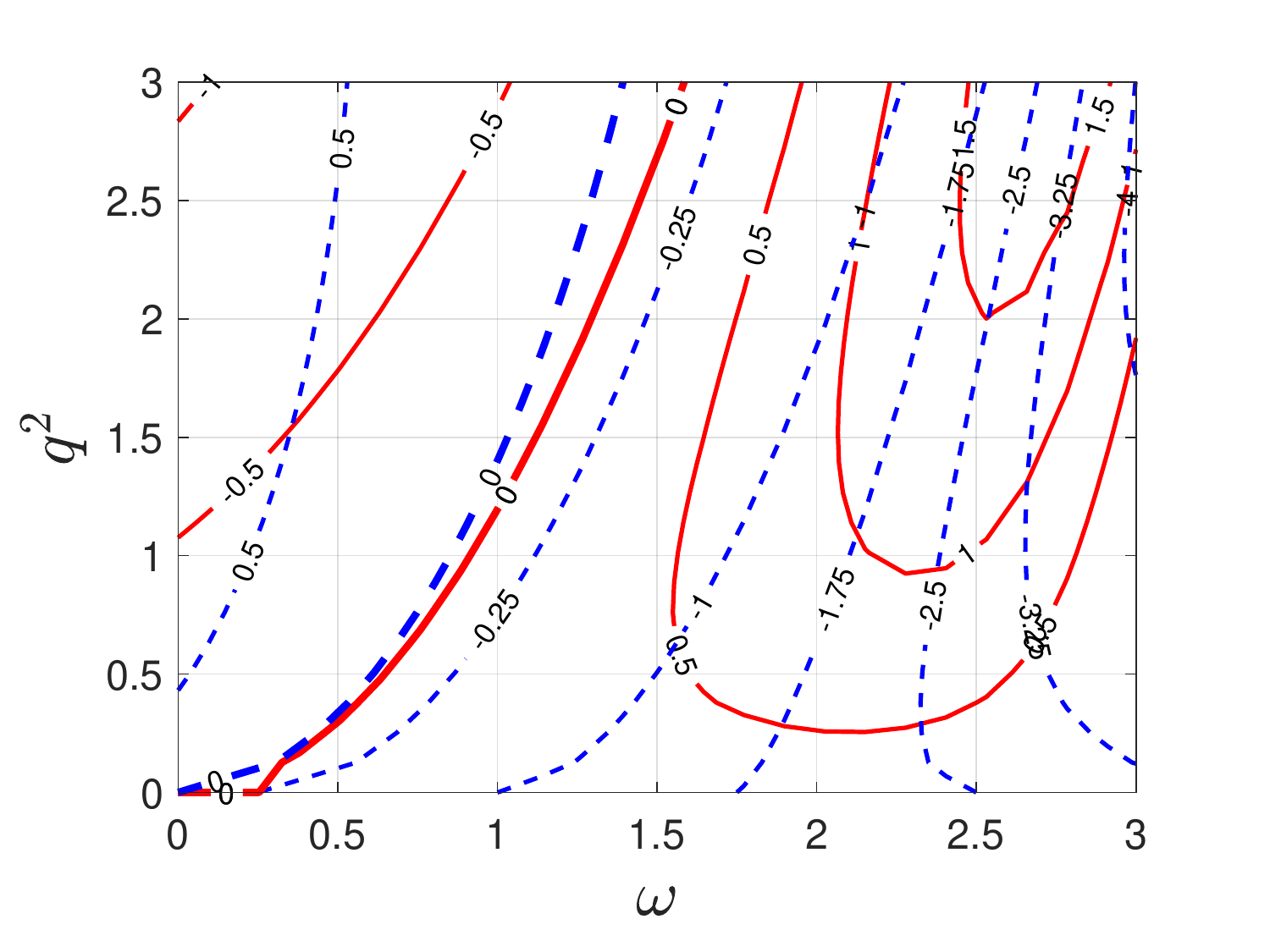}
        \caption{}
        \label{f7a}
    \end{subfigure}
    ~ 
    \begin{subfigure}[h]{0.485\textwidth}
        \includegraphics[width=\textwidth]{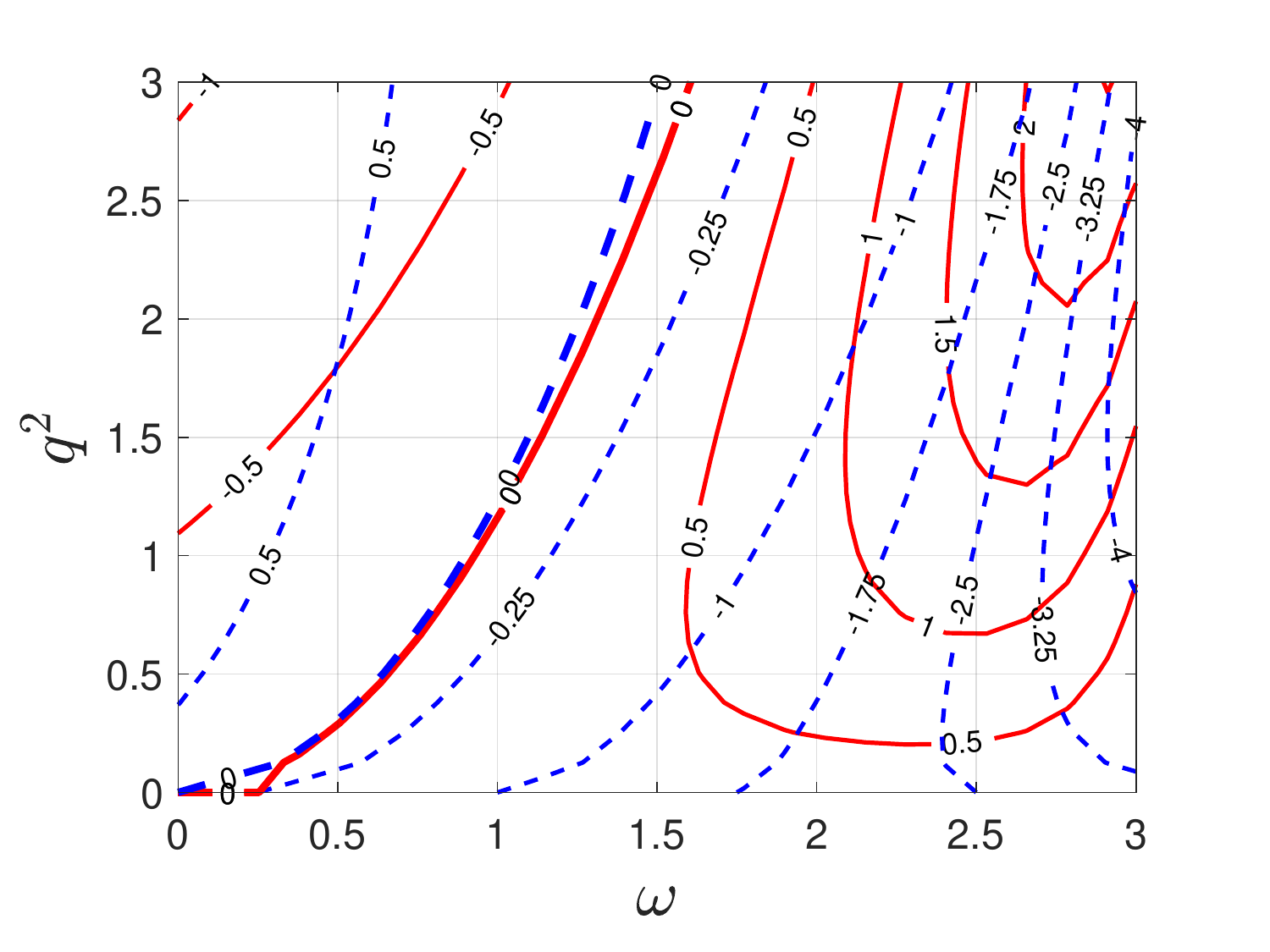}
        \caption{}
        \label{f7b}
    \end{subfigure}
    ~ 
     \begin{subfigure}[h]{0.485\textwidth}
        \includegraphics[width=\textwidth]{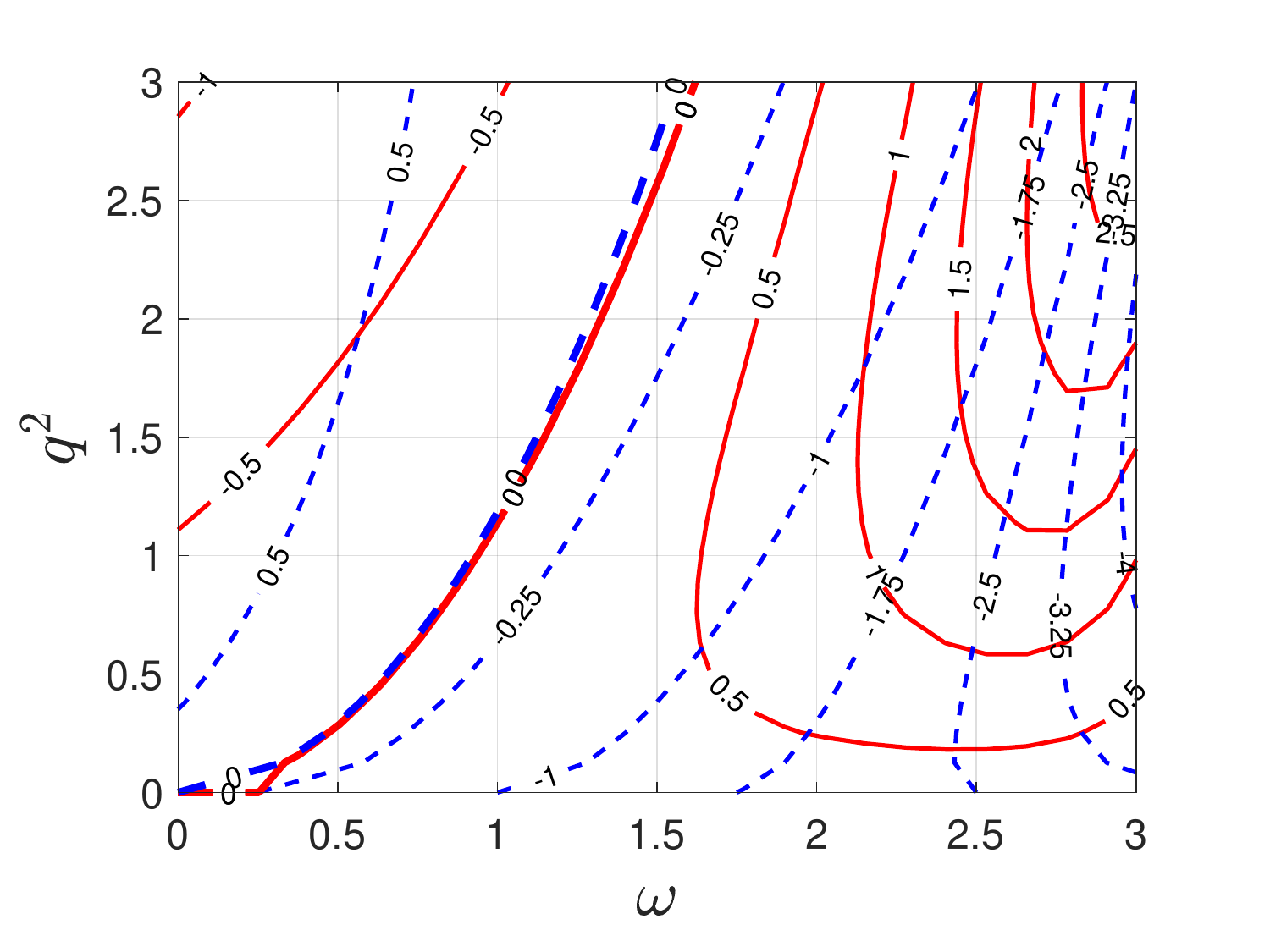}
        \caption{}
        \label{f7c}
    \end{subfigure}
    \begin{subfigure}[h]{0.485\textwidth}
        \includegraphics[width=\textwidth]{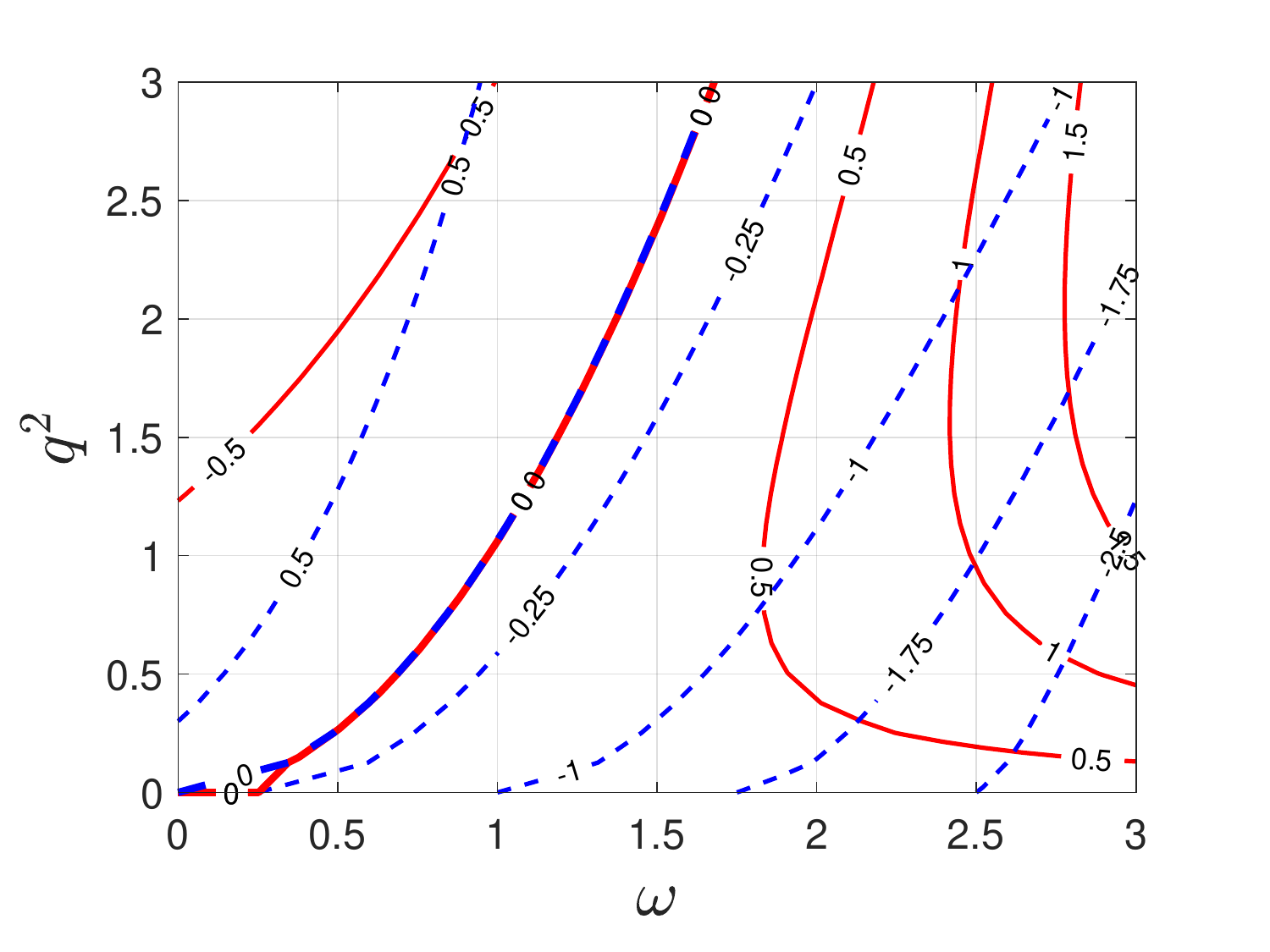}
        \caption{}
        \label{f7d}
    \end{subfigure}
    ~ 
    ~ 
    ~ 
    \caption{Contour plots for the functions $\phi_R$ (blue dashed) and $\phi_I$ (red solid) at (a) $\kappa {\bf B}=0.25$, (b) $\kappa {\bf B}=0.30$, (c) $\kappa {\bf B}=0.33$, (d) $\kappa {\bf B}=0.5$. }\label{res_fig7}
\end{figure}

In Figure \ref{res_fig7}, the functions $\phi_R$ and $\phi_I$ (with  upper plus sign in (\ref{PRPI})) are shown as contour plots in $(\omega, q^2)$ space for representative values of the magnetic field: (a) $\kappa\mathbf B=0.25$, (b) $\kappa\mathbf B=0.30$, (c) $\kappa\mathbf B=0.33$, and (d) $\kappa\mathbf B=0.5$. The dashed
(blue) and solid (red) curves denote  $\phi_R$ and $\phi_I$ respectively. The numbers indicated on
the curves are the values of these functions along the curves. Therefore, any crossing point of $\phi_R=0$ (bold dashed blue curve) and $\phi_I=0$ (bold solid red curve) implies that both functions vanish simultaneously. This is the desired solution.


 In \cite{Bu:2018drd},  such a crossing occurred at a ``single point'' in the $(\omega, q^2)$ space, for each $\kappa \mathbf B \ge 0.33$,
 similarly to the case shown in Figure \ref{f7a}. However, beyond the  weak field approximation, new branches emerge as demonstrated in Figures \ref{f7b}, \ref{f7c} and \ref{f7d}: the contours $\phi_R=0$ and $\phi_I=0$ coincide within a ``continuous interval'' in the $(\omega, q^2)$ space, starting from  a large enough $\kappa \mathbf B$. This implies a continuum  of non-dissipative CMW modes.
 In Figure \ref{f7d}, it seems like the contours  $\phi_R=0$  and $\phi_I=0$  coincide completely, though the Figure is somewhat misleading.
  Indeed, by zooming in $(\omega,q^2)$ space, we observe that the solution exists only on piece-wise intervals in the $(\omega,q^2)$ space.
It is also worth adding that there are additional solution branches at larger $\omega$, which we do not display here.

Finally,  the results above were obtained for non-dissipative CMW propagating along the magnetic field direction. One can ask the question if similar wave could propagate
at an angle  $\alpha$ with respect to the magnetic field, or even orthogonal to it.
We  briefly report that changing $\alpha$ from $\vec{q}\parallel\vec{\mathbf{B}}$ to $\vec{q}\perp\vec{\mathbf{B}}$ leads to separation of $\phi_R=0$ and $\phi_I=0$ contours.
Non-dissipative solutions disappear at  a critical value of $\alpha=\alpha_c[\kappa\mathbf{B]}$  (e.g. $\alpha_c\simeq\pi/6$  for $\kappa\mathbf{B}=0.33$).

\section{Concluding remarks}\label{conclusion}

In this work we focused on the influence of strong background e/m fields  on the chiral anomaly-induced transport phenomena for a holographically defined thermal plasma.
The constitutive relations for the vector and axial currents $\vec J, \vec J_5$ were evaluated
 within two complementary approximation schemes: fixed order gradient expansion (up to the first order) and the all-order gradient resummation (linear in the charge densities
 $\rho,\rho_5$). A summary of all the results could be found in the introductory section.
The main highlights of our study are:

\noindent $\bullet$  There are three types of gapless modes propagating in the chiral plasma: the chiral magnetic wave (CMW) \cite{Kharzeev:2010gd},
the chiral electric wave (CEW) \cite{Pu:2014fva} and the chiral Hall density wave (CHDW) \cite{Bu:2018psl,Bu:2018drd},
which could be searched experimentally in heavy ion collisions or,  more likely, in condensed matter experiments where the external fields are under better control.

\noindent $\bullet$  While most of the transport coefficients are found to be suppressed by the external fields and vanish at asymptotically strong fields,
the Ohmic conductivity gets enhanced in parallel electrical and magnetic fields, which is  an experimentally interesting phenomenon to be searched for.

\noindent $\bullet$ Some anomaly-induced transport phenomena display noticeable dependence on the relative angle between  the external fields
$\vec{\mathbf E}$ and $ \vec{\mathbf B}$.
This  sensitivity could be used in real experiments to  zoom into one or another anomaly-induced phenomena.






\noindent $\bullet$ When $\vec{\bf E}=0$, the all-order resummed constitutive relations (\ref{resumjj5}, \ref{resumjj5b}) are parameterised by four independent TCFs, which are functions of $\omega,\vec q$ and $\vec{\bf B}$. Intriguingly, these TCFs are found to show a common singularity at certain value of real $\omega$ when $\kappa {\bf B}$ is strong enough. Moreover, this singularity is identified as QNM frequency and obey Landau level behavior (\ref{QNM fit}) as a function of $\kappa {\bf B}$.

\noindent $\bullet$ In \cite{Bu:2018drd}  we discovered a discrete set of entirely non-dissipative and thus long-lived CMW modes emerging
in a weak magnetic field.  Present work examined CMW exactly without the weak field approximation.
We found that, depending on the magnetic field,  the discrete set extends into several continuous intervals in the $(\omega,q^2)$ space.
This effect should have a clear experimental signature worth exploring.

Real experiments involving chiral plasma, such as the one produced in heavy ion collisions or the primordial plasma in early Universe, involve strong dynamical fields
with non-homogeneous profiles.  While the study reported above was limited to constant background fields, it revealed important new anomaly-induced phenomena and
serves as a step towards development of a self-consistent  chiral MHD.  Any dynamical simulations of the latter are beyond the scope of this paper.

\appendix

\section*{Acknowledgements}

We would like to thank Umut G\"{u}rsoy and Shu Lin for useful discussions.
YB would like to thank the hospitality of the Department of Physics at Ben-Gurion University of the Negev where  this work was initiated. YB was supported by the Fundamental Research Funds for the Central Universities under grant No.122050205032 and the Natural Science Foundation of China (NSFC) under the grant No.11705037.
TD and ML were supported by the Israeli Science Foundation (ISF) grant \#1635/16 and the BSF grant  \#2014707.


\providecommand{\href}[2]{#2}\begingroup\raggedright\endgroup

\end{document}